%% file: main.tex
\documentclass[12pt,reqno]{amsart}
\usepackage{amsmath, amssymb, amsthm, graphicx, tikz, hyperref, lipsum, stylesheet, qcircuit, subcaption, algorithm, algpseudocodex, subfiles}
\usepackage[most]{tcolorbox}
\tcbuselibrary{skins}
\newtcolorbox{grayalgbox}{
    colback=gray!10,
    colframe=gray!50,
    boxrule=0.5pt,
    arc=2pt,
    left=1.7em
}
\algrenewcommand\algorithmicprocedure{\textbf{TM}}
\theoremstyle{definition}
\newtheorem{theorem}{Theorem}[section]
\newtheorem{prop}[theorem]{Proposition}
\newtheorem{lemma}[theorem]{Lemma}
\newtheorem{cor}[theorem]{Corollary}
\newtheorem{definition}[theorem]{Definition}

\title[Power of Clifford and Ising Computing]{Computational Power of Clifford and Ising Quantum Computers}

\author{Sawyer Dobson}
\address{Department of Mathematics\\ University of California\\ Santa Barbara, CA 93106}
\email{sad@math.ucsb.edu}

\author{Zhenghan Wang}
\address{Department of Mathematics\\ University of California\\ Santa Barbara, CA 93106}
\email{zhenghwa@math.ucsb.edu}

\begin{document}
\begin{abstract}
We extend the efficient classical simulation of Ising quantum computers from braidings to measurements. We investigate several different definitions of efficient Clifford quantum computation by varying the power of the circuit generator. In case the circuit generator can see the entire input, we give a detailed proof that efficient simulation of classical computation would imply the separation of complexity classes \textbf{P} and \textbf{PSPACE}. If the circuit generator sees only the input length, we prove that Clifford quantum computers decide what we call affine languages.
\end{abstract}

\maketitle
\tableofcontents

\subfile{sections/1}

\subfile{sections/2}

\subfile{sections/3}

\subfile{sections/4}

\subfile{sections/5}

\section*{Acknowledgments}
The second author is partially supported by \texttt{ARO W911NF261A263} and thanks Wim van Dam for pointing out the online discussions of Clifford simulation of classical computation.

\bibliographystyle{amsplain}
\bibliography{biblio.bib}

\end{document}

%% file: sections/1.tex
\section{Introduction}
With the validation of quantum error correction theory by multiple hardware platforms (see e.g. \cite{lukin24,google25, ms26}), quantum computing enters the engineering era from scientific research.  The standard route to the construction of a universal quantum computer consists of two steps: first to build a Clifford quantum computer, and then add magic to achieve universality.  It is well-known that a Clifford quantum computer can be efficiently simulated by a classical quantum computer (see e.g. \cite{NC10}).  But the converse is not as trivial as we might think: the efficient simulation of classical computation by Clifford quantum computers implies the difficult complexity separation that \textbf{P} is not \textbf{PSPACE}---a folklore result scattered among various online discussions.

Closely related to Clifford quantum computing is quantum computing with Majoranas in topological quantum computing (see e.g. \cite{majorana25}).  The mathematical underpinning of Majorana quantum computing is the Ising anyon model (see e.g. \cite{wangbook}).  Braidings in the Ising theory can be simulated efficiently classically as they are Clifford gates.  While it is widely believed that measurements will not add any computational power to the Majorana/Ising quantum computer, a mathematical proof is still missing, especially for the more powerful interferometry measurement \cite{FL15}.\footnote{It is not clear to the authors if the simulation of fusion measurement is explicitly in the literature as well.}

In this paper, we provide details for both results and explore the computational power of Clifford and Ising quantum computation. We aim to answer the following questions:
\begin{itemize}
    \item What is a reasonable model of a Clifford quantum computer, and what class of problems can it solve efficiently?
    \item What is a reasonable model of an Ising quantum computer, and what class of problems can it solve efficiently?
    \item How do these classes of problems compare to each other? How do they compare to the classical complexity classes $\bP$ and $\bL$?
\end{itemize}

Section \ref{sec:prelims} covers the preliminary material necessary for our later results. Section \ref{subsec:complexity} introduces the complexity class $\oplus\bL$, which is critical for our study of Clifford quantum computers. The connection between $\oplus\bL$ and Clifford computation was initially developed by Aaronson and Gottesman in \cite{AG04}, where they showed that when $\oplus \bL$ is defined as the class of problems which are log-space Turing reducible to the simulation of \texttt{CNOT} circuits, then the problem of simulating stabilizer circuits lies in $\oplus \bL$. The usual definition of $\oplus \bL$, which is based on the parity of accepting computation branches of a nondeterministic log-space Turing machine, is mentioned, but its connection with the \texttt{CNOT} circuit definition is not elucidated. Thus, we develop a novel self-contained proof that the two definitions are equivalent. The main content of the proof is showing that \texttt{CNOT} circuit is \textit{complete} for $\oplus \bL$ under log-space reductions, which is our Theorem \ref{thm:CNOT_complete}. We then show using a standard line of reasoning that this implies that the class of languages log-space reducible to the simulation of \texttt{CNOT} circuits is exactly $\oplus \bL$. We also provide a straightforward proof that $\oplus\bL\subseteq \mathbf{SPACE}\P{\P{\log n}^2}$. (Theorem \ref{thm:parityL_in_log2}, which is a recursive argument based on the proof of Savitch's theorem).

In Section \ref{subsec:clifford_qc}, we explain the basics of the Clifford group and define \textit{stabilizer circuits}, which are the kinds of circuits used for Clifford quantum computation. We delay further exploration of Clifford computation to Section \ref{sec:clif}.

In Section \ref{subsec:ising_qc}, we introduce the building blocks of topological quantum computation with Ising anyons, including anyons, unitary modular categories (UMCs), the Ising UMC, and how computational states are encoded in the category's hom spaces. We begin with a physics-based introduction to anyons, explaining how in 2 spatial dimensions, it's possible for particles to have non-abelian \textit{exchange statistics}. We then show how to model such particles as simple objects in a UMC, and we introduce the inverse operations of \textit{fusion} and \textit{splitting}, as well as what it means to \textit{braid} anyons. Then, we explain the graphical calculus for UMCs, which allows us to perform calculations involving morphisms visually, which is very useful for when calculations involve more than just a couple anyons. Finally, we give the specific example of the Ising UMC, which we focus on in the remainder of the paper. Certain hom spaces of the Ising category have particularly nice tensor decompositions, which is why it's easy to develop a theory of quantum computation using Ising anyons. Another reason for focusing on Ising anyons is that they are perhaps the non-abelian anyon that is closest to experimental realization. We show how computational states can be encoded and manipulated via braiding and measurement. Further discussion of the computational power of Ising anyons is delayed to Section \ref{sec:ising_complexity}.

Section \ref{sec:clif} tackles the question of how to reasonably model Clifford quantum computation. This question is not as straightforward as it may seem because we have to use some amount of classical computation to generate stabilizer circuits of the appropriate size. We end up with different complexity classes depending on how much power we grant the circuit generator. We either allow the circuit generator to see the entire input, or only see the input length. Independently, the circuit generator is granted arbitrary polynomial time computation, or only arbitrary logarithmic space computation. This leads us to four different complexity classes. Such a discrepancy doesn't arise when defining the class \textbf{BQP} essentially because arbitrary polynomial time quantum computation already contains all of classical polynomial time computation, but the same isn't true for Clifford computation. Aaronson and Gottesman assumed the circuit generator could see the entire input, and when limited to logarithmic space, this assumption leads to the complexity class $\oplus \bL$. When the circuit generator can only see the input length, the power of Clifford computation is dramatically reduced. We show that such circuit families can only decide \textit{affine} languages, and we handle the issue of how non-binary languages are encoded into binary to be used in a circuit. Again, this nuance doesn't arise when defining \textbf{BQP} because unrestricted quantum circuits can perform conversions between all ``reasonable'' kinds of encodings. Yet when restricted to Clifford gates, only affine encoding transformations are possible.

The connection between Clifford gates, stabilizer circuits, and affine spaces was first investigated in \cite{DD03}, and later in \cite{CGW17}. However, we are not aware of anywhere in the literature where our definition of affine languages is investigated and compared to the class of languages that uniform families of stabilizer circuits can decide.

In our attempt to capture the power of Clifford computation when the circuit generator only sees the input length, we define two new complexity classes \textbf{PUSC} and \textbf{LUSC}, which stand for \textbf{P}olynomial time \textbf{U}niform \textbf{S}tabilizer \textbf{C}ircuit and \textbf{L}ogarithmic space \textbf{U}niform \textbf{S}tabilizer \textbf{C}ircuit, respectively. We also introduce the complexity class \textbf{AFF}, the class of affine languages, and in Theorem \ref{thm:usc_classification}, we show how \textbf{PUSC} and \textbf{LUSC} are related to \textbf{AFF}. This is the first main result of the paper, and it gives a complete characterization of \textbf{PUSC}, and a characterization of \textbf{LUSC} up to separation between $\bL$ and $\oplus \bL$. We also provide several standard additional results aimed at helping differentiate which languages are affine and which aren't.

Finally, in Section \ref{sec:ising_complexity}, we investigate the power of computation via Ising anyons. We first show that braiding can be classically simulated, which is a known result. Then, we show that all forms of measurement can be simulated, including arbitrary interferometric measurements. In particular, we show that all interferometric measurements of an Ising anyonic chain in the dense qubit encoding are Pauli measurements, which is our Theorem \ref{thm:measurements_pauli} and the second main result of the paper. This gives us an analog of the Gottesman-Knill theorem for computation with Ising anyons.

%% file: sections/2.tex
\section{Preliminaries}\label{sec:prelims}
\subsection{Complexity Theory}\label{subsec:complexity}
We assume basic familiarity with computational complexity theory (\cite{Sipser13} is a standard reference). This section introduces the complexity class $\oplus \bL$, which will be particularly relevant to the study of Clifford quantum computers. We then define a simple problem, called \texttt{CNOT-CIRCUIT}, and show that it is complete for $\oplus \bL$ under log-space reductions. An efficient polynomial time algorithm for \texttt{CNOT-CIRCUIT} proves that $\oplus \bL\subseteq \textbf{P}$. We'll also show that $\oplus \bL\subseteq \textbf{SPACE}\P{(\log n)^2}$. This implies a separation between $\textbf{P}$ and $\textbf{PSPACE}$ if Clifford quantum computers can efficiently solve all problems in $\textbf{P}$, which was an approach mentioned in \cite{deBeaudrap10}. A more extended discussion lies in Section \ref{sec:clif}.

We will also need a particular closure property of $\oplus \bL$ to fully explain the connection between Clifford quantum computers and $\oplus \bL$. We elaborate on the approach taken by Aaronson and Gottesman in \cite{AG04} to show that the problem of simulating a Clifford quantum computer lies in $\oplus \bL$. The notion of an oracle Turing machine is introduced for this purpose.

\begin{definition}
Given a deciding nondeterministic Turing machine $M$, we define $\msA_M(\sigma)$ to be equal to the number of accepting computation branches of $M$ on $\sigma$. The class $\oplus \bL$ (pronounced ``parity L'') is the class of all languages \texttt{A} such that there exists a deciding nondeterministic log-space Turing machine $M$ such that, for all inputs $\sigma$, $\msA_M(\sigma)$ is odd iff $\sigma\in \texttt{A}$.
\end{definition}

Compare this definition of $\oplus \bL$ to that of $\textbf{NL}$ -- they're similar. The difference lies in the fact that the Turing machines for $\oplus \bL$ must have an odd number of accepting computation branches when the input string is in the language, whereas for $\textbf{NL}$, the Turing machines must have at least one accepting computation branch. There is no obvious way to transform Turing machines to convert the ``odd number'' condition into the ``at least one'' condition, or vice-versa, so it is unknown if $\oplus \bL\subseteq \textbf{NL}$, $\textbf{NL}\subseteq \oplus\bL$, or neither.

To get a better understanding of $\oplus \bL$ and situate it in the complexity hierarchy, it's useful to know some problems which are complete for $\oplus \bL$. We say a language \texttt{A} is \textit{complete} for $\oplus \bL$ if the following two conditions hold:
\begin{itemize}
    \item $\texttt{A}\in \oplus\bL$.
    \item $\texttt{A}$ is $\oplus \bL$-hard. In other words, for any other language $\texttt{B}\in \oplus \bL$, there exists a log-space Turing machine $M_\texttt{B}$ that outputs an instance of \texttt{A} when it receives an instance of \texttt{B} as input, and it outputs a non-instance of \texttt{A} when it receives a non-instance of \texttt{B} as input. We say that \texttt{B} is \textit{log-space mapping reducible} to \texttt{A}, or simply that \texttt{B} \textit{reduces} to \texttt{A} in log-space.
\end{itemize}

Note that the log-space requirement does not constrain the output of $M_\texttt{B}$ to $\mcO(\log n)$ many symbols. On the contrary, we assume $M_\texttt{B}$ has three different tapes -- a read-only input tape, a read-write work tape, and a write-only output tape (where the tape head can only move to the right). The only tape constrained to logarithmic space is the work tape. The reason we use log-space machines to compute the reduction from \texttt{B} to \texttt{A}, rather than poly-time machines, is because, as we'll see later, poly-time machines are too capable -- any problem in $\oplus \bL$ can be solved by a deterministic poly-time Turing machine. Thus, every language in $\oplus \bL$ is poly-time reducible to any other (non-empty, non-trivial) language.

Several examples of problems complete for $\oplus\bL$ are listed below \cite{Damm90}:
\begin{itemize}
    \item $\texttt{DET}_2$: given a matrix $A$ over $\bbF_2$, compute its determinant.
    \item $\texttt{EQUATIONS}_2$: given a matrix $A$ and vector $b$ over $\bbF_2$, determine whether or not $b$ is in the column space of $A$.
    \item $\texttt{MATPOW}_2$: given a matrix $A$ over $\bbF_2$ and integers $i,j,k$, compute $(A^k)_{ij}$.
\end{itemize}

Intuitively, if a problem is complete for $\oplus \bL$, then that problem is the ``hardest'' any problem in $\oplus \bL$ can be because a solution to that problem would yield a solution to any other problem in $\oplus \bL$ using equivalent computational resources. The $\oplus \bL$ complete problem we're most interested in is called \texttt{CNOT-CIRCUIT}, defined below. Later, we'll see that it is closely related to the problem of classically simulating Clifford quantum computers.

\begin{definition}
A \texttt{CNOT} circuit is a quantum circuit that uses only the not gate $X$ and the controlled-not gate \texttt{CNOT}. We define \texttt{CNOT-CIRCUIT} to be the collection of all descriptions of \texttt{CNOT} circuits with the property that, when all qubits are initialized to 0 and the circuit is executed, an output of 1 is obtained on the first qubit. By a description of a circuit, we mean a list of the gates that are applied along with which wires they act on, in order. The description of a circuit $C$ is denoted $\V{C}$.
\end{definition}

Even though the concept of a \texttt{CNOT} circuit is purely classical, we phrase it using quantum computing language to more easily make parallels between \texttt{CNOT} circuits and Clifford circuits, which we'll define later, and say that a \texttt{CNOT} circuit is a Clifford circuit. It's also clear that \texttt{CNOT} circuits are reversible and are treated so that the number of output bits/qubits is the same as the number of input bits/qubits.

\begin{figure}[t]
    \centering
    \includegraphics[width=\linewidth]{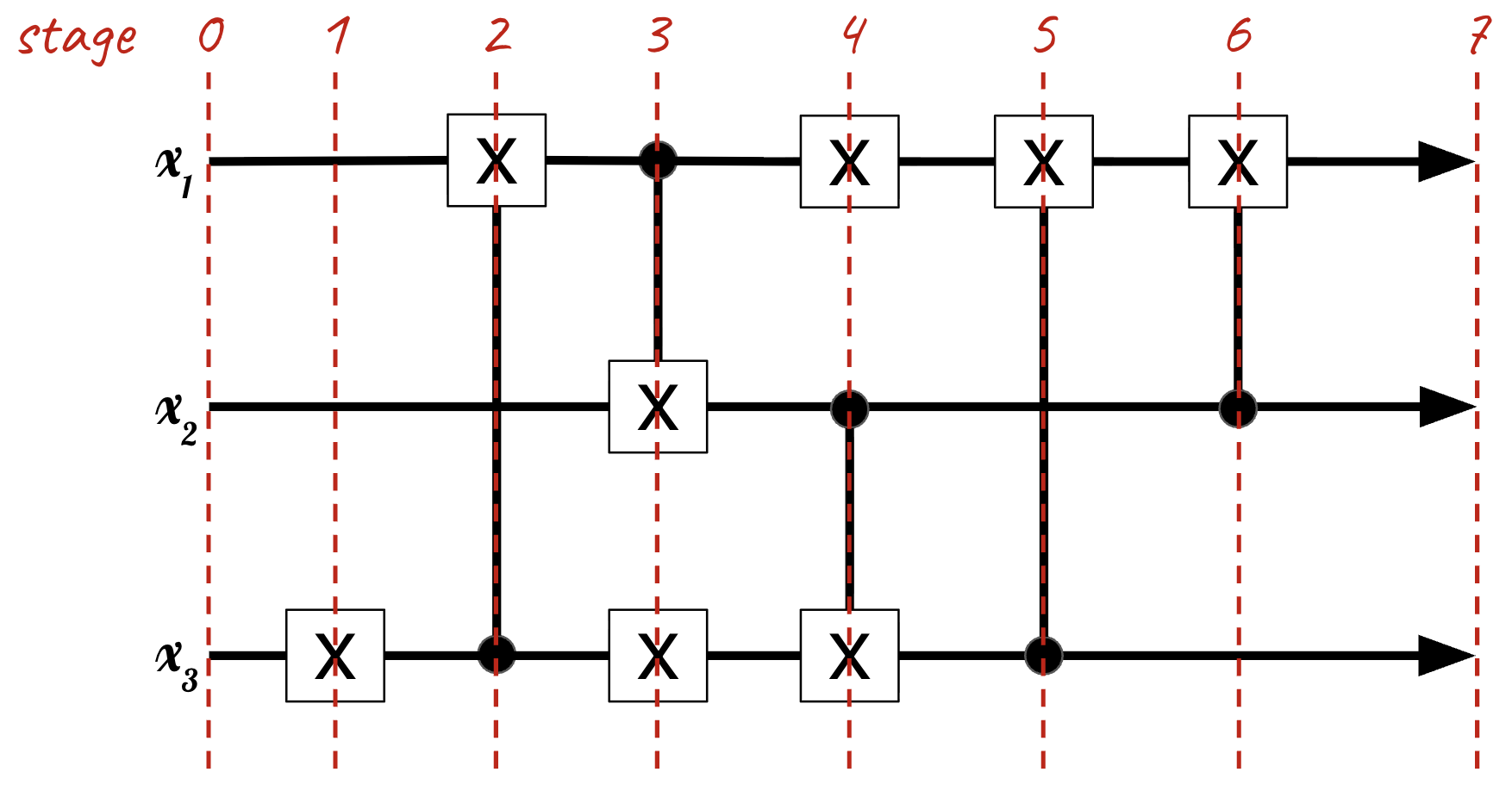}
    \caption{A \texttt{CNOT} circuit with three wires and stages indicated.}
    \label{fig:circuit_with_stages}
\end{figure}

We can think of a \texttt{CNOT} circuit as a collection of horizontal wires connected by \texttt{CNOT} gates from one wire to another, possibly with \texttt{NOT} gates along the wires, as in Figure \ref{fig:circuit_with_stages}. The symbol $\boxed{X}$ alone along a wire is a \texttt{NOT} gate, and a $\boxed{X}$ connected via a vertical line to another wire is a \texttt{CNOT} gate. The inputs of the circuit appear on the left, denoted $x_1$, \ldots, $x_n$, where $n$ is the number of wires. The outputs appear on the right, denoted by a right arrow.

Every \texttt{CNOT} circuit can be arranged into a series of \textit{stages} so that on each stage, each wire is part of a \texttt{CNOT} gate, has a \texttt{NOT} gate, or has nothing. For instance, the circuit in Figure \ref{fig:circuit_with_stages} can be arranged into stages 0 through 7. The value of each wire on stage 0 is equal to the boolean value it is given as input, and the value of each wire on the final stage is the boolean output of that wire. The value of a wire on any intermediate stage is the boolean value it has \emph{after} all gates of that stage have been applied.

Thinking about \texttt{CNOT} circuits in this way gives us a slick proof that \texttt{CNOT-CIRCUIT} is complete for $\oplus \bL$. The proof relies on the notion of a \textit{stage digraph}, which connects \texttt{CNOT} circuits to the idea of parity of a number of paths, which is how we relate it to $\oplus \bL$.

\begin{definition}\label{def:stage_digraph}
Suppose $C$ is a \texttt{CNOT} circuit with $n$ horizontal wires arranged into $m$ stages. The \emph{stage digraph} $G$ of $C$ is formed as follows. For each $i$ and $j$ such that $1\leq i\leq n$ and $0\leq j\leq m$, $G$ has a vertex labeled $s_{i,j}$. In addition, $G$ has a ``start vertex'' $s$, so in total, $G$ has $n(m+1)+1$ vertices. For all $i$ and all $j\geq 1$, $G$ has an edge from $s_{i,j-1}$ to $s_{i,j}$. For each \texttt{CNOT} gate in $C$, if that gate occurs in the $j^\text{th}$ stage from wire $i_1$ to wire $i_2$, then $G$ has an edge from $s_{i_1,j}$ to $s_{i_2,j}$. Finally, for each \texttt{NOT} gate in $C$, if that gate occurs on the $i^\text{th}$ wire in the $j^\text{th}$ stage, then $G$ has an edge from $s$ to $s_{i,j}$. We call the vertex $s_{1,m}$ the \textit{terminal vertex} $t$.
\end{definition}

\begin{figure}[t]
    \centering
    \includegraphics[width=\linewidth]{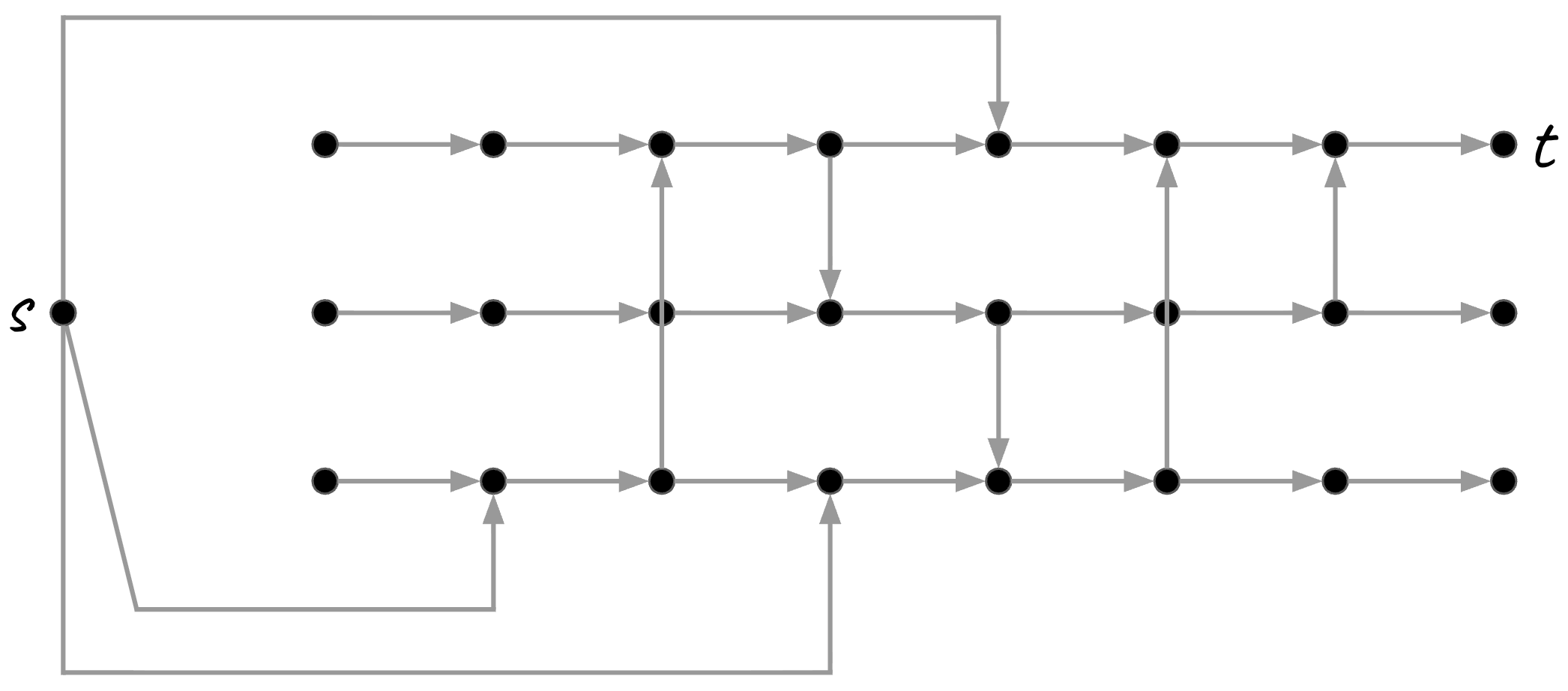}
    \caption{The stage digraph of the \texttt{CNOT} circuit in Figure \ref{fig:circuit_with_stages}.}
    \label{fig:ex_stage_digraph}
\end{figure}

Figure \ref{fig:ex_stage_digraph} gives an example of a stage digraph. There is an important lemma concerning the relationship between a circuit and its stage digraph, which we'll need in order to prove that \texttt{CNOT-CIRCUIT} is complete for $\oplus \bL$.

\begin{lemma}\label{lem:stage_digraph}
Suppose $C$ is a \texttt{CNOT} circuit with $n$ horizontal wires arranged into $m$ stages, and $G$ is the corresponding stage digraph of $C$. Then $\V{C}\in \texttt{CNOT-CIRCUIT}$ if and only if the number of paths from $s$ to $t$ in $G$ is odd.
\end{lemma}
\begin{proof}
We prove this lemma via induction. Let $v_{i,j}\in \bbF_2$ be the value of the $i^\text{th}$ wire of $C$ on the $j^\text{th}$ stage when every input wire is assigned the value 0. The inductive hypothesis is that for all $i$, $v_{i,j}=1$ iff the number of paths from $s$ to $s_{i,j}$ is odd. The base case $j=0$ holds because for all $i$, $v_{i,0}=0$ and there are no paths from $s$ to $s_{i,0}$.

For the inductive step, suppose the inductive hypothesis holds up to $j-1$, and consider what's on the $i^\text{th}$ wire of $C$ in the $j^\text{th}$ stage. There are four possibilities: either it's the control of a \texttt{CNOT} gate, it's the target of a \texttt{CNOT} gate, it's a \texttt{NOT} gate, or no gate is there.
\begin{enumerate}
    \item If it's the control of a \texttt{CNOT} gate, then $v_{i,j}=v_{i,j-1}$ because a \texttt{CNOT} gate has no effect on the control wire. Moreover, there are the same number of paths from $s$ to $s_{i,j}$ as there are from $s$ to $s_{i,j-1}$ because the only in-neighbor of $s_{i,j}$ is $s_{i,j-1}$.
    \item If it's the target of a \texttt{CNOT} gate, suppose the $(i')^\text{th}$ wire has the corresponding control. Then $v_{i,j}\equiv v_{i,j-1}+v_{i',j} \pmod 2$. Moreover, since the only in-neighbors of $s_{i,j}$ are $s_{i,j-1}$ and $s_{i',j}$, the number of paths from $s$ to $s_{i,j}$ is the sum of the number of paths from $s$ to $s_{i,j-1}$ and the number of paths from $s$ to $s_{i',j}$. Thus, the inductive hypothesis and case (1) imply $v_{i,j}=1$ iff the number of paths from $s$ to $s_{i,j}$ is odd.
    \item If it's a \texttt{NOT} gate, then $v_{i,j}=v_{i,j-1}+1$. Moreover, since the only in-neighbors of $s_{i,j}$ are $s_{i,j-1}$ and $s$, the number of paths from $s$ to $s_{i,j}$ is 1 more than the number of paths from $s$ to $s_{i,j-1}$.
    \item If no gate is present, Then $v_{i,j}=v_{i,j-1}$ and the number of paths from $s$ to $s_{i,j}$ equals the number of paths from $s$ to $s_{i,j-1}$.
\end{enumerate}
This completes the inductive step. With $i=1$ and $j=m$, we get that $v_{1,m}=1$ iff the number of paths from $s$ to $s_{1,m}=t$ is odd.
\end{proof}

Another technical lemma we'll use is the following, which tells us how we can modify a Turing machine so that it has uniform properties upon halting. It relies on the notion of a \textit{configuration} of a Turing machine. For a log-space machine $T$ on input $\sigma$, a configuration is a tuple $(q,i,j,w)$ where $q$ is a state of $T$, $i\leq \A{\sigma}$ is the index of the input tape head, $j$ is the index of the work tape head, and $w$ is the contents of the work tape. Since $T$ operates in log-space, the number of configurations of $T$ is bounded by some polynomial in the input length $\A{\sigma}$. The \textit{configuration digraph} of $T$ is the directed graph whose vertices are configurations of $T$ and whose edges represent legal transitions.

\begin{lemma}\label{lem:uniform_halting}
Suppose $T'$ is a nondeterministic log-space Turing machine that halts on every input along every computation branch. Then there exists another nondeterministic log-space machine $T$ and function $\msR_T\in \mcO(\text{poly}\,n)$ with the following properties:
\begin{itemize}
    \item $\msA_T=\msA_{T'}$.
    \item $T$ has exactly two halting configurations -- one accepting and one rejecting.
    \item On every input of length $n$, every computation branch of $T$ halts in \textit{exactly} $\msR_T(n)$ time steps, where $\msR_T(n)$ can be computed by a Turing machine in space $\mcO(\log n)$.
\end{itemize}
\end{lemma}
\begin{proof}
We design $T$ to exactly simulate $T'$, with the following changes. Since $T'$ operates in log-space and always halts, there exists a polynomial function $\msR_{T'}$ such that, on every input of length $n$, every computation branch of $T'$ halts in \textit{at most} $\msR_{T'}(n)$ steps. $T$ begins by computing $R=\msR_{T'}\P{\A{\sigma}}$, which can be done in space $\mcO(\log n)$. A ``step counter'' is initialized with the value of $R$, and then $T$ begins simulating $T'$. After every step of the simulation, $T$ decreases the step counter and keeps it adjacent to the work tape head. Once the simulation of $T'$ is complete, $T$ continues decreasing the step counter until it reaches 0. Then, $T$ clears the work tape and moves the input tape head and work tape head to the beginning of their respective tapes. Finally, $T$ enters the unique accept state or the unique reject state depending on whether or not $T'$ accepted or rejected $\sigma$ in the simulation.

Since $T$ only accepts or rejects once the work tape is cleared and the tape heads are reset, $T$ has only one accepting configuration and only one rejecting configuration. Moreover, on every input of length $n$, every computation branch of $T$ halts in \textit{exactly} $\msR_T(n)=r_1(n)+r_2(n)\msR_{T'}(n)$ time steps, where $r_1(n)$ is the time needed to compute $\msR_{T'}(n)$, and $r_2(n)$ is the logarithmic overhead due to decrementing and moving the step counter. Note that both $r_1(n)$ and $r_2(n)$ can be computed by a Turing machine in space $\mcO(\log n)$, and $\msR_{T'}$ is a polynomial. Hence, $\msR_T(n)$ can be computed in space $\mcO(\log n)$ as well. Also, $\msA_T=\msA_{T'}$ and $T$ operates in log-space.
\end{proof}

\begin{figure}[b]
    \centering
    \includegraphics[width=\linewidth]{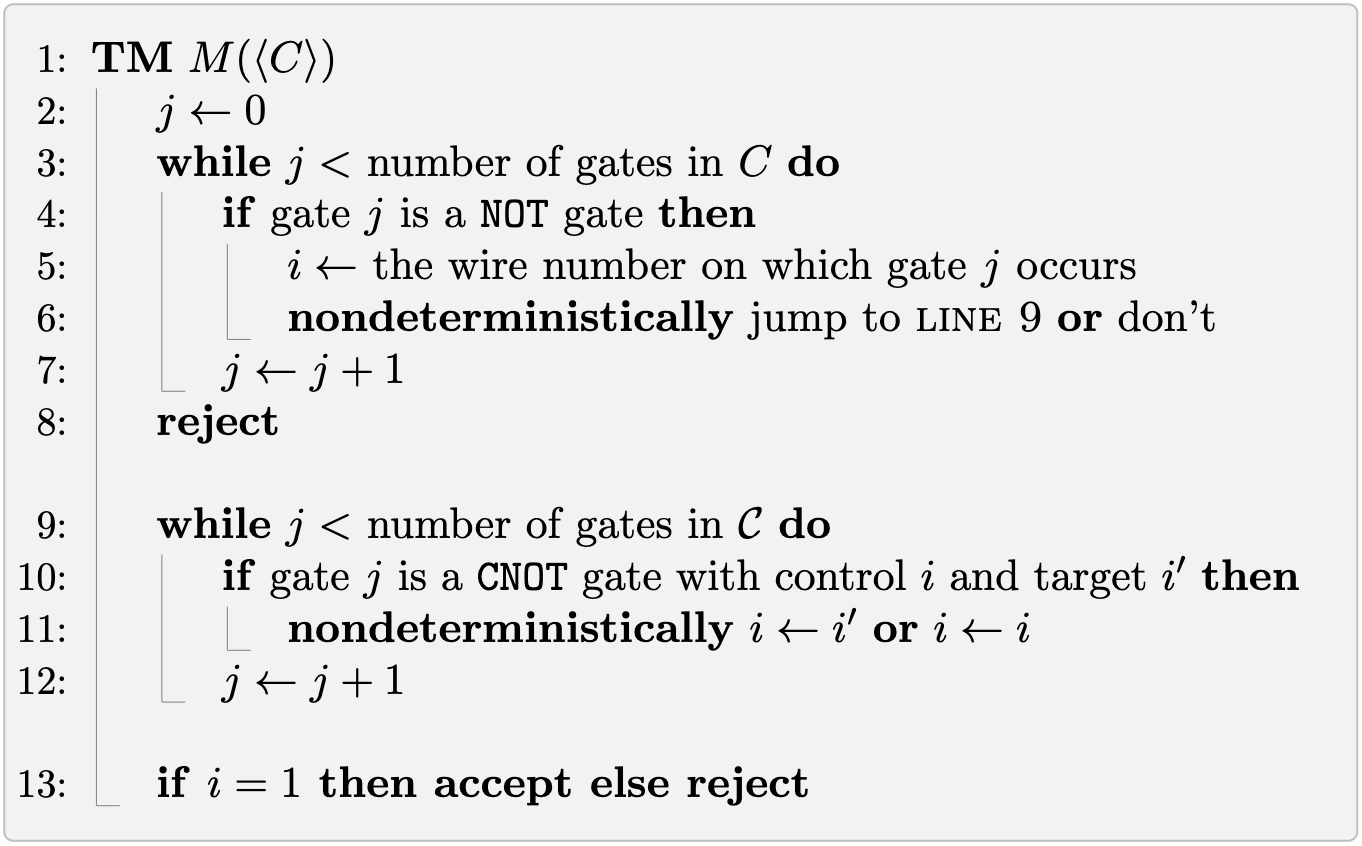}
    \caption{Pseudocode for the algorithm $M$ that traverses the stage digraph of $C$ and accepts if the path it takes ends at $t$.}
    \label{fig:stage_digraph_alg}
\end{figure}

\begin{theorem}\label{thm:CNOT_complete}
\texttt{CNOT-CIRCUIT} is complete for $\oplus \bL$ under log-space reductions.
\end{theorem}
\begin{proof}

First, let's show $\texttt{CNOT-CIRCUIT}\in \oplus \bL$. To do so, we'll design a nondeterministic log-space Turing machine $M$ such that $\msA_M(\V{C})$ is odd iff $\V{C}\in \texttt{CNOT-CIRCUIT}$. The idea is that $M$ will nondeterministically take a path starting at $s$ through a the stage digraph of $C$. $M$ accepts if the path it takes ends at $t$, and rejects otherwise. See Figure \ref{fig:stage_digraph_alg}.

Throughout the algorithm, $M$'s work tape contains only the numbers $i$ and $j$ in binary (with possibly some constant overhead), so $M$ runs in logarithmic space. Moreover, $\msA_M(\V{C})$ is the number of paths from $s$ to $t$ in $G$, so by Lemma \ref{lem:stage_digraph}, $\msA_M(\V{C})$ is odd iff $\V{C}\in \texttt{CNOT-CIRCUIT}$. Hence, $\texttt{CNOT-CIRCUIT}\in \oplus \text{L}$.

Now, we'll show \texttt{CNOT} is $\oplus \bL$-hard; that is, every language $\texttt{B}\in \oplus \bL$ reduces to \texttt{CNOT-CIRCUIT} in log-space. The overall idea is simple, but there are some technical details. Let $T$ be a nondeterministic log-space Turing machine such that $\msA_T(\sigma)$ is odd iff $\sigma\in \texttt{B}$. By Lemma \ref{lem:uniform_halting}, we can assume WLOG that $T$ has exactly one accepting configuration, and on every input of length $n$, $T$ halts in exactly $\msR_T(n)$ time steps, where $\msR_T\in \mcO(\text{poly}\,n)$ and $\msR_T(n)$ can be computed by a Turing machine in space $\mcO(\log n)$. We want to design a deterministic log-space Turing machine $M_\texttt{B}$ that, given some $\sigma$ as input, outputs a description of a \texttt{CNOT} circuit $C$ that lies in \texttt{CNOT-CIRCUIT} iff $\sigma\in \texttt{B}$. The idea is to have a wire in $C$ for each configuration of $T$ on $\sigma$, and use \texttt{CNOT} gates to represent legal transitions between configurations.

\begin{figure}[t]
    \centering
    \includegraphics[width=\linewidth]{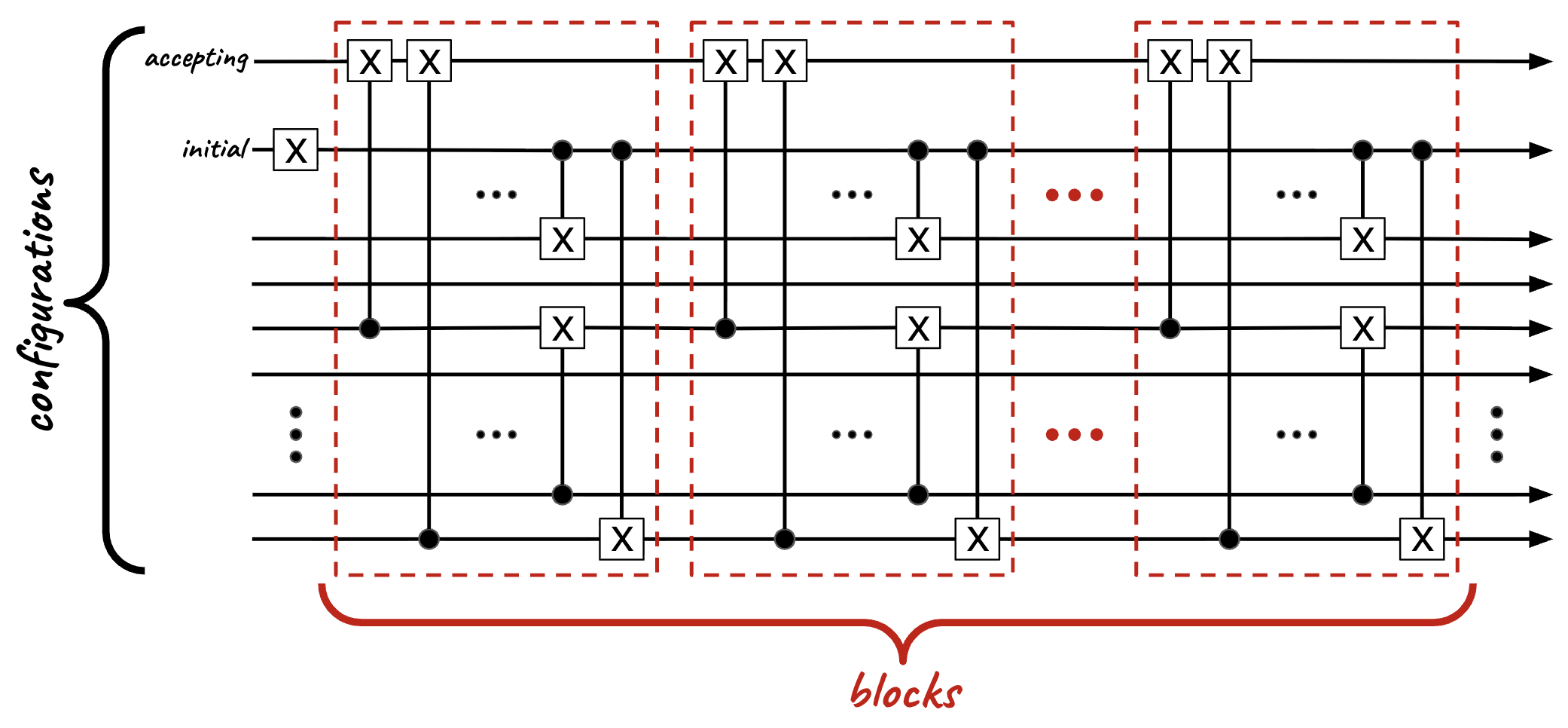}
    \caption{The \texttt{CNOT} circuit $C$ constructed from some input $\sigma$ so that $\V{C}\in \texttt{CNOT-CIRCUIT}$ iff $\sigma\in \texttt{B}$. There are exactly $\msR_T\P{\A{\sigma}}$ blocks of \texttt{CNOT} gates. Each block is control-prioritized.}
    \label{fig:CNOT_reduction}
\end{figure}

The high-level construction is shown in Figure \ref{fig:CNOT_reduction}. $T$'s unique accepting configuration is represented by the first wire, and the second wire represents $T$'s initial configuration. There is a single \texttt{NOT} gate at the beginning of the second wire. Then, there are $\msR_T\P{\A{\sigma}}$ blocks of \texttt{CNOT} gates, boxed in red. Each block has one \texttt{CNOT} gate for each legal transition between configurations that $T$ can make (which, of course, depend on $\sigma$). More specifically, if $T$ can transition from configuration $c_1$ to configuration $c_2$, there is a $\texttt{CNOT}$ gate from the wire representing $c_1$ to the wire representing $c_2$. Note that $T$ can never loop, since all computation branches are finite, so it's possible to arrange the \texttt{CNOT} gates within each block so that along each wire, all controls precede all targets. If this holds, we say the block is \textit{control-prioritized}. Each block being control-prioritized is important so that accepting computation branches of $T$ aren't double-counted.

There are two things to prove: $C$ can be described by a log-space Turing machine $M_\texttt{B}$ given input $\sigma$, and $\V{C}\in \texttt{CNOT-CIRCUIT}$ iff $\sigma\in \texttt{B}$. The latter is straightforward in view of Lemma \ref{lem:stage_digraph}. Indeed, if $G$ is a stage digraph of $C$, then there is a one-to-one correspondence between paths in $G$ from $s$ to $t$ and paths in the configuration digraph of $T$ from the initial configuration to the accepting configuration. This follows from the fact that each block of \texttt{CNOT} gates is control-prioritized, and the fact that $T$ halts in exactly $\msR_T\P{\A{\sigma}}$ time steps on every computation branch, which is the same as the number of blocks.

To see how $M_\texttt{B}$ should operate, the only difficulty is ensuring that $M_\texttt{B}$ outputs \texttt{CNOT} gates in an order that guarantees the \texttt{CNOT} blocks are control-prioritized.

Given a configuration $c$ of $T$ on $\sigma$, we define the \textit{halting distance} of $c$ to be the length of a longest path from from $c$ to any halting configuration in the configuration digraph of $T$. Note that every path from $c$ to a halting configuration has the same length because every computation branch of $T$ halts in the same number of steps. Thus, it's straightforward for $M_\texttt{B}$ to compute the halting distance of any configuration $c$ in log-space -- just follow transitions arbitrarily, keeping track of how many time steps elapse, until a halting state is reached.

With this idea in mind, we can see how to implement $M_\texttt{B}$. $M_\texttt{B}$ starts by writing a $\texttt{NOT}$ gate to wire 2 (on the output tape), then computing $R=\msR_T\P{\A{\sigma}}$. Then, $M_\texttt{B}$ writes $R$ \texttt{CNOT} blocks to the output tape in the same way, as follows. It first finds all the configurations with halting distance 1, and writes down the \texttt{CNOT} gates corresponding to all the transitions out of those configurations. Then, it does the same thing for configurations with halting distance 2, then 3, and so on up to $R$. Writing down the \texttt{CNOT} gates in order of smallest halting distance to largest ensures that all controls precede all targets in each block, so the blocks are control-prioritized.

Lastly, note that $M_\texttt{B}$ uses only $\mcO(\log n)$ cells on its work tape, since it only has to store variables representing the current block index, the current halting distance, and the current configuration (plus some logarithmic overhead to compute the halting distance of that configuration). Thus, \texttt{B} is log-space reducible to \texttt{CNOT-CIRCUIT}.
\end{proof}

Note that there's a polynomial time deterministic algorithm that decides \texttt{CNOT-CIRCUIT} -- simply keep track of the bit values on every wire and apply the gates of the circuit sequentially. Hence, as a corollary of Theorem \ref{thm:CNOT_complete} and the fact that all log-space reductions are necessarily poly-time reductions, we get the following.

\begin{cor}\label{cor:parityL_in_P}
$\oplus\bL \subseteq \textbf{P}$.
\end{cor}

This gives a bound on $\oplus \bL$ in terms of time complexity, but it's possible to do something similar in terms of space complexity. The following bound on $\oplus \bL$ in terms of deterministic space complexity will be relevant later when we discuss a possible approach to separating $\textbf{P}$ and $\textbf{PSPACE}$. The proof follows a similar argument to the proof of Savitch's Theorem in \cite[Chapter 8.1]{Sipser13}.

\begin{theorem}\label{thm:parityL_in_log2}
$\oplus \bL\subseteq \textbf{SPACE}\P{\P{\log n}^2}$.
\end{theorem}
\begin{proof}
Let $\texttt{B}\in \oplus \bL$ and $T$ be a nondeterministic log-space Turing machine such that $\msA_T(\sigma)$ is odd iff $\sigma\in \texttt{B}$. By Lemma \ref{lem:uniform_halting}, we can assume WLOG that $T$ has exactly one accepting configuration on a given input, and that on every input of length $n$, every computation branch of $T$ halts in \textit{exactly} $\msR_T(n)$ time steps, where $\msR_T\in \mcO(\text{poly}\,n)$ and $\msR_T(n)$ can be computed by a Turing machine in space $\mcO(\log n)$. We design a deterministic machine $M$ operating in space $\mcO\P{\P{\log n}^2}$ that decides \texttt{B}.

$M$'s task is to analyze the execution of $T$ on $\sigma$ and determine whether an even or odd number of computation branches accept. The key insight that makes this possible in space $\mcO\P{\P{\log n}^2}$ is the realization that the quantity $\rho(c_1,c_2,r)$, which we define as the parity of the number of ways $T$ can get from a configuration $c_1$ to a configuration $c_2$ in exactly $r$ time steps, can be computed recursively. Note that
\begin{equation}\label{eq:parity_comp}
\rho(c_1,c_2,r)\equiv \sum_{c_3}\rho\P{c_1,c_3,\F{\frac{r}{2}}} \rho\P{c_3,c_2,\C{\frac{r}{2}}} \pmod 2
\end{equation}
where the sum is over all configurations $c_3$ of $T$ on $\sigma$. The quantity $M$ must compute is $\rho\P{c_I(\sigma),c_A(\sigma),\msR_T\P{\A{\sigma}}}$, where $c_I(\sigma)$ is the initial configuration of $T$ on $\sigma$ and $c_A(\sigma)$ is the accepting configuration of $T$ on $\sigma$. The pseudocode in Figure \ref{fig:savitch_alg} describes $M$'s execution in detail. We let $\msC_T(\sigma)$ be the set of configurations of $T$ on $\sigma$.

\begin{figure}[t]
    \centering
    \includegraphics[width=\linewidth]{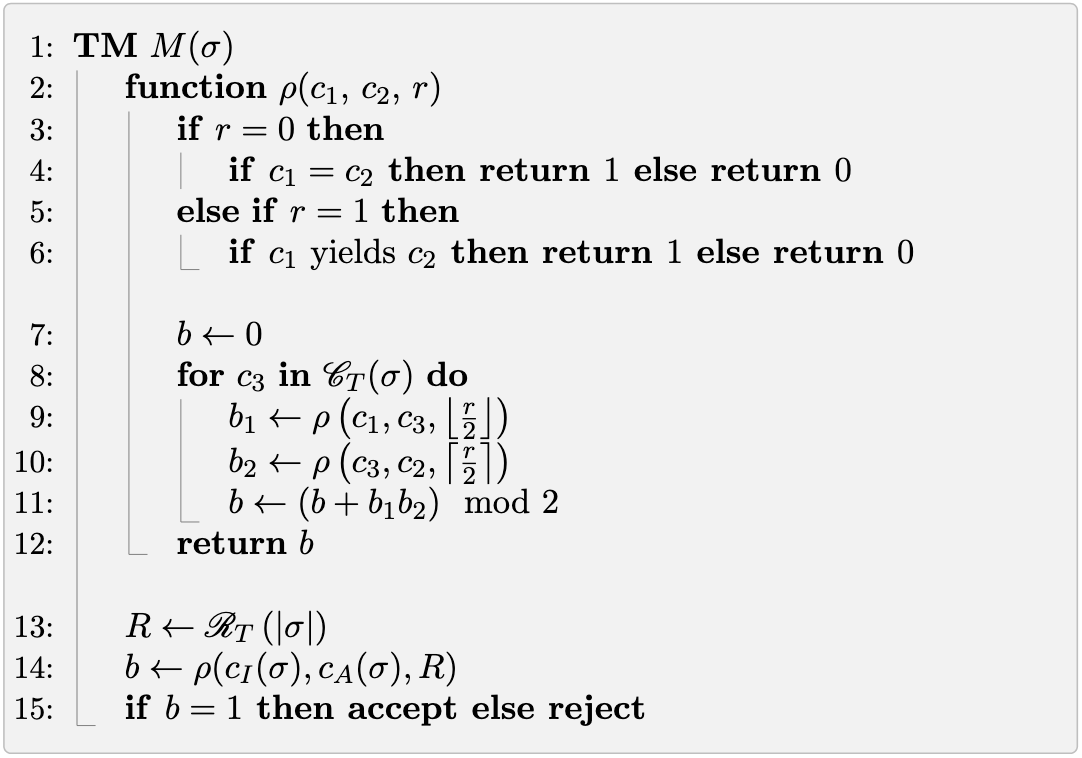}
    \caption{Pseudocode for a deterministic machine $M$ that decides \texttt{B}.}
    \label{fig:savitch_alg}
\end{figure}

In light of Equation (\ref{eq:parity_comp}), it's easy to see that $M$ correctly decides \texttt{B}. To analyze $M$'s space complexity, note that on an input $\sigma$ of length $n$, $M$ computes $R=\msR_T\P{\A{\sigma}}$ in space $\mcO(\log n)$ and then executes a recursive function $\rho$ which has a recursion depth of $\mcO\P{\log_2 R}=\mcO(\log n)$. Each recursive call to $\rho$ requires storing two configurations, an integer $r\leq R$, and some constant overhead, which totals $\mcO(\log n)$ space. Thus, the total space used by $M$ on its work tape is $\mcO\P{\P{\log n}^2}$.
\end{proof}

We conclude this section by introducing oracle Turing machines, which are necessary for defining relativized complexity classes. This will become relevant later in Section \ref{sec:clif}, where we'll need to know that there's a way to convert an algorithm that relies on solving multiple instances of \texttt{CNOT-CIRCUIT} into an algorithm that only needs to solve a single instance. The idea is to give Turing machines an extra capability allowing them to instantaneously solve a particular problem.

\begin{definition}
Let \texttt{A} be a language. An \textit{\texttt{A}-oracle Turing machine} is a Turing machine with an extra read/write tape called the oracle tape and three special states -- the oracle query state, the oracle accept state, and the oracle reject state. Whenever the oracle query state is entered, the machine transitions to the oracle accept state if the current string written on the oracle tape lies in \texttt{A}, and it transitions to the oracle reject state otherwise. This whole procedure occurs in a single time-step, and the tape heads do not move. Sometimes, an \texttt{A}-oracle Turing machine is simply called an oracle Turing machine if the language \texttt{A} is clear from context or irrelevant. A log-space oracle Turing machine has four tapes: the input tape, work tape, output tape, and oracle tape. The oracle tape has no space constraint, but it is write-only, and the oracle tape is magically cleared right after a query.
\end{definition}

If a log-space \texttt{A}-oracle Turing machine decides a particular problem \texttt{B}, then we say \texttt{B} is \textit{log-space Turing reducible} to \texttt{A}. The class of all languages \texttt{B} which are log-space Turing reducible to \texttt{A} is denoted $\bL^\texttt{A}$. If $\mcC$ is a class of languages, then $\bL^\mcC$ is the union over all classes $\bL^\texttt{B}$ with $\texttt{B}\in \mcC$. We will particularly be interested in the class $\bL^{\oplus \bL}$, which, by the following proposition, is equal to $\bL^\texttt{B}$ for any language \texttt{B} complete for $\oplus \bL$.

\begin{prop}\label{prop:recomp}
Suppose \texttt{B} is complete for $\oplus \bL$ under log-space reductions. Then $\bL^{\oplus \bL}=\bL^\texttt{B}$.
\end{prop}
\begin{proof}
By definition, $\bL^\texttt{B}\subseteq \bL^{\oplus \bL}$. For the reverse inclusion, let $\texttt{A}\in \oplus \bL$, and suppose $\texttt{C}\in \bL^\texttt{A}$. Hence, there exists a log-space \texttt{A}-oracle Turing machine $\Theta$ that decides \texttt{C}. Since \texttt{B} is complete for $\oplus\bL$ under log-space mapping reductions, there exists a log-space Turing machine $M$ that outputs an element of \texttt{B} iff its input is an element of \texttt{A}. We design a \texttt{B}-oracle Turing machine $\Lambda$ deciding \texttt{C}.

A naive first approach is to have $\Lambda$, on input $\sigma$, simulate $\Theta$, but whenever a query is made, $\Lambda$ first converts the string on the oracle tape to a different string by simulating $M$. This would work, except for the fact that the oracle tape is write-only and the string it contains might be longer than $\mcO\P{\log \A{\sigma}}$. Unfortunately, we can't write the query on the work tape instead.

Instead, we use a technique called \textit{recomputation}. $\Lambda$ simulates $\Theta$ as usual, but ignores any writes to the oracle tape. When $\Theta$ makes a query, $\Lambda$ simulates $M$ to get the query string. But how does $M$ access its input if we haven't printed it anywhere? The key is to recompute the individual symbols of the desired input of $M$ as they're accessed. For instance, suppose $M$ accesses the third symbol of its desired input. $\Lambda$ repeats the simulation of $\Theta$ starting from the pervious oracle query until $\Theta$ prints the third symbol to the oracle tape. $\Lambda$ then uses that symbol in the simulation of $M$. Anything $M$ writes to its output tape, $\Lambda$ writes to its oracle tape. Finally, when the simulation of $M$ is complete, $\Lambda$ makes a query and then continues with the simulation of $\Theta$.

By the definition of a log-space mapping reduction, the results of the queries that $\Theta$ obtains from its \texttt{A}-oracle are the same as the results that $\Lambda$ obtains from its \texttt{B}-oracle. Hence, $\Lambda$ decides the same language that $\Theta$ does, which is \texttt{C}. Therefore, $\texttt{C}\in \bL^\texttt{B}$.
\end{proof}

\begin{cor}\label{cor:CNOT_complete}
$\bL^{\oplus \bL}=\bL^\texttt{CNOT-CIRCUIT}$.
\end{cor}
\begin{proof}
Immediate from Theorem \ref{thm:CNOT_complete} and Proposition \ref{prop:recomp}.
\end{proof}

In fact, a result by Hertrampf, Reith, and Vollmer (\cite{HRV00}) says that $\bL^{\oplus \bL}$ is simply $\oplus \bL$.

\begin{theorem}\label{thm:parityL_closure}
$\bL^{\oplus \bL}=\oplus \bL$.
\end{theorem}

One way of interpreting Theorem \ref{thm:parityL_closure} is that a language is log-space Turing reducible to \texttt{CNOT-CIRCUIT} if and only if that language is log-space mapping reducible to \texttt{CNOT-CIRCUIT}. This is true because, by Corollary \ref{cor:CNOT_complete}, $\bL^{\oplus \bL}$ is exactly the collection of problems log-space Turing reducible to \texttt{CNOT-CIRCUIT}, and by Theorem \ref{thm:CNOT_complete}, $\oplus \bL$ is exactly the collection of problems log-space mapping reducible to \texttt{CNOT-CIRCUIT}. Thus, any language that can be decided by a log-space \texttt{CNOT-CIRCUIT}-oracle Turing machine that makes multiple oracle queries can also be decided by a log-space \texttt{CNOT-CIRCUIT}-oracle Turing machine that only makes a single query at the end of its computation and directly outputs the query result.

\subsection{Clifford Quantum Computers}\label{subsec:clifford_qc}
We assume familiarity with basic quantum computing principles, for which \cite{NC10} is a standard reference. In this section, we introduce the Clifford quantum computer, which is a quantum circuit that uses a certain limited class of gates called \textit{Clifford gates}. These gates generate what's known as the \textit{Clifford group}, which was first studied in the context of building quantum error correction codes according to the stabilizer formalism (see \cite{Gottesman97} and \cite{CRSS97}). For this reason, Clifford circuits are also called \textit{stabilizer circuits}.

The $n$-qubit Clifford group $\mcC_n$ is most easily defined as the normalizer of the $n$-qubit Pauli group $\mcP_n$ (the group generated by $n$-fold tensor products of the gates $I,X,Y,Z\in \mcU(2)$) inside $\mcU\P{2^n}$. Since $\mcP_n$ includes a choice of phase $\S{\pm 1, \pm i}$, it has order $4^{n+1}$. It is well-known that, up to global phase, $\mcC_n$ is generated by tensor products of the Hadamard gate $H$, the quarter phase gate $S$, and the controlled-not gate \texttt{CNOT}, shown below.
\begin{equation*}
H=\frac{1}{\sqrt{2}}\begin{bmatrix}1 & 1\\1 & -1\end{bmatrix},\qquad
S=\begin{bmatrix}1 & 0\\0 & i\end{bmatrix},\qquad
\texttt{CNOT}=\begin{bmatrix}1 & 0 & 0 & 0\\0 & 1 & 0 & 0\\0 & 0 & 0 & 1\\0 & 0 & 1 & 0\end{bmatrix}
\end{equation*}
For a proof that these gates generate the Clifford group, see \cite{Renes24}. We thus obtain the definition of a stabilizer circuit.

\begin{definition}
A \textit{Clifford circuit}, also known as a \textit{stabilizer circuit}, is a quantum circuit that uses only the Hadamard gate $H$, the quarter-phase gate $S$, and the controlled-not gate CNOT.
\end{definition}

The reason why the word ``stabilizer'' is used comes from the following theorem, which characterizes the quantum states that can be attained at the output of a Clifford circuit when initialized with a computational basis state. We say that an operator ``stabilizes'' a state if the state lies in the 1-eigenspace of the operator.

\begin{theorem}[\cite{AG04}]\label{thm:stabilizer_theorem}
Given any $n$-qubit state $\ket{\psi}$, the following are equivalent:
\begin{itemize}
    \item[(i)] $\ket{\psi}$ can be obtained from $\ket{0}^{\otimes n}$ by $H$, $S$, and CNOT gates only.
    \item[(ii)] $\ket{\psi}$ can be obtained from $\ket{0}^{\otimes n}$ by $H$, $S$, CNOT, and measurement gates only.
    \item[(iii)] $\ket{\psi}$ is stabilized by all operators in a group generated by $n$ independent Pauli operators.
    \item[(iv)] $\ket{\psi}$ is uniquely determined (up to phase) by the group of Pauli operators that stabilize $\ket{\psi}$.
\end{itemize}
If these conditions hold, $\ket{\psi}$ is called a \textit{stabilizer state}, and the group of Pauli operators stabilizing $\ket{\psi}$ is called the stabilizer of $\ket{\psi}$.
\end{theorem}

We want to answer the question: What is the class of problems that Clifford quantum computers can solve efficiently? There are several reasonable answers to this question, which we'll explore in Section \ref{sec:clif}.

Some facts about stabilizer states and stabilizer groups are useful to know when analyzing Clifford circuits. For instance, it's easy to show that the stabilizer of a stabilizer state is always an Abelian group that doesn't contain $-I$. We can generalize the notion and call any Abelian subgroup of $\mcP_n$ that doesn't contain $-I$ a \textit{stabilizer group}. Stabilizer groups always have order a power of two. A stabilizer group which acts on $n$-qubits and has order $2^n$ is called a \textit{maximal stabilizer}, because no strictly larger stabilizer group contains it. A maximal stabilizer is generated by $n$ independent Pauli operators and uniquely determines (up to phase) a stabilizer state $\ket{\psi}$ that is stabilized by every operator in the group.

Another important fact, which we'll use later, is the following theorem (see \cite{Gottesman26}).

\begin{theorem}\label{thm:logical_ops}
Suppose $\mcS\subseteq \mcP_n$ is a stabilizer of order $2^r$. Then
\begin{itemize}
    \item The centralizer and normalizer of $\mcS$ in $\mcP_n$ coincide. This group is denoted $N(\mcS)$.
    \item The quotient group $N(\mcS)/\mcS$ is isomorphic to $\mcP_{n-r}$.
\end{itemize}
\end{theorem}

\begin{cor}\label{cor:maximal_stabilizer_normalizer}
Suppose $\mcS$ is a maximal stabilizer and $A\in N(\mcS)$. Then exactly one of $A$, $iA$, $-A$, and $-iA$ lies in $\mcS$.
\end{cor}
\begin{proof}
Note that $\mcS$, $i\mcS$, $-\mcS$, and $-i\mcS$ are four distinct cosets of $\mcS$ in $N(\mcS)$. These must be all the elements of $N(\mcS)/\mcS$ because $\A{\mcS}=2^n$, so Theorem \ref{thm:logical_ops} implies that
$$\A{N(\mcS)/\mcS}= \A{\mcP_0}=4.$$
Hence, every $A\in N(\mcS)$ lies in exactly one of $\mcS$, $i\mcS$, $-\mcS$, and $-i\mcS$.
\end{proof}

\subsection{Ising Quantum Computers}\label{subsec:ising_qc}
In this section, we introduce the Ising quantum computer, which is a topological quantum computer that uses Ising anyons. We give a brief review of topological quantum computation here; for a more in-depth overview, see \cite{NSSFD08} for a physical treatment and \cite{RW18} and \cite{Wang10} for a mathematical one. We assume the reader is familiar with the basic principles of quantum mechanics.

We begin by introducing the physics of \textit{anyons} -- the physical phenomenon that provides the Hilbert space in which computation is performed. Then, we see how anyonic systems can be represented mathematically by \textit{unitary modular categories} (UMCs), and we introduce the Ising category. Finally, we'll show how computational states are encoded in the Ising category in two different ways, the \textit{sparse encoding} and the \textit{dense encoding}. In Section \ref{sec:ising_complexity}, we'll investigate the computational power of such Ising quantum computers in detail. We'll show that the sparse and dense encoding are equivalent in computational power, and that Clifford quantum computers can simulate Ising quantum computers.

\subsubsection{Anyons}\label{subsubsec:anyons}
In theory, a topological quantum computer encodes information in the degenerate ground state space of a collection of quasiparticles called anyons. The property that makes anyons useful for quantum computation, namely their non-trivial exchange statistics, was first described by Leinaas and Myrheim in \cite{LM77}, and the term ``anyon'' was coined by Wilczek in \cite{Wilczek82}. Later, the concept of non-abelian anyons was introduced (see \cite{MR91} and \cite{Wen91}), which are anyons with high dimensional exchange statistics, and they can be used for computation within Hilbert spaces of exponentially large dimension.

The exchange statistics of a particle is one of its fundamental quantum mechanical properties. The concept comes from studying the symmetry of a wave function $\ket{\psi}$ in some Hilbert space $\mcH$ describing the state of $n$ identical particles. When the particles are exchanged according to some permutation $\sigma\in S_n$ (along continuous paths in space), $\ket{\psi}$ evolves to some new state $\rho(\sigma)\ket{\psi}$, where $\rho(\sigma)$ is a unitary operator on $\mcH$. The exact paths the particles take during the exchange is inconsequential, since in 3 spatial dimensions, any two collections of paths inducing the same permutation can be continuously transformed into each other without any particle collisions.\footnote{More formally, this is the statement that the $n$ particle configuration space $\text{Conf}_n\P{\bbR^3}=\S{(x_1,\ldots,x_n)\in \P{\bbR^3}^n\,\vert\,x_i\text{ all distinct}}$ is simply-connected. See \cite{FN62}.} Hence, $\rho(\sigma)$ is well-defined, and we get a representation $\rho:S_n\to \mcU(\mcH)$, called the statistics of the particles.

If $\text{dim }\rho=1$, the particles are said to obey abelian statistics, and if $\text{dim }\rho>1$, then the particles obey \textit{parastatistics}. However, it is known that, in 3 dimensions, particles obeying parastatistics are physically equivalent to particles obeying abelian statistics with additional, unobservable internal degrees of freedom \cite{DHR70}. Thus, it is enough to consider the case $\text{dim }\rho=1$. There are only two one-dimensional representations of $S_n$ -- the trivial representation (corresponding to bosons), and the sign representation (corresponding to fermions). So, at least in 3 dimensions, particle statistics are straightforward.

\begin{figure}[b]
    \centering
    \includegraphics[width=\linewidth]{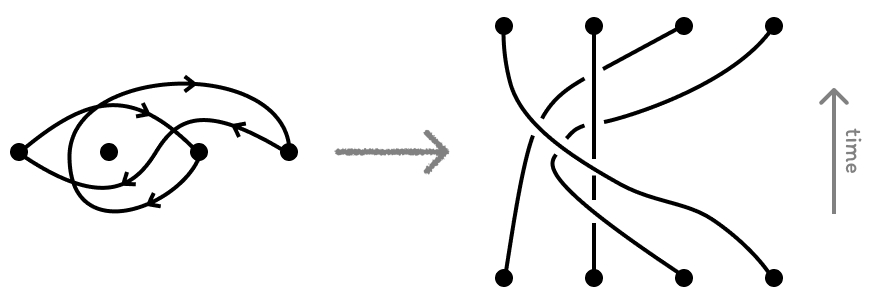}
    \caption{Exchange of particles along paths in 2 spatial dimensions becomes an element of the braid group in (2+1) dimensional spacetime.}
    \label{fig:exchange_to_braiding}
\end{figure}

Things are much more interesting in 2 spatial dimensions, where the reduction from parastatistics to abelian statistics is not possible. Moreover, it's no longer the case that the choice of path is inconsequential.\footnote{The fundamental group of the 2 dimensional configuration space $\text{Conf}_n\P{\bbR^2}=\S{(x_1,\ldots,x_n)\in \P{\bbR^2}^n\,\vert\,x_i\text{ all distinct}}$ is the pure braid group on $n$ strands \cite{Birman74}. This space covers the unordered configuration space $\text{UConf}_n\P{\bbR^2}:=\text{Conf}_n\P{\bbR^2}/S_n$ with deck transformation group $S_n$, so
$$\pi_1\P{\text{UConf}_n\P{\bbR^2}}/\pi_1\P{\text{Conf}_n\P{\bbR^2}}\cong S_n$$
It follows that $\pi_1\P{\text{UConf}_n\P{\bbR^2}}$ is the $n$ strand braid group.} A spacetime diagram of the paths taken by the particles is an element of the $n$ strand braid group $B_n$ (see Figure \ref{fig:exchange_to_braiding}). Two collections of paths corresponding to distinct elements of $B_n$ cannot be continuously transformed into each other without their braids intersecting in spacetime, representing a particle collision, so we can't conclude that different braids lead to the same evolution in the wave function. Thus, for each $\sigma\in B_n$, we get potentially distinct operators $\rho(\sigma)$ that send $\ket{\psi}$ to $\rho(\sigma)\ket{\psi}$ when the particles are braided according to $\sigma$. This is a unitary representation $\rho:B_n\to \mcU\P{\mcH}$ of $B_n$. These hypothetical particles living in 2 dimensions whose statistics are unitary representations of $B_n$ are called \textit{anyons}. If $\text{dim }\rho=1$, they are \textit{abelian anyons}, and if $\text{dim }\rho>1$, they are \textit{non-abelian anyons}.

While all of this is sound in theory, experimentally demonstrating particles in 2 dimensions with non-trivial braiding statistics is difficult. Even when confined to 2 dimensions, electrons, protons, neutrons, and photons are all either bosons or fermions. However, it's possible for many such particles to give rise to anyons as local excitations or disturbances of their collective ground state wave function, for instance, in fractional quantum Hall states (\cite{Stern08}, \cite{MR91}, \cite{Wen91}, \cite{Laughlin83}, \cite{Halperin84}, \cite{ASW84}). These excitations are called \textit{quasiparticles}, and it's possible for quasiparticles to exhibit non-trivial braiding statistics. Nevertheless, non-abelian anyons have remained experimentally elusive.

Beyond braiding, there is another operation we can perform with anyons called \textit{fusion}. Fusion occurs when two distant anyons are brought nearby one another so they can be treated as a single anyon. However, the anyon resulting from the fusion is not, in general, uniquely determined. In the next section, we'll model this situation mathematically. The trivial anyon, denoted $\mathbf{1}$, represents the vacuum, or the absence of any quasiparticle. Any other anyon is unchanged under fusion with $\mathbf{1}$.

\subsubsection{Unitary Modular Categories}
A \textit{unitary modular category} (UMC) is the mathematical framework we'll use to describe anyonic systems. In this framework, physical states are represented by morphisms in the category's hom spaces. Since the full definition of a UMC is quite involved, we'll only cover the aspects of that we need specifically; for a full discussion on UMCs and their variants, we direct the reader to \cite{Wang10}, \cite{RW18}, \cite{Turaev16}, and \cite{EGNO15}. We assume the reader is familiar with the basics of category theory \cite{MacLane98}.

At a foundational level, a UMC is a \textit{fusion category}, and anyons are \textit{simple objects} in the category. A fusion category $\mcC$ is $\bbC$-linear, meaning all hom spaces are finite-dimensional vector spaces over $\bbC$, and the simple objects in $\mcC$ are exactly those objects whose space of endomorphisms is one-dimensional. In other words, $c\in \mcC$ is simple if $\Hom\P{c,c}\cong \bbC$. These simple objects, or anyons, are the backbone for the entire category in the sense that every object in $\mcC$ must be a finite direct sum of simple objects, i.e. $\mcC$ is \textit{semisimple}. We also require that there are only finitely many types of anyons, which means there must be finitely many isomorphism classes of simple objects.

In light of our earlier discussion about anyons, we know there are two primary physical operations we can do with them: braiding and fusion. In a fusion category, fusion of anyons is modeled by a bifunctor $\otimes: \mcC\times \mcC\to \mcC$. Since the category is semisimple, for any two objects $a,b\in\mcC$, we have
\begin{equation}\label{eq:fusion_rule}
a\otimes b\cong \bigoplus_{c\in L} N^{a,b}_c c
\end{equation}
where $L$ is a finite set of representatives of all the simple objects in $\mcC$, called a \textit{label set}. The numbers $N^{a,b}_c$ are nonnegative integers, equal to the dimension of $\Hom\P{a\otimes b,c}$ as a vector space over $\bbC$. If $N^{a,b}_c\in\S{0,1}$ for all $a,b,c$, then $\mcC$ is called \textit{multiplicity-free}. Moving forward, we will assume all fusion categories are multiplicity-free. If $N^{a,b}_c=1$, then the triple $(a,b,c)$ is called \textit{admissible}. Equation (\ref{eq:fusion_rule}) is called a fusion rule.

The trivial anyon $\mathbf{1}$ satisfies the trivial fusion rules $\mathbf{1}\otimes a\cong a\cong a\otimes\mathbf{1}$ for any $a\in\mcC$, so we'll refrain from writing them out. We also require that $\mathbf{1}$ is simple, i.e. $\Hom\P{\mathbf{1},\mathbf{1}}\cong \bbC$.

In addition to being a fusion category, a UMC is unitary, which means that, among other conditions, its hom spaces come with an inner product, making them Hilbert spaces. This condition allow us to treat suitable hom spaces as physical state spaces. In practice, anyonic systems can be created by starting with a single anyon, which is the \textit{topological charge} of the system, and \textit{splitting} that anyon to obtain more, which is the inverse process to fusion. For example, suppose we start with an anyon $a$ and split it twice to end up with three $a$ anyons. The Hilbert space of the system would be modeled by the hom space $V=\Hom(a\otimes a\otimes a, a)$. We can obtain a basis for this Hilbert space in two different ways, which correspond to the two different ways of performing the fusion -- either fuse the first two anyons, then fuse the result with the third, or fuse the last two anyons, then fuse the result with the first. For example, let's assume $a\otimes a\cong \bone\oplus a$. First, for each admissible triple $(a,b,c)$, pick a unit-length basis vector $e^{a,b}_c\in \Hom(a\otimes b,c)$ (only one is necessary since $N^{a,a}_\bone=1$). Then
\begin{equation}\label{eq:left_fusion}
\S{e^{\bone,a}_a\circ \P{e^{a, a}_\bone\otimes \text{Id}_a}, e^{a, a}_a\circ \P{e^{a,a}_a\otimes \text{Id}_a}}
\end{equation}
is the basis for $V$ corresponding to fusing the first two $a$'s first, and
\begin{equation}\label{eq:right_fusion}
\S{e^{a,\bone}_a\circ \P{\text{Id}_a\otimes e^{a, a}_\bone}, e^{a, a}_a\circ \P{\text{Id}_a\otimes e^{a,a}_a}}
\end{equation}
is the basis corresponding to fusing the last two $a$'s first. There is a unitary change of basis matrix, denoted $F^{aaa}_a$, called an $F$-matrix, that converts between these two bases, and the entries of the matrix are called $F$-symbols. This will become more clear later when we discuss the graphical calculus for UMCs.

\begin{figure}[b]
    \centering
    \begin{subfigure}{0.45\textwidth}
        \centering
        \includegraphics[width=\textwidth]{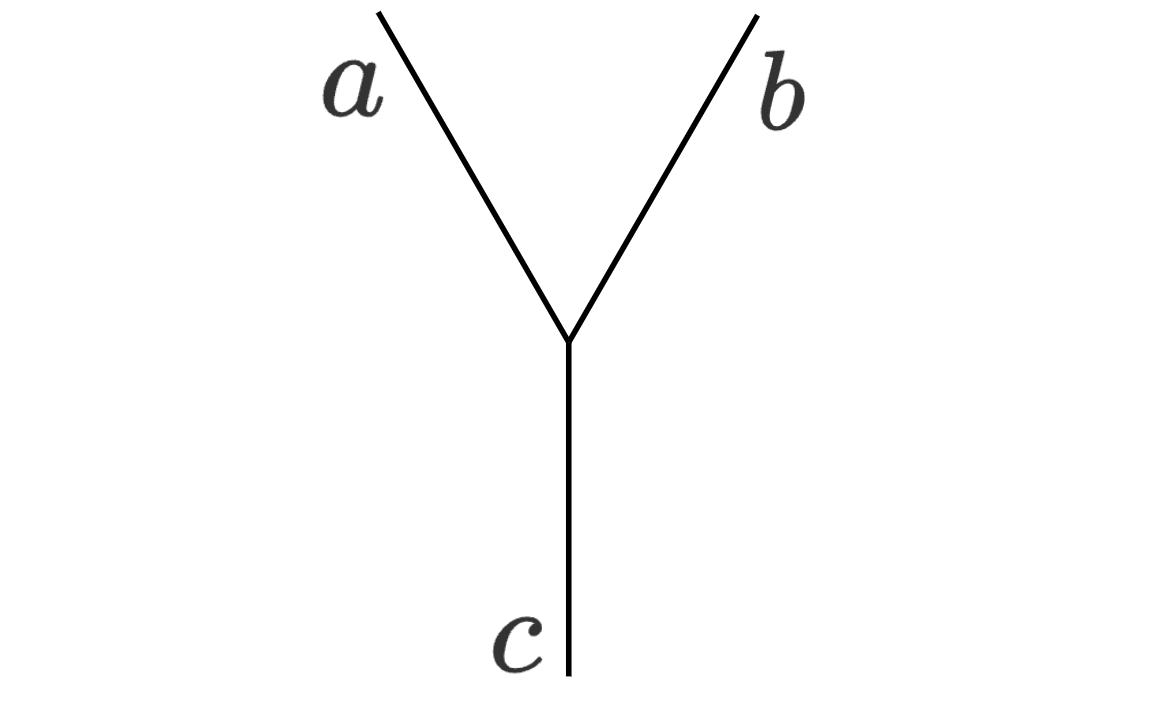}
        \caption{Diagrammatic representation of $e^{a,b}_c\in \Hom(a\otimes b,c)$.}
        \label{fig:e_ab}
    \end{subfigure}
    \hfill
    \begin{subfigure}{0.45\textwidth}
        \centering
        \includegraphics[width=\textwidth]{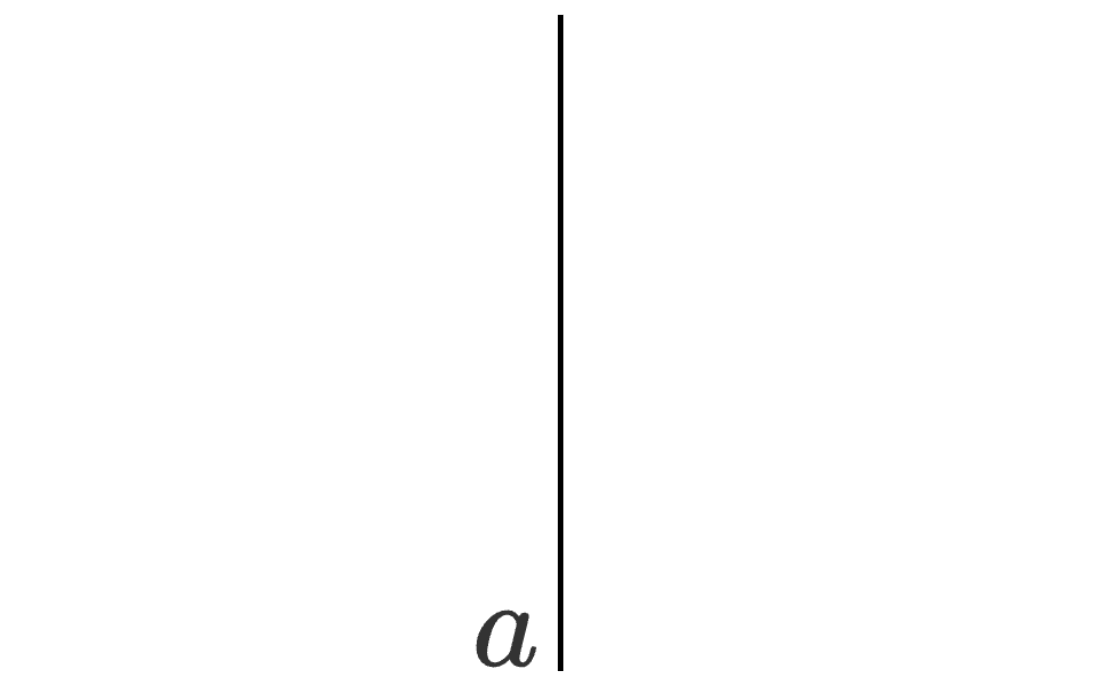}
        \caption{Diagrammatic representation of $\text{Id}_a\in \Hom(a,a)$.}
        \label{fig:id_a}
    \end{subfigure}
    \caption{Building blocks for the graphical calculus on morphisms.}
    \label{fig:graphical_bulding_blocks}
\end{figure}

Other than fusing two anyons together, anyons can be braided. Mathematically, this is accomplished by the notion of a braiding in a fusion category, which is a natural isomorphism $\beta_{a,b}:a\otimes b\xrightarrow{\sim} b\otimes a$ for every pair $a,b\in\mcC$. Braidings induce unitary maps on the hom spaces, so we can view braiding as a physical process. For example, suppose we have anyons $a$, $b$, and $c$ with $N^{a,b}_c =1$. Let $V^{a,b}_c=\Hom(a\otimes b,c)$, which is one-dimensional. The braiding $\beta_{a,b}:a\otimes b\to b\otimes a$ induces a unitary isomorphism
$$R_{b,a}:V^{b,a}_c\to V^{a,b}_c$$
sending a morphism $f$ to $f\circ \beta_{a,b}$. Once normalized basis vectors $e^{a,b}_c$ and $e^{b,a}_c$ are chosen for $V^{a,b}_c$ and $V^{b,a}_c$, respectively, the matrix for $R_{b,a}$ is simply a unit complex number, denoted $R^{a,b}_c$. Thus, we have
\begin{equation}\label{eq:braiding}
R_{b,a}\P{e^{b,a}_c}=R^{a,b}_ce^{a,b}_c
\end{equation}
The numbers $R^{a,b}_c$ for admissible triples $(a,b,c)$ are called the $R$-symbols.

\begin{figure}[t]
    \centering
    \begin{subfigure}{0.45\textwidth}
        \centering
        \includegraphics[width=\textwidth]{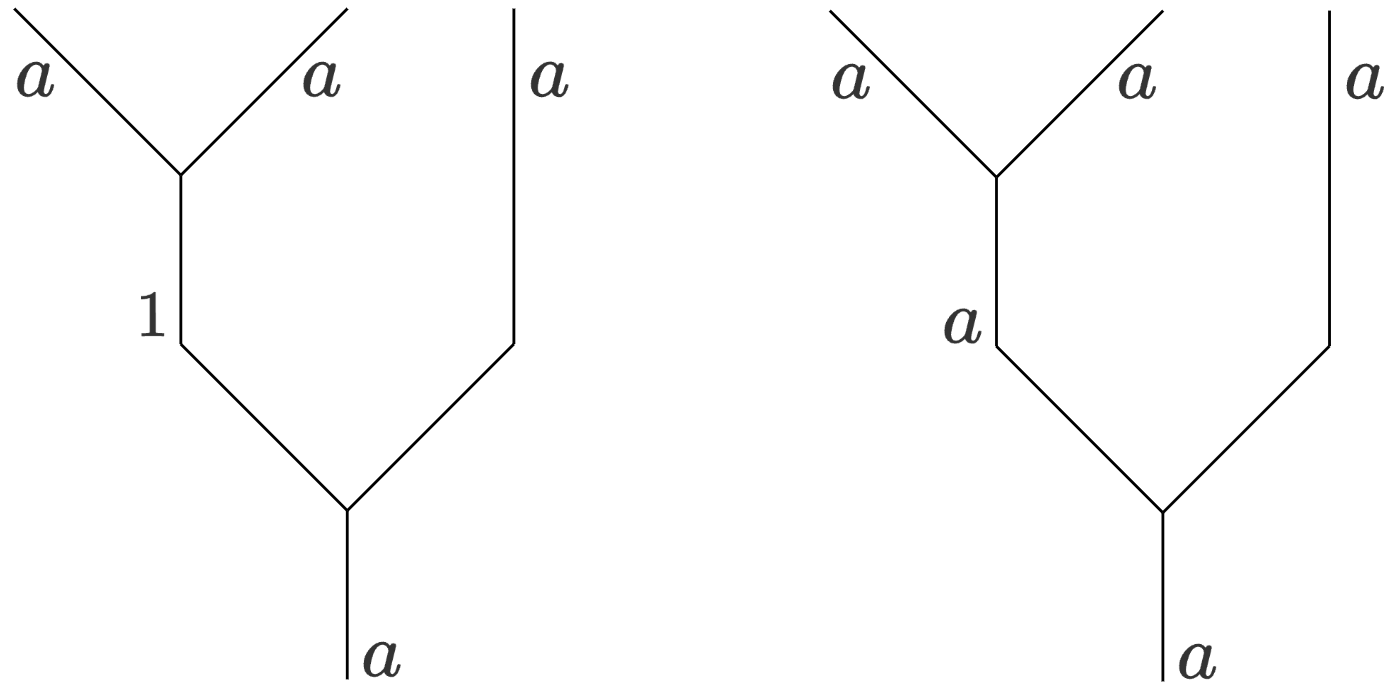}
        \caption{Diagrammatic representation of the basis in Equation (\ref{eq:left_fusion}).}
        \label{fig:left_fusion}
    \end{subfigure}
    \hfill
    \begin{subfigure}{0.45\textwidth}
        \centering
        \includegraphics[width=\textwidth]{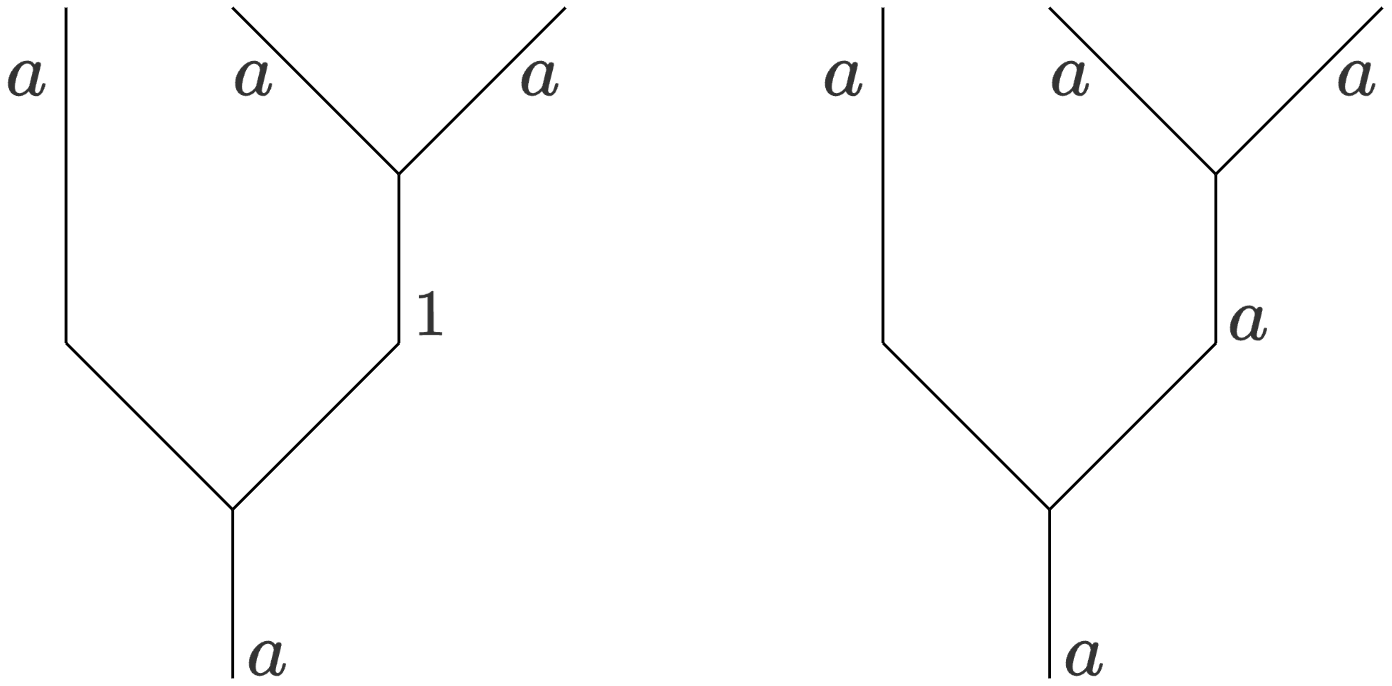}
        \caption{Diagrammatic representation of the basis in Equation (\ref{eq:right_fusion}).}
        \label{fig:right_fusion}
    \end{subfigure}
    \caption{Two bases of $\Hom(a\otimes a\otimes a,\bone)$ that are related by an $F$-move.}
    \label{fig:f_move}
\end{figure}

As mentioned earlier, there is a convenient graphical calculus for aiding with computation in the hom spaces of a UMC. For each admissible triple $(a,b,c)$, we have a unit-length basis vector $e^{a,b}_c\in \Hom(a\otimes b,c)$ represented by the picture in Figure \ref{fig:e_ab}. Each edge in the diagram is labeled with an anyon. The top of the picture represents the morphism's domain, and the bottom represents the codomain. The trivalent vertex represents the fusion of two anyons into one. The identity morphism on an object is a single vertical segment labeled with that object, as in Figure \ref{fig:id_a}. The tensor product stacks pictures horizontally, and composition stacks pictures vertically. For instance, the basis in Equation (\ref{eq:left_fusion}) is shown pictorially in Figure \ref{fig:left_fusion}, and the basis in Equation (\ref{eq:right_fusion}) is shown in Figure \ref{fig:right_fusion}. These pictures are also called \textit{fusion trees}. The $F$-matrix converts between the two pictures, and such a change is called an $F$-move. This is also possible when the change is part of a larger diagram.

\begin{figure}
    \centering
    \includegraphics[width=0.6\linewidth]{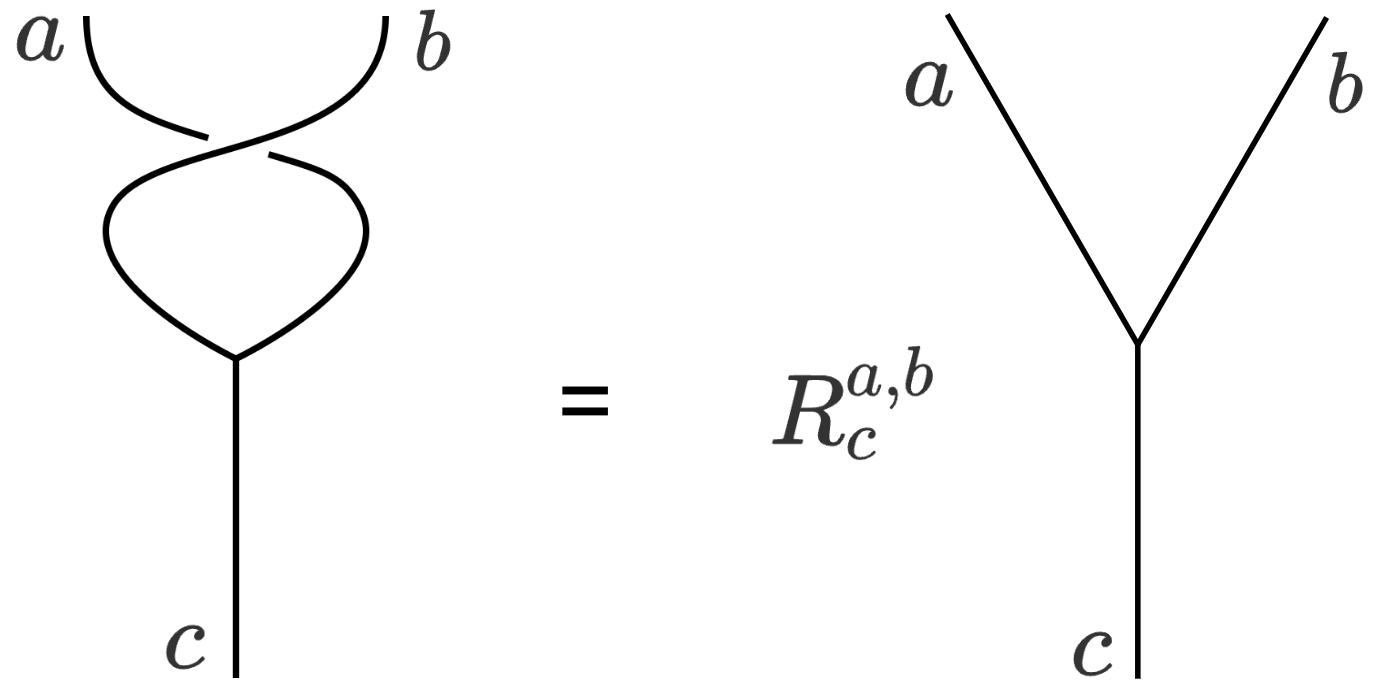}
    \caption{Diagrammatic representation of Equation (\ref{eq:braiding}).}
    \label{fig:braiding}
\end{figure}

We represent braiding in the graphical calculus by a strand crossing over another, as in Figure \ref{fig:braiding}. The $R$-symbols allow us to write a diagram containing braids in terms of one without braids.

\subsubsection{Encoding Computational States}
A topological quantum computer is a quantum computer whose qubits are encoded in the states of an anyonic system. The UMC which is perhaps the most experimentally promising is the Ising UMC. We will define this category and show how to encode computational states inside its hom spaces.

\begin{definition}
The Ising UMC has three simple objects $\bone$, $\sigma$, and $\psi$. The anyon $\sigma$ is called the Ising anyon, and $\psi$ is the Majorana fermion. The non-trivial fusion rules are as follows.
$$\sigma\otimes \sigma\cong \bone \oplus \psi,\quad \sigma\otimes \psi\cong \psi\otimes \sigma\cong\sigma, \quad \psi\otimes \psi\cong\bone$$
The only multidimensional $F$-matrix is
$$F^{\sigma\sigma\sigma}_\sigma=\frac{1}{\sqrt 2}\begin{bmatrix}1 & 1\\ 1 & -1\end{bmatrix}$$
The $1\times 1$ $F$-matrices all have entry 1 except
$$F^{\sigma\psi\sigma}_\psi=F^{\psi\sigma\psi}_{\sigma}=\B{-1}$$
Finally, we have the following $R$-symbols
$$R^{\sigma\sigma}_\bone=e^{-\pi i/8},\quad R^{\psi\psi}_\bone=-1, \quad R^{\psi\sigma}_\sigma=R^{\sigma\psi}_\sigma=-i, \quad R^{\sigma\sigma}_\psi=e^{3\pi i/8}$$
\end{definition}

\begin{figure}[b]
    \centering
    \begin{subfigure}{0.4\textwidth}
        \centering
        \includegraphics[width=\textwidth]{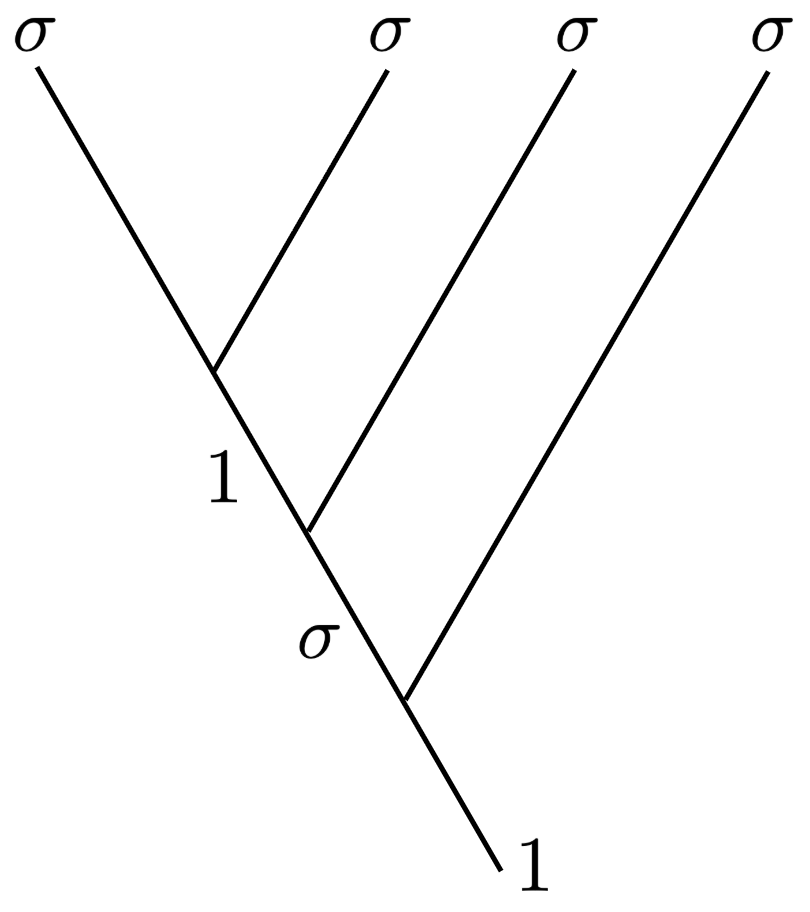}
    \end{subfigure}
    \hfill
    \begin{subfigure}{0.4\textwidth}
        \centering
        \includegraphics[width=\textwidth]{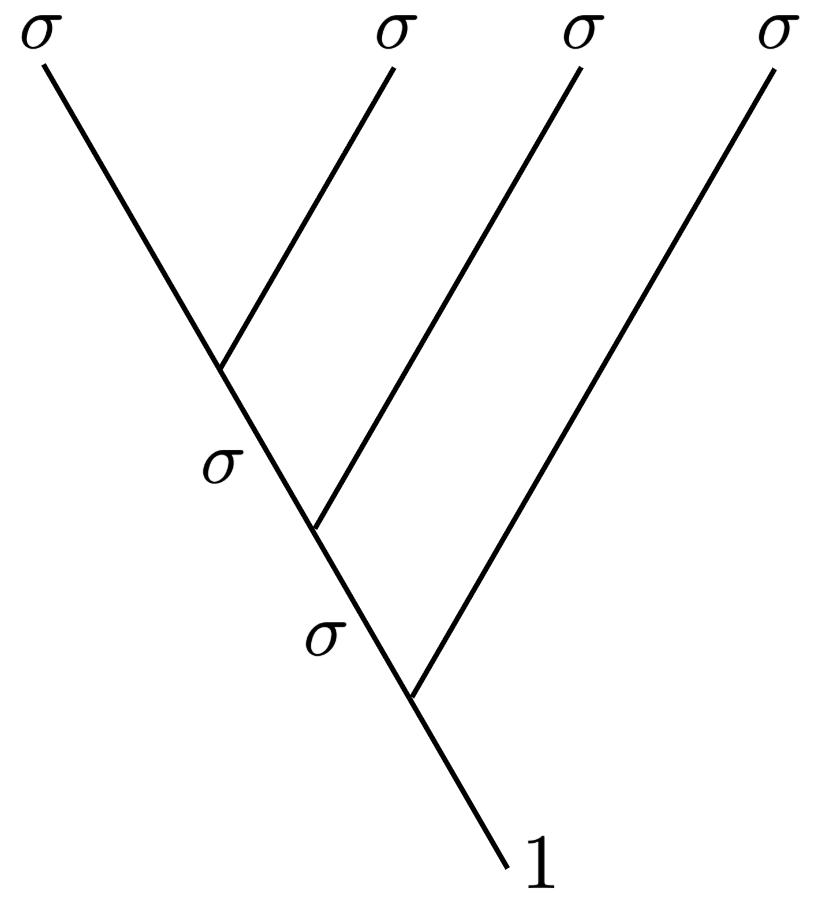}
    \end{subfigure}
    \caption{The ``standard'' basis for $\Hom\P{\sigma^{\otimes 4},\bone}$.}
    \label{fig:sparse_tilted}
\end{figure}

\begin{figure}[t]
    \centering
    \begin{subfigure}[b]{0.4\textwidth}
        \centering
        \includegraphics[width=\textwidth]{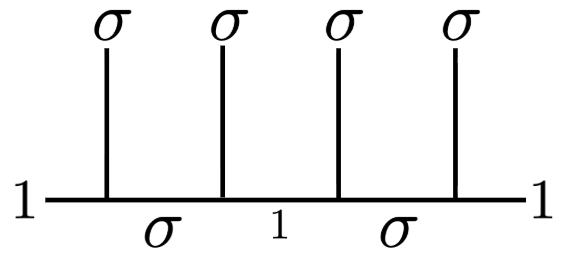}
        \caption{Fusion tree corresponding to the state $\ket{0}$.}
        \label{fig:sparse0}
    \end{subfigure}
    \hfill
    \begin{subfigure}[b]{0.4\textwidth}
        \centering
        \includegraphics[width=\textwidth]{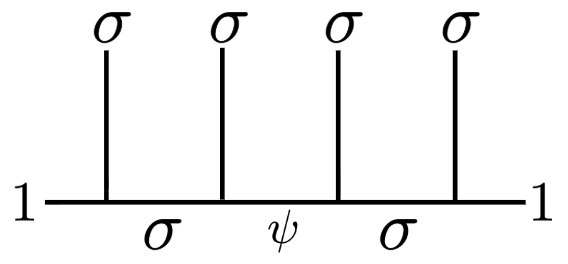}
        \caption{Fusion tree corresponding to the state $\ket{1}$.}
        \label{fig:sparse1}
    \end{subfigure}
    \caption{Rotated versions of Figure \ref{fig:sparse_tilted}.}
    \label{fig:sparse}
\end{figure}

Since $\sigma$ is the only non-abelian anyon in the Ising theory, we use collections of $\sigma$ anyons to encode computational states for quantum computation. There are two main ways of doing this, called the sparse encoding and the dense encoding. In this paper, we will restrict our attention to the dense encoding, but similar complexity theoretic results can be obtained when the sparse encoding is used instead.

We form a single qubit by splitting the vacuum $\bone$ to create four $\sigma$ anyons, which have state space $\Hom(\sigma\otimes \sigma\otimes \sigma\otimes \sigma,\bone)$. In the graphical calculus, a basis for this space is given by the fusion trees in Figure \ref{fig:sparse_tilted}. We will typically ``rotate'' these fusion trees and give them two extra edges, as in Figure \ref{fig:sparse}. These two basis elements are orthogonal, which is why we can think of them as the $\ket{0}$ and $\ket{1}$ in an abstract qubit space $\bbC^2$.

\begin{figure}[b]
    \centering
    \includegraphics[width=\linewidth]{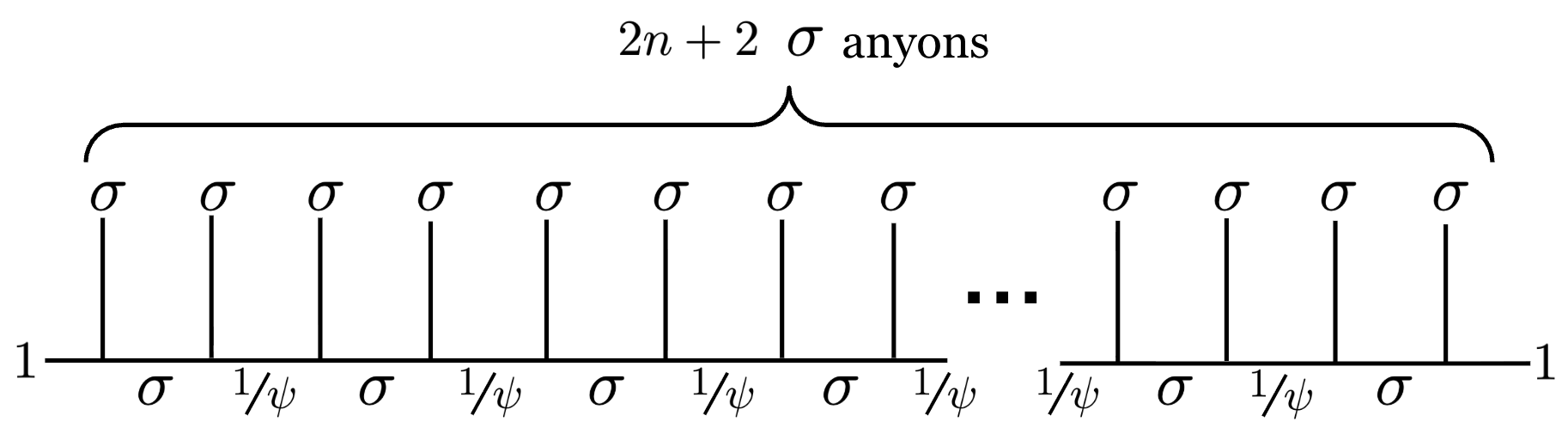}
    \caption{An anyonic chain of $2n+2$ $\sigma$ anyons. This is the dense encoding of $n$ qubits.}
    \label{fig:dense}
\end{figure}

To get a $2^n$ dimensional space representing $n$ qubits, we split the vacuum to create $2n+2$ $\sigma$ anyons, which have state space $\Hom\P{\sigma^{\otimes 2n+2},\bone}$. This space is represented pictorially in Figure \ref{fig:dense}, which is called an \textit{anyonic chain}. Edges in the diagram labeled by $\bone/\psi$ are called \textit{qubit edges}, and there are $n$ of them. The space is spanned by the $2^n$ fusion trees whose qubit edges have been assigned a definitive value of either $\bone$ or $\psi$, which are represented algebraically by kets of the form $\ket{x_1\cdots x_n}$ with $x_j\in \bbF_2$. Each qubit edge corresponds to a single bit in $\ket{x_1\cdots x_n}$, where a $\bone$ on the edge gives a 0 in the ket, and a $\psi$ on the edge gives a 1 in the ket.

In Section \ref{sec:ising_complexity}, we'll see how to use braiding and a process called interferometric measurement to manipulate the state of an Ising quantum computer and perform computation.

%% file: sections/3.tex
\section{Complexity of Clifford Quantum Computers}\label{sec:clif}
In this section, we answer the question, ``What problems can Clifford quantum computers efficiently solve?'' We propose four reasonable answers.

What does it mean for a Clifford quantum computer to solve a problem? After all, a single stabilizer circuit can only handle inputs of a fixed length. In order to tackle arbitrary inputs, we need some way of generating circuits of the appropriate size from the given input. Typically, this is done by an oracle Turing machine, called the circuit generator, that writes circuit descriptions to its oracle tape and asks its oracle a question about the circuit's output. Different complexity classes come from different choices for how much power we give the circuit generator.\footnote{One may wonder why this particular nuance does not arise in the definition of \textbf{BQP}. The answer is that universal quantum circuits are powerful enough to simulate arbitrary poly-time classical computation. On the other hand, stabilizer circuits are not. Even arbitrary log-space classical computation is intractable for stabilizer circuits in a precise sense.}

\subsection{Unrestricted Input}
In the most powerful case, the circuit generator is allowed unrestricted access to the input and arbitrary poly-time computation. The circuit generator constructs stabilizer circuits and asks its oracle about the output to those circuits. More precisely, we define the language \texttt{STAB} as the set of descriptions of stabilizer circuits which ``accept'' the input $\ket{0^n}$, and then the computational power of Clifford quantum computing is the relativized class $\bP^\texttt{STAB}$, i.e. the set of problems that are poly-time Turing reducible to \texttt{STAB}.

\begin{definition}
Let \texttt{STAB} be the collection of all descriptions of stabilizer circuits with the property that, when all qubits are initialized to 0 and the circuit is executed, an output of 1 on the first qubit is measured with certainty.
\end{definition}

In this case, the oracle ability is redundant. Since stabilizer circuits can be simulated by poly-time classical machines (\cite{AG04}, \cite{Gottesman98}), allowing such a machine oracle access to \texttt{STAB} has no effect. In other words, $\bP^\texttt{STAB}=\bP$. Thus, when arbitrary classical pre- and post-processing is allowed, Clifford quantum computers offer no additional power. Things get more interesting when more restrictions are placed on the circuit generator in an attempt to isolate the computational power of stabilizer circuits from that of classical computers.

Suppose that instead of arbitrary polynomial time pre- and post-processing, we restrict our circuit generator to use logarithmic space.\footnote{Only the machine's work tape is restricted to $\mcO(\log n)$ symbols. The oracle tape has no restrictions, but in practice it's limited to $\mcO(\text{poly}\, n)$ symbols because a log-space machine only has polynomially many configurations.} This was the approach taken by Aaronson and Gottesman in \cite{AG04}, and it leads to the complexity class $\bL^\texttt{STAB}$, i.e. those problems that are log-space Turing reducible to \texttt{STAB}. They showed that $\texttt{STAB}\in \bL^\texttt{CNOT-CIRCUIT}$ and then argued that this implies $\bL^\texttt{STAB}=\oplus\bL$; however they omitted some details, which we include here.

\begin{theorem}[\cite{AG04}]\label{thm:clif=parityL}
$\bL^\texttt{STAB}=\oplus\bL$.
\end{theorem}

\begin{proof}
In Section \ref{sec:prelims}, we showed that $\bL^\texttt{CNOT-CIRCUIT}=\oplus\bL$. Thus, $\texttt{STAB}\in \oplus\bL$.

Moreover, notice that $X=HS^2H$, so if $\V{C}\in \texttt{CNOT-CIRCUIT}$, then by replacing every Pauli $X$ gate by the sequence of gates $HSSH$, we get a stabilizer circuit $C'$ equivalent to $C$. Hence, $\V{C'}\in \texttt{STAB}$. Since \texttt{CNOT-CIRCUIT} is $\oplus\bL$-hard, \texttt{STAB} is $\oplus\bL$-hard as well. This means \texttt{STAB} is $\oplus\bL$-complete. By Proposition \ref{prop:recomp}, $\bL^\texttt{STAB}=\oplus\bL$.
\end{proof}

If we interpret $\bL^\texttt{STAB}$ as the class of problems that Clifford quantum computers can solve efficiently, then we get an interesting corollary of Theorem \ref{thm:clif=parityL}, which was suggested in \cite{deBeaudrap10}. Namely, if Clifford quantum computers are as powerful as classical computers, then $\bP\neq \textbf{PSPACE}$. A more rigorous formulation is the following.

\begin{cor}[\cite{deBeaudrap10}]\label{cor:PnotPSPACE}
Suppose that every language in $\bP$ were log-space Turing reducible to $\texttt{STAB}$. Then $\bP\neq \textbf{PSPACE}$.
\end{cor}

\begin{proof}
The premise of the corollary is that $\bP\subseteq \bL^\texttt{STAB}=\oplus\bL$. Theorem \ref{thm:parityL_in_log2} tells us that $\oplus\bL\subseteq \textbf{SPACE}\P{\P{\log n}^2}$, and the space hierarchy theorem (see \cite[Chapter 9.1]{Sipser13}) implies that
$$\textbf{SPACE}\P{\P{\log n}^2}\subsetneq \textbf{PSPACE}$$
\end{proof}

This provides a potential way to separate $\bP$ and \textbf{PSPACE} -- show how Clifford quantum computers can solve arbitrary problems in $\bP$.

\subsection{Restricted Input}
Some may argue that arbitrary log-space computation is still too much power to grant the circuit generator, and it doesn't really isolate the computational power of the generator from the power of stabilizer circuits themselves. Perhaps, to isolate the power of stabilizer circuits better, we should only give the circuit generator access to the \textit{length} of the input, rather than every individual symbol. The input length is necessary to generate circuits with an appropriate number of wires, but beyond that, the generator gets no other information. The actual input is fed into the circuits, possibly with some ancilla qubits initialized to $\ket{0}$. However, the circuit generator should not be allowed to ``see'' the outputs of the circuits it generates, otherwise it could completely recover the input, and the resulting class would be identical to the unrestricted input case. A circuit generator which is blind to the outputs of the circuits it produces is equivalent to a circuit generator that only produces a single circuit. Thus, we limit the generator to producing a single stabilizer circuit solely from the input length, and that circuit gets executed on the input to determine the answer. This seems like a fairly natural definition because it is analogous to the circuit definition of \textbf{BQP}, but using stabilizer circuits instead of universal quantum circuits. We call the resulting classes $\textbf{PUSC}$ (\textbf{P}oly-time \textbf{U}niform \textbf{S}tabilizer \textbf{C}ircuit) and $\textbf{LUSC}$ (\textbf{L}og-space \textbf{U}niform \textbf{S}tabilizer \textbf{C}ircuit) depending on whether the generator is allowed arbitrary poly-time computation or if it has a log-space constraint.

To be more precise, we need the notion of a \textit{circuit family}, which is simply a sequence of quantum circuits $\S{C_n}_{n\in\bbN}$ such that $C_n$ has at least $n$ wires. These circuits should be uniformly computable in the sense that a single Turing machine (with some computational resource restriction) can compute them solely by looking at the size of the input. This motivates the following definition.

\begin{definition}\label{def:uniform_circuit_family}
A family $\S{C_n}$ of quantum circuits is \textit{poly-time uniform} if there exists a polynomial time Turing machine that outputs a description of $C_n$ when given $n$ 1's as input. By a description of the circuit, we mean a list of all the circuit's gates in the order they're applied. Similarly, $\S{C_n}$ is \textit{log-space uniform} if there exists a log-space Turing machine that outputs a description of $C_n$ when given $n$ 1's as input.\footnote{This does not imply that $C_n$ must have $\mcO(\log n)$ gates. On the contrary, we grant the Turing machine an unlimited write-only output tape, which is separate from its read-write work tape that's limited to $\mcO(\log n)$ cells. Hence, $C_n$ may have $\mcO(\text{poly}\,n)$ gates.}
\end{definition}

Notice that, in Definition \ref{def:uniform_circuit_family}, the input length $n$ is provided in unary to the Turing machine, not binary. This ensures the machine enjoys the same computational resources as in the unrestricted input case, it's just blind to what that input is.

Another technical issue we run into is the challenge posed by non-binary languages, i.e. those languages whose alphabet isn't $\S{0,1}$. Thinking about non-binary languages as lying in \textbf{BQP} is not an issue, since it's possible to encode such languages in binary, and the choice of encoding is inconsequential because we can freely convert between encodings. However, with stabilizer circuits, only affine transformations are possible, so it's necessary to discuss in detail how non-binary languages are encoded into binary for use in quantum circuits.

\begin{definition}
Suppose \texttt{A} is a language over an alphabet $\Sigma$. A \textit{binary encoding function} for \texttt{A} is an injective map $\epsilon:\Sigma\to \bbF_2^k$ for some $k\geq \C{\log_2 \A{\Sigma}}$. If $\sigma=\sigma_1\cdots \sigma_n\in \Sigma^*$ is a string over $\Sigma$, then $\epsilon(\sigma)=\epsilon(\sigma_1)\cdots \epsilon(\sigma_n)\in \bbF_2^{kn}$. Moreover, we define
$$\epsilon(\texttt{A})=\S{\epsilon(\sigma)\,\vert\,\sigma\in \texttt{A}}$$
\end{definition}

Now what does it mean for a circuit family to solve a computational problem?

\begin{definition}\label{def:circuit_acceptance}
For a quantum circuit $C$ and a binary string $x$ whose length is at most the number of wires that $C$ has, suppose $C$ is initialized with the computational basis state $\ket{x00\cdots}$, and then $C$ is executed.
\begin{itemize}
    \item If the probability of measuring an output of 1 on the first qubit is at least 2/3, then $C$ \textit{accepts} $x$.
    \item If the probability of measuring an output of 1 on the first qubit is at most 1/3, then $C$ \textit{rejects} $x$.
\end{itemize}
In addition, if $\S{C_n}$ is a circuit family, \texttt{A} is a language, and $\epsilon$ is a binary encoding function for \texttt{A}, we say that $\S{C_n}$ \textit{decides} \texttt{A} with respect to $\epsilon$ if there exists $N\in \bbN$ such that, for all strings $\sigma\in \Sigma^n$ with $n\geq N$, $C_n$ has at least $kn$ wires and $C_n$ accepts $\epsilon(\sigma)$ if $\sigma\in \texttt{A}$ and $C_n$ rejects $\epsilon(\sigma)$ otherwise.
\end{definition}

This is equivalent to the standard approach when defining \textbf{BQP}, where some bounded error probability tolerance is necessary for the class to be robust. But, as a corollary of Proposition \ref{prop:5050}, it turns out that we can assume stabilizer circuits always succeed and give the correct answer with probability 1. This is because if $C$ is a stabilizer circuit initialized with any computational basis state, the probability of measuring 1 on the output of the first qubit is either 0, 1/2, or 1. Thus, if $\S{C_n}$ is a family of stabilizer circuits that decides a language \texttt{A} with respect to $\epsilon$, then for lengthy enough $\sigma$, determining membership $\sigma\in \texttt{A}$ is as simple as running $C_{|\sigma|}$ a single time, because you will always get the ``correct answer''.

\begin{prop}\label{prop:5050}
Given a stabilizer state $\ket{\psi}$ and a Pauli operator $P$, $\ket{\psi}$ is either an eigenstate of $P$ or an equal superposition of two eigenstates with differing eigenvalues.
\end{prop}
\begin{proof}
Let $\mcS$ be the group of Pauli operators which stabilize $\ket{\psi}$. By Corollary \ref{cor:maximal_stabilizer_normalizer}, the only Pauli operators that commute with all elements of $\mcS$ are the elements of $\mcS$ themselves, up to a phase of $\pm 1,\pm i$. Now we have two cases to consider:
\begin{enumerate}
    \item $P$ commutes with every element of $\mcS$. Then $\lambda P\in \mcS$ for some $\lambda\in \S{\pm 1,\pm i}$. Hence, $\lambda P\ket{\psi}=\ket{\psi}$, so $\ket{\psi}$ is an eigenstate of $P$, with eigenvalue $\lambda^{-1}$.
    \item $P$ doesn't commute with some operator $Q\in \mcS$. Two elements of the Pauli group which don't commute must anti-commute, so $PQ=-QP$. Suppose the eigenvalues of $P$ are $\pm \lambda$. Then the projectors onto $P$'s eigenspaces are $\Pi_0=\frac12\P{I+\lambda^{-1}P}$ and $\Pi_1=\frac12\P{I-\lambda^{-1}P}$.
    For $j\in\bbF_2$,
    \begin{align*}
        \bra{\psi}\Pi_j^\dagger\Pi_j\ket{\psi} &= \bra{\psi}\Pi_j\ket{\psi}\\
        &= \frac12+\frac12(-1)^j \lambda^{-1}\bra{\psi}P\ket{\psi}\\
        &= \frac12+\frac12(-1)^j \lambda^{-1}\bra{\psi}PQ\ket{\psi}\\
        &= \frac12-\frac12(-1)^j \lambda^{-1}\bra{\psi}QP\ket{\psi}\\
        &= \frac12-\frac12(-1)^j \lambda^{-1}\bra{\psi}P\ket{\psi}\\
        &= \frac12
    \end{align*}
    The last line follows by comparing the second line with the second to last line and noting $\bra{\psi}P\ket{\psi}=0$. Hence, $\Pi_0\ket{\psi}$ and $\Pi_1\ket{\psi}$ are both eigenstates of $P$ for different eigenvalues, they have equal magnitude $\frac{1}{\sqrt{2}}$, and $\ket{\psi}=\Pi_0\ket{\psi}+\Pi_1\ket{\psi}$.
\end{enumerate}
\end{proof}

Notice that in Definition \ref{def:circuit_acceptance}, we only require sufficiently long inputs $\sigma$ to be correctly decided by the appropriate circuit, rather than all $\sigma$. When defining \textbf{BQP}, this distinction is irrelevant, since a circuit family which doesn't have the correct output on sufficiently small inputs can be modified to always be correct without any cost to long input time complexity. However, this is not the case for stabilizer circuits, as we'll see below. Hence, if we want the complexity classes we define to only depend on long input behavior, our definition should be forgiving for small inputs.

Now we can finally formally define the classes $\textbf{PUSC}$ and $\textbf{LUSC}$. Their definitions are analogous to the circuit definition of \textbf{BQP}, in light of the fact that the bounded error condition on Clifford circuits is irrelevant, we only care about behavior on sufficiently long inputs, and we must be explicit about encoding non-binary languages into binary.

\begin{definition}\label{def:USC}
Suppose $A$ is a language. We say that a binary encoding function $\epsilon$ for \texttt{A} is \textit{poly-time admissible} if there exists a poly-time uniform family of stabilizer circuits that decides \texttt{A} with respect to $\epsilon$. Similarly, $\epsilon$ is \textit{log-space admissible} if there exists a log-space uniform family of stabilizer circuits that decides \texttt{A} with respect to $\epsilon$. We let $\textbf{PUSC}$ be the class of languages that have a poly-time admissible encoding, and similarly $\textbf{LUSC}$ is the class of languages that have a log-space admissible encoding.
\end{definition}

The above definition may seem difficult to work with because of the fact that the encoding is existentially quantified. Unfortunately, the choice of encoding function is not inconsequential. It's possible that some $\epsilon$ are poly-time admissible, and some aren't. However, any binary encoding function $\epsilon$ that sends letters to \textit{linearly independent} binary strings is universal in the sense that if \texttt{A} has any poly-time admissible encodings, then $\epsilon$ is one of them (and similarly for log-space admissible encodings). We prove the particular case of the \textit{one-hot encoding}, defined below. This makes it easier to determine whether or not a particular language is in $\textbf{PUSC}$ or $\textbf{LUSC}$.

\begin{definition}
A \textit{one-hot encoding} of a language \texttt{A} over $\Sigma$ is an encoding that sends letters to the ``standard basis vectors'' in $\bbF_2^r$, where $r=\A{\Sigma}$. In other words, if we let $e_j\in \bbF_2^r$ be the bit string of all 0s except a 1 in the $j^\text{th}$ position, and $\Sigma=\S{\sigma_1,\ldots,\sigma_r}$, then a one-hot encoding of \texttt{A} sends $\sigma_j$ to $e_j$. There are $r!$ possible one-hot encodings.
\end{definition}

\begin{prop}\label{prop:universal_encoding_func}
Let $\mcE$ be a one-hot encoding of a language \texttt{A}. Then $\texttt{A}\in \textbf{PUSC}$ if and only if $\mcE$ is poly-time admissible. Similarly, $\texttt{A}\in \textbf{LUSC}$ if and only if $\mcE$ is log-space admissible.
\end{prop}

\begin{proof}
\begin{figure}
    \centering
    \includegraphics[width=0.8\linewidth]{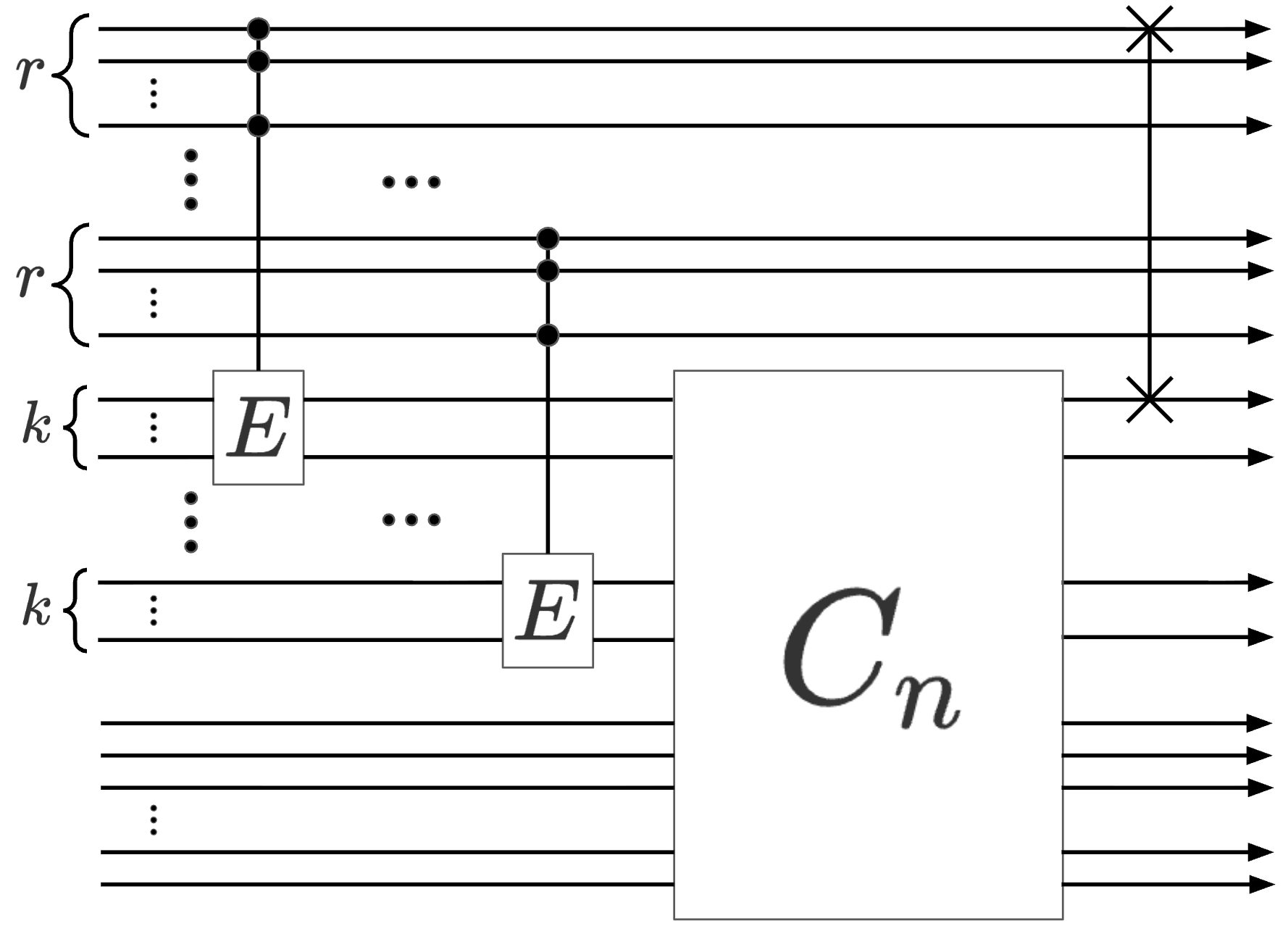}
    \caption{The circuit $\tilde{C}_n$. The first $rn$ wires are initialized with $\ket{\mcE(\sigma)}$, where $\sigma$ is a length $n$ string over $\Sigma$. Then $\ket{\mcE(\sigma)}$ is converted to $\ket{\epsilon(\sigma)}$ via the $E$ gates. When the output of the first wire is measured, 1 is obtained if and only if $\sigma\in \texttt{A}$.}
    \label{fig:encoding_conversion}
\end{figure}

Let $\Sigma$ be the alphabet of \texttt{A} and $r=\A{\Sigma}$. Suppose $\texttt{A}\in \textbf{LUSC}$, so there exists a binary encoding function $\epsilon:\Sigma\to \bbF_2^k$, a log-space Turing machine $M$, and a family of stabilizer circuits $\S{C_n}$ such that $M$ outputs $\V{C_n}$ on input $1^n$, and for all $\sigma\in \Sigma^n$, when $C_n$ is initialized with $\ket{\epsilon(\sigma)00\cdots}$ and executed, a measurement of 1 is obtained from the first qubit if and only if $\sigma\in \texttt{A}$.

To show that $\mcE$ is log-space admissible, the idea is to use a \texttt{CNOT} circuit to convert $\ket{\mcE(\sigma)}$ into $\ket{\epsilon(\sigma)}$, and then execute $C_n$ on the converted output. We design a new Turing machine $\tilde{M}$ that, given input $1^n$, outputs a description of the circuit $\tilde{C}_n$, depicted in Figure \ref{fig:encoding_conversion}. In the figure, there are $n$ groups of $r$ wires each, followed by $n$ groups of $k$ wires each, followed by some ancilla wires. Between corresponding groups of $r$ wires and $k$ wires, there is a circuit $E$ (consisting entirely of \texttt{CNOT} gates) that converts a block of $r$ bits of $\ket{\mcE(\sigma)}$ to a block of $k$ bits of $\ket{\epsilon(\sigma)}$, where $\sigma\in \Sigma^n$. An example of how $E$ is implemented for a specific $\epsilon$ is shown in Figure \ref{fig:E_implementation}. Then, the original circuit $C_n$ is executed on the converted input, and a final swap gate puts the desired output on the first wire.

It's easy to see that the circuit family $\S{\tilde{C}_n}$ correctly decides \texttt{A} with respect to $\mcE$, and that $\tilde{M}$ can describe $\tilde{C}_n$ in space $\mcO(\log n)$. An identical proof also works with a poly-time restriction instead of the log-space restriction.

\begin{figure}
    \centering
    \includegraphics[width=\linewidth]{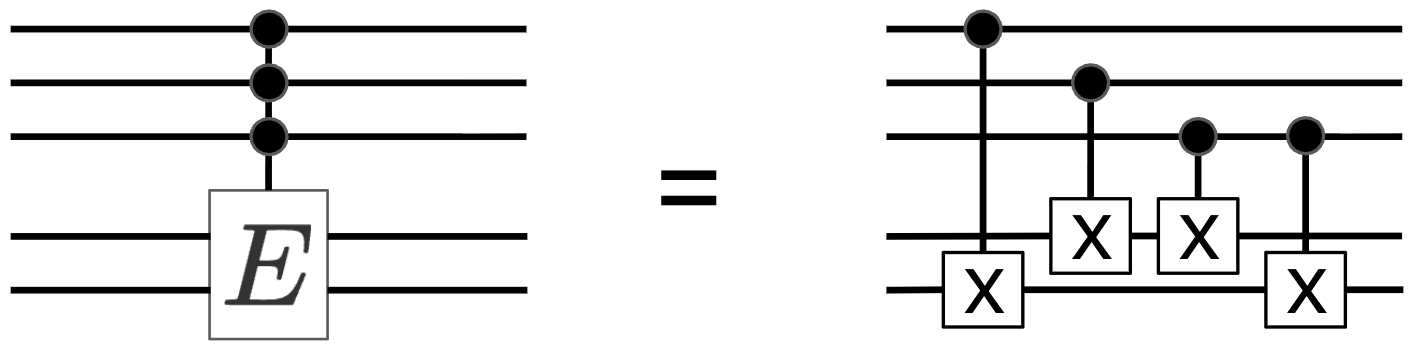}
    \caption{An example implementation of $E$ for use in $\tilde{C}_n$ (see Figure \ref{fig:encoding_conversion}). In this example, $\A{\Sigma}=3$, $\epsilon(\sigma_1)=01$, $\epsilon(\sigma_2)=10$, and $\epsilon(\sigma_3)=11$. This collection of \texttt{CNOT} gates computes the transformation $\ket{\mcE(\sigma_j)}\ket{00}\mapsto \ket{\mcE(\sigma_j)}\ket{\epsilon(\sigma_j)}$. In general, there is a \texttt{CNOT} gate from wire $i$ to wire $\P{j+\A{\Sigma}}$ if the $j^\text{th}$ bit of $\epsilon(\sigma_i)$ is 1.}
    \label{fig:E_implementation}
\end{figure}
\end{proof}

The remainder of this section is dedicated to giving an alternative characterization of $\textbf{PUSC}$ and $\textbf{LUSC}$ in terms of what we call \textit{affine} languages. Theorem \ref{thm:usc_classification} shows that $\textbf{PUSC}$ is the collection of affine languages in $\bP$, and $\textbf{LUSC}$ contains all affine languages in $\bL$ but no affine languages outside of $\oplus\bL$.

\begin{definition}
Let \texttt{A} be a language over $\Sigma$ and $\epsilon:\Sigma\to \bbF_2^k$ be a binary encoding function for \texttt{A}. We say \texttt{A} is \textit{affine with respect to $\epsilon$} if there exist a vectors $\ba_n\in \bbF_2^{kn}$ and a bits $b_n\in \bbF_2$ such that, for all sufficiently large $n$ and $\sigma\in \Sigma^n$, $\sigma\in \texttt{A}$ if and only if $\ba_n^T\epsilon(\sigma)+b_n=1$. We call \texttt{A} \textit{affine} if \texttt{A} is affine with respect to some binary encoding function.
\end{definition}

As with Definition \ref{def:USC}, the choice of encoding function is not inconsequential. It's possible for a language \texttt{A} to be affine with respect to one encoding, but not another. The following proposition shows that, similar to Proposition \ref{prop:universal_encoding_func}, if a language is affine, then it is also affine with respect to any one-hot encoding. Hence, it allows us to more easily verify whether or not a language is affine.

\begin{prop}
Suppose $\mcE$ is a one-hot encoding for the language \texttt{A}. Then \texttt{A} is affine iff \texttt{A} is affine with respect to $\mcE$.
\end{prop}

\begin{proof}
Let $\Sigma=\S{\sigma_1,\ldots,\sigma_r}$, and suppose $\mcE(\sigma_j)=e_j$. Assume \texttt{A} is affine, so there exists a binary encoding function $\epsilon:\Sigma\to \bbF_2^k$ for \texttt{A}, vectors $\ba_n\in \bbF_2^{kn}$, and bits $b_n\in \bbF_2$ such that for all $\sigma\in \Sigma^n$ with $n$ sufficiently large, $\sigma\in \texttt{A}$ iff $\ba_n^T\epsilon(\sigma)+b_n = 1$. We define new vectors $\tilde{\ba}_n$ and bits $\tilde{b}_n$ as follows.

First, set $\tilde{b}_n=b_n$. We will construct $\tilde{\ba}_n\in \bbF_2^{rn}$ from $\ba_n\in \bbF_2^{kn}$ by transforming each consecutive block of $k$ bits into a consecutive block of $r$ bits. More precisely, let $E$ be the $r\times k$ matrix over $\bbF_2$ whose $j^\text{th}$ row is the vector $\epsilon(\sigma_j)$. Think of $\tilde{\ba}_n$ as $n$ blocks of $r$ bits, and likewise think of $\ba_n$ as $n$ blocks of $k$ bits. We define the $j^\text{th}$ block of $\tilde{\ba}_n$ to be $E$ times the $j^\text{th}$ block of $\ba_n$.

Since $E^T$ converts the $\mcE$ representation of a letter to its $\epsilon$ representation, then for a given string $\sigma\in \Sigma^n$, we have
$$\ba_n^T\epsilon(\sigma)=\tilde{\ba}_n^T\mcE(\sigma)$$
Hence, for sufficiently large $n$, $\tilde{\ba}_n^T\mcE(\sigma) +\tilde{b}_n=\ba_n^T\epsilon(\sigma) + b_n=1$. This implies \texttt{A} is affine with respect to $\mcE$.
\end{proof}

We can also give a characterization of affine languages that is encoding-independent, called the \textit{rectangle condition}, which relies on the notion of a \textit{rectangle}.

\begin{definition}
A length $n$ \textit{rectangle} over an alphabet $\Sigma$ is a set of four strings $\sigma^{00}$, $\sigma^{01}$, $\sigma^{10}$, and $\sigma^{11}$ of length $n$ that are all identical except at two distinct indices $i$ and $j$. At index $i$ there appear two letters $a_0,a_1\in \Sigma$, and at index $j$ there appear two letters $b_0,b_1\in \Sigma$, satisfying the following:
\vspace{2mm}
\begin{center}
\begin{tabular}{c c}
    $\sigma^{00}_i=\sigma^{01}_i=a_0\qquad$ & $\sigma^{10}_i=\sigma^{11}_i=a_1$\\
    &\\
    $\sigma^{00}_j=\sigma^{10}_j=b_0\qquad$ & $\sigma^{01}_j=\sigma^{11}_j=b_1$
\end{tabular}
\end{center}
\end{definition}
\vspace{2mm}

\begin{theorem}\label{thm:encoding_ind}
The following two properties of a language \texttt{A} are equivalent to \texttt{A} being affine. Let $\Sigma$ be the alphabet of \texttt{A}, and let $\bbone_\texttt{A}:\Sigma^*\to \bbF_2$ be the indicator that a string $\sigma$ lies in \texttt{A} (i.e. $\sigma\in \texttt{A}$ iff $\bbone_\texttt{A}(\sigma)=1$).
\begin{itemize}
    \item For sufficiently large $n$ and integers $l$ between 1 and $n$, there exist bits $b_n\in \bbF_2$ and functions $f_{n,l}:\Sigma\to \bbF_2$ such that, for all $\sigma=\sigma_1\cdots \sigma_n\in \Sigma^n$,
    \begin{equation}\label{eq:xor_sep}
    \bbone_\texttt{A}(\sigma)=b_n+\sum_{l=1}^n f_{n,l}(\sigma_l)
    \end{equation}
    If \texttt{A} has this property, we say \texttt{A} is \textit{XOR separable}.
    \item For sufficiently large $n$, an even number of the strings in any length $n$ rectangle over $\Sigma$ lie in \texttt{A}. If \texttt{A} has this property, we say \texttt{A} satisfies the \textit{rectangle condition}.
\end{itemize}
\end{theorem}

\begin{proof}
The proof that \texttt{A} being XOR separable is equivalent to \texttt{A} being affine is straightforward. We will show that \texttt{A} being XOR separable is equivalent to \texttt{A} satisfying the rectangle condition.

First, suppose \texttt{A} is XOR separable. For large enough $n$, let $\sigma^{00}$, $\sigma^{01}$, $\sigma^{10}$, $\sigma^{11}$ be a length $n$ rectangle over $\Sigma$. Then using the fact that the four strings in the rectangle are identical except at indices $i$ and $j$, we have
\begin{align*}
    &\bbone_\texttt{A}\P{\sigma^{00}} + \bbone_\texttt{A}\P{\sigma^{01}} + \bbone_\texttt{A}\P{\sigma^{10}} + \bbone_\texttt{A}\P{\sigma^{11}}\\
    &= 4b_n + \sum_{l=1}^n f_{n,l}\P{\sigma^{00}_l} + \sum_{l=1}^n f_{n,l}\P{\sigma^{01}_l} + \sum_{l=1}^n f_{n,l}\P{\sigma^{10}_l} + \sum_{l=1}^n f_{n,l}\P{\sigma^{11}_l}\\
    &=4b_n+4\sum_{l\neq i,j}f_{n,l}\P{\sigma^{00}_l}+2\sum_{l=i,j} \P{f_{n,l}\P{\sigma^{00}_l} + f_{n,l}\P{\sigma^{11}_l}}\\
    &=0
\end{align*}
The last line follows from the fact that all terms in the previous line are even. This implies \texttt{A} satisfies the rectangle condition.

Conversely, suppose \texttt{A} satisfies the rectangle condition. Fix some $s_0\in \Sigma$. For large enough $n$, let $\sigma^n$ be the length $n$ string in which every letter is $s_0$, and define $b_n=\bbone_\texttt{A}(\sigma^n)$. Also, for any $l\leq n$ and $s\in \Sigma$, let $s_0^{l-1}ss_0^{n-l}$ be the length $n$ string in which every letter is $s_0$, except for the $l^\text{th}$ letter which is $s$. Define
$$f_{n,l}(s)=b_n+\bbone_\texttt{A}\P{s_0^{l-1}ss_0^{n-l}}$$
Note that $f_{n,l}(s_0)=0$, a fact we will use later.

Suppose, for contradiction, that there exist length $n$ strings $\sigma^{00}$ and $\sigma^{10}$ which differ only at some index $i$ and satisfy
\begin{equation}\label{eq:rect_lem}
\bbone_\texttt{A}\P{\sigma^{00}}+\bbone_\texttt{A}\P{\sigma^{10}}\neq f_{n,i}\P{\sigma^{00}_i} + f_{n,i}\P{\sigma^{10}_i}
\end{equation}
Among all such $\sigma^{00}$ and $\sigma^{10}$ satisfying (\ref{eq:rect_lem}), choose a pair that has a maximal number of $s_0$'s at indices other than $i$. Note that we can't have every letter besides the one at index $i$ be $s_0$, otherwise $\bbone_\texttt{A}\P{\sigma^{00}}=f_{n,i}\P{\sigma^{00}_i}$ and $\bbone_\texttt{A}\P{\sigma^{10}} = f_{n,i}\P{\sigma^{10}_i}$ by definition, so (\ref{eq:rect_lem}) doesn't hold. Thus, for some index $j\neq i$, $\sigma^{00}_j$ and $\sigma^{10}_j$ are equal to each other but not equal to $s_0$. We define $\sigma^{01}$ to be $\sigma^{00}$ except with $s_0$ at index $j$, and similarly we define $\sigma^{11}$ to be $\sigma^{10}$ except with $s_0$ at index $j$. Then $\sigma^{00}$, $\sigma^{10}$, $\sigma^{01}$, and $\sigma^{11}$ form a rectangle, and the rectangle condition implies
\begin{align*}
    \bbone_\texttt{A}\P{\sigma^{01}}+\bbone_\texttt{A}\P{\sigma^{11}} &=\bbone_\texttt{A}\P{\sigma^{00}}+\bbone_\texttt{A}\P{\sigma^{10}}\\
    &\neq f_{n,i}\P{\sigma^{00}_i} + f_{n,i}\P{\sigma^{10}_i}=f_{n,i}\P{\sigma^{01}_i} + f_{n,i}\P{\sigma^{11}_i}
\end{align*}
Hence, $\sigma^{01}$ and $\sigma^{11}$ satisfy (\ref{eq:rect_lem}), but they each have one more instance of $s_0$, contradicting our choice of $\sigma^{00}$ and $\sigma^{10}$. Thus, the negation of (\ref{eq:rect_lem}) always must hold, i.e. for any two strings $\sigma$ and $\sigma'$ that only differ potentially at some index $i$, we must have
\begin{equation}\label{eq:rect_lem_complete}
\bbone_\texttt{A}\P{\sigma}+\bbone_\texttt{A}\P{\sigma'} = f_{n,i}\P{\sigma_i} + f_{n,i}\P{\sigma'_i}
\end{equation}

In particular, suppose we fix a length $n$ string $\sigma$, and we let $\sigma^l$ be the string whose first $l$ letters are $s_0$ and whose last $n-l$ letters match the last $n-l$ letters of $\sigma$ (so $\sigma=\sigma^0$ and $\sigma^n$ is consistent with how we already defined it). We will complete the proof by showing Equation (\ref{eq:xor_sep}) holds for $\sigma$. Indeed, for $l>0$, $\sigma^{l-1}$ and $\sigma^l$ differ only potentially at index $l$, so we can apply Equation (\ref{eq:rect_lem_complete}) to get
\begin{align*}
    \bbone_\texttt{A}\P{\sigma^{l-1}}+\bbone_\texttt{A}\P{\sigma^l} &= f_{n,l}\P{\sigma^{l-1}_l} + f_{n,l}\P{\sigma^l_l}\\
    \bbone_\texttt{A}\P{\sigma^{l-1}} &= \bbone_\texttt{A}\P{\sigma^l} + f_{n,l}\P{\sigma_l} + f_{n,l}\P{s_0}\\
    \bbone_\texttt{A}\P{\sigma^{l-1}} &= \bbone_\texttt{A}\P{\sigma^l} + f_{n,l}\P{\sigma_l}\\
\end{align*}
Fulling expanding out this recurrence starting with $l=1$ yields
$$\bbone_\texttt{A}\P{\sigma^0} = \bbone_\texttt{A}\P{\sigma^n} + \sum_{l=1}^n f_{n,l}\P{\sigma_l}$$
which is exactly Equation (\ref{eq:xor_sep}).
\end{proof}

Theorem \ref{thm:encoding_ind} gives us a more straightforward way to show that a given language isn't affine.

\begin{cor}
Suppose \texttt{A} is a language over $\Sigma$ and there exists an infinite sequence of rectangles over $\Sigma$ of strictly increasing length that each have odd intersection with \texttt{A}. Then \texttt{A} is not affine.
\end{cor}

\begin{proof}
If \texttt{A} were affine, then by Theorem \ref{thm:encoding_ind}, sufficiently lengthy rectangles would all have even intersection with \texttt{A}.
\end{proof}

Finally, we will make the connection between \textbf{AFF} and \textbf{PUSC} and \textbf{LUSC}. We are able to characterize \textbf{PUSC} exactly and give upper and lower bounds on \textbf{LUSC}.

\begin{theorem}\label{thm:usc_classification}
We have the following.
\begin{equation*}
\textbf{PUSC}=\textbf{AFF}\cap \bP
\end{equation*}
\begin{equation*}
\textbf{AFF}\cap \bL\subseteq \textbf{LUSC}\subseteq \textbf{AFF}\cap \oplus \bL
\end{equation*}
\end{theorem}

\begin{proof}
First, suppose $\texttt{A}\in \textbf{PUSC}$, so there exists a binary encoding function $\epsilon:\Sigma\to \bbF_2^k$ for $\texttt{A}$ and a poly-time uniform family of stabilizer circuits $\S{C_n}$ that decides \texttt{A} with respect to $\epsilon$. We know $\texttt{A}\in \bP$ since, given an input $\sigma$ of length $n$, a poly-time machine can generate and evaluate $C_n$ on input $\ket{\epsilon(\sigma)00\cdots}$. To see that $\texttt{A}\in \textbf{AFF}$, take $\sigma$ sufficiently lengthy and let $\bx=\epsilon(\sigma)$ and $\ket{\psi}= \ket{\epsilon(\sigma)00\cdots}$. We investigate the acceptance condition of $C_n$, which is that $C_n\ket{\psi}$ is a $-1$-eigenvector of $Z_1$ (the Pauli $Z$ operator acting on the first qubit). In other words, $Z_1C_n\ket{\psi}=-C_n\ket{\psi}$, or
$$C_n^\dagger Z_1 C_n \ket{\psi}=-\ket{\psi}$$
so $\ket{\psi}$ is a $-1$-eigenvector of $C_n^\dagger Z_1 C_n$. But $C_n$ is Clifford, so $C_n^\dagger Z_1 C_n$ is a Pauli operator, and moreover we know $\ket{\psi}$ is a computational basis state. The only way a computational basis state can be a $-1$-eigenvector of a Pauli operator is if that Pauli operator takes the form
$$C_n^\dagger Z_1 C_n=(-1)^{b_n} Z_1^{a_{n,1}}\otimes Z_2^{a_{n,2}}\otimes \cdots $$
where $b_n,a_{n,1},a_{n,2}, \ldots \in \bbF_2$. Multiplying both sides by $C_n$ on the left and $\ket{\psi}$ on the right, we see that $C_n\ket{\psi}$ is an eigenvector of $Z_1$ with eigenvalue
$$(-1)^{b_n+a_{n,1}x_1+a_{n,2}x_2+\cdots +a_{n,kn}x_{kn}}$$
where $x_j$ is the $j^\text{th}$ bit of $\bx$. Thus, if $\ba_n=(a_{n,1},a_{n,2},\ldots,a_{n,kn})\in \bbF_2^{kn}$, this eigenvalue is $-1$ exactly when $\ba_n^T\bx+b_n=1$. Hence, \texttt{A} is affine, so $\textbf{PUSC}\subseteq \textbf{AFF}\cap \bP$.

For the reverse direction, assume $\texttt{A}\in \textbf{AFF}\cap \bP$, so in particular \texttt{A} is XOR separable. That means, for sufficiently large $n$, there exist bits $b_n\in \bbF_2$ and functions $f_{n,l}:\Sigma\to \bbF_2$ such that, for all $\sigma=\sigma_1\cdots \sigma_n\in \Sigma^n$, we have
\begin{equation}\label{eq:xor_decomp}
\bbone_\texttt{A}\P{\sigma}=b_n+\sum_{l=1}^n f_{n,l}\P{\sigma_l}
\end{equation}
We normalize these functions by picking a reference symbol $s_0\in \Sigma$ and letting $\tilde{b}_n=\bbone_\texttt{A}\P{s_0^n}$ (where $s_0^n$ is the length $n$ string over $\Sigma$ in which every letter is $s_0$) and $\tilde{f}_{n,l}(s)=f_{n,l}(s_0)+f_{n,l}\P{s}$. By Equation (\ref{eq:xor_decomp}), we have $b_n=\tilde{b}_n+\sum_{l=1}^n f_{n,l}(s_0)$. Thus, for any $\sigma\in \Sigma^n$,
\begin{align*}
    \tilde{b}_n+\sum_{l=1}^n \tilde{f}_{n,l}\P{\sigma_l} &= \tilde{b}_n+\sum_{l=1}^n f_{n,l}\P{s_0}+\sum_{l=1}^n f_{n,l}\P{\sigma_l}\\
    &= b_n+\sum_{l=1}^n f_{n,l}\P{\sigma_l}\\
    &= \bbone_\texttt{A}\P{\sigma}
\end{align*}
Moreover, note that $\tilde{f}_{n,l}\P{s_0}=f_{n,l}\P{s_0}+f_{n,l}\P{s_0}=0$, and
\begin{align*}
    \bbone_\texttt{A}\P{s_0^{l-1}ss_0^{n-l}} &= \tilde{b}_n+\sum_{j=1}^{n-1} \tilde{f}_{n,j}\P{s_0} + \tilde{f}_{n,l}\P{s} + \sum_{j=l+1}^n\tilde{f}_{n,j} \P{s_0}\\
    &= \tilde{b}_n + \tilde{f}_{n,l}\P{s}
\end{align*}
so
\begin{equation}\label{eq:compute_fnl}
\tilde{f}_{n,l}\P{s}=\bbone_\texttt{A}\P{s_0^{l-1}ss_0^{n-l}}+\tilde{b}_n
\end{equation}

Let $\Sigma=\S{\sigma_1,\ldots,\sigma_r}$, and let $\mcE$ be the one-hot encoding of \texttt{A} that sends $\sigma_j$ to $e_j=0^{j-1}10^{n-j}$. We design a poly-time Turing machine $M$ that, on input $1^n$, outputs a description of a stabilizer circuit $C_n$ with $rn+1$ wires that accepts $\mcE(\sigma)$ iff $\sigma\in \texttt{A}$ (for sufficiently large $n$). This would prove $\mcE$ is poly-time admissible, so $\texttt{A}\in \textbf{PUSC}$.

The machine begins by computing $\tilde{b}_n$, which is possible in polynomial time since $\tilde{b}_n$ is the indicator that $s_0^n\in \texttt{A}$, and $\texttt{A}\in \bP$. Also, $M$ can compute $\tilde{f}_{n,l}(s)$ for any $s\in \Sigma$ in polynomial time using the same logic applied to Equation (\ref{eq:compute_fnl}). The circuit $C_n$ that $M$ describes will have $rn+1$ wires; $rn$ wires for the input and a single ancilla to store the computation's output. If $\tilde{b}_n=1$, $M$ writes a \texttt{NOT} gate on the output wire. Then, $M$ loops through all $l\leq n$ and all $j\leq r$, computes $\tilde{f}_{n,l}\P{\sigma_j}$, and if $\tilde{f}_{n,l}\P{\sigma_j}=1$, $M$ writes a \texttt{CNOT} gate from wire $(l-1)r+j$ to the last wire (the ancilla/output wire). This ensures that when the $l^\text{th}$ symbol of the input is $\sigma_j$, the parity of the output is switched. Finally, a swap gate puts the output on the first wire. Clearly $M$ operates in polynomial time, and the circuit $C_n$ accepts a string $\sigma$ exactly when $\tilde{b}_n+\sum_{l=1}^n \tilde{f}_{n,l}\P{\sigma_l}=1$, which is exactly when $\sigma\in \texttt{A}$.

This concludes the proof that $\textbf{PUSC}=\textbf{AFF}\cap \bP$. The bounds on $\textbf{LUSC}$ are shown in a similar way, the only difference being that we don't know how a log-space machine can simulate a stabilizer circuit, so we can't conclude $\textbf{LUSC}\subseteq \bL$. Thus, the best upper bound we can prove for $\textbf{LUSC}$ is $\textbf{AFF}\cap \oplus \bL$.
\end{proof}

This characterization of $\textbf{PUSC}$ and $\textbf{LUSC}$ in terms of affine languages shows that these two classes are quite limited. It is straightforward to come up with examples of languages that are provably non-affine. For instance, the language \texttt{A} over $\Sigma=\S{\text{x},\text{y}}$ consisting of all strings that start with xx is not affine since we have the length $n+2$ rectangle $\S{\text{xxx}^n, \text{xyx}^n, \text{yxx}^n, \text{yyx}^n}$ for arbitrarily large $n$, which has odd intersection with \texttt{A}. Despite being decidable in constant time by a standard Turing machine, no family of stabilizer circuits whose circuit generator only sees the input length can decide \texttt{A}.

%% file: sections/4.tex
\section{Complexity of Ising Quantum Computers}\label{sec:ising_complexity}
In the last section of this paper, we will investigate the computational power of Ising quantum computers as they were defined in Section \ref{subsec:ising_qc}. Recall that in an abstract quantum computer, we can manipulate the state either by applying unitary gates or by performing non-unitary measurements. The situation is similar with a topological quantum computer; we enact unitary transformations of the state by braiding anyons, and we can also perform measurements of topological charge. We first show how braidings are performed in the anyonic chain formalism, and we prove that they can be simulated classically. Then, we do the same for measurement. This shows that topological quantum computers with Ising anyons are no more powerful than classical computers, a sort of analog to the Gottesman-Knill theorem for Ising computation.

\subsection{Simulating Braiding}\label{subsec:sim_braiding}
\begin{figure}[b]
    \centering
    \begin{subfigure}{0.45\linewidth}
        \centering
        \includegraphics[width=\linewidth]{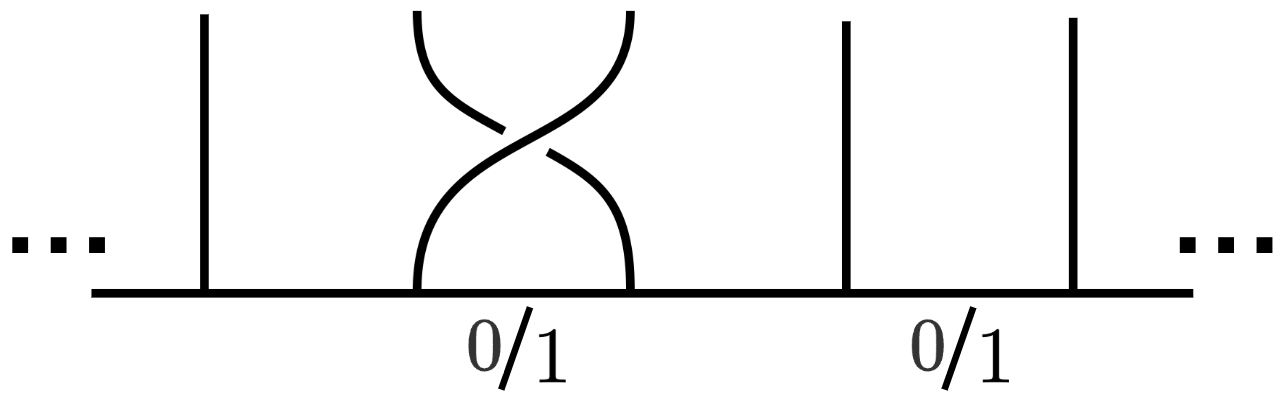}
        \caption{Braiding above a qubit.}
        \label{fig:braiding_above}
    \end{subfigure}
    \hfill
    \begin{subfigure}{0.45\linewidth}
        \centering
        \includegraphics[width=\linewidth]{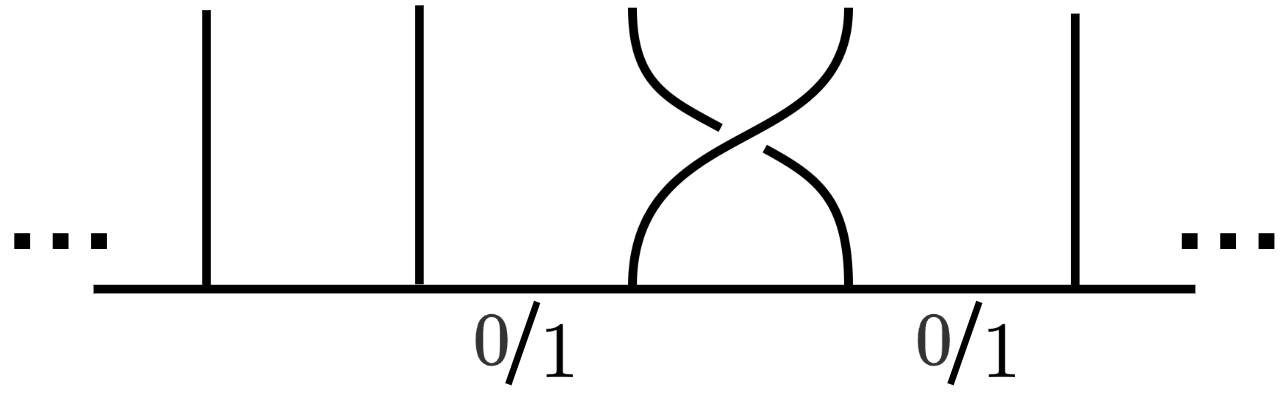}
        \caption{Braiding between two qubits.}
        \label{fig:braiding_between}
    \end{subfigure}
    \caption{Braiding anyons in an anyonic chain.}
    \label{fig:braiding_anyonic_chain}
\end{figure}

Braiding anyons, in the lab, means physically moving them around each other as was described in subsection \ref{subsubsec:anyons}. In the graphical calculus, this is achieved by a strand representing an anyon crossing over another strand. Figure \ref{fig:braiding_anyonic_chain} shows how braiding looks in the context of anyonic chains. For another reference on this computation, as well as a discussion on which Clifford gates can be attained from braiding, see \cite{AGW09}.

In the dense encoding, every braid can be classified into one of two kinds: it's either a braid \textit{above} a qubit, as in Figure \ref{fig:braiding_above}, or it's a braid \textit{between} two qubits, as in Figure \ref{fig:braiding_between}. Notice that we've simplified the labels on the anyonic chain; any edges that should be labeled with $\sigma$ are unlabeled, any edges that should be labeled with $\bone$ are labeled with 0, and any edges that should be labeled with $\psi$ are labeled with 1. Thus, qubit edges are labeled $0/1$ instead of $\bone/\psi$. We'll keep this convention moving forward. The diagrams in Figure \ref{fig:braiding_anyonic_chain} can be resolved into linear combinations of the standard basis vectors using F-moves and R-moves, as shown in the proof of the following proposition.

\begin{prop}
In the dense qubit encoding, braiding above a qubit is equivalent to applying the gate 
$e^{-i\pi/8}\sqrt{X}$ to that qubit, and braiding between two qubits is equivalent to applying the gate $e^{-i\pi/8}(S\otimes S)\boxed{\bC Z}$ to those two qubits, where $\boxed{\bC Z}$ is the controlled Pauli $Z$ gate.
\end{prop}

\begin{proof}

\begin{figure}[b]
    \centering
    \includegraphics[width=\linewidth]{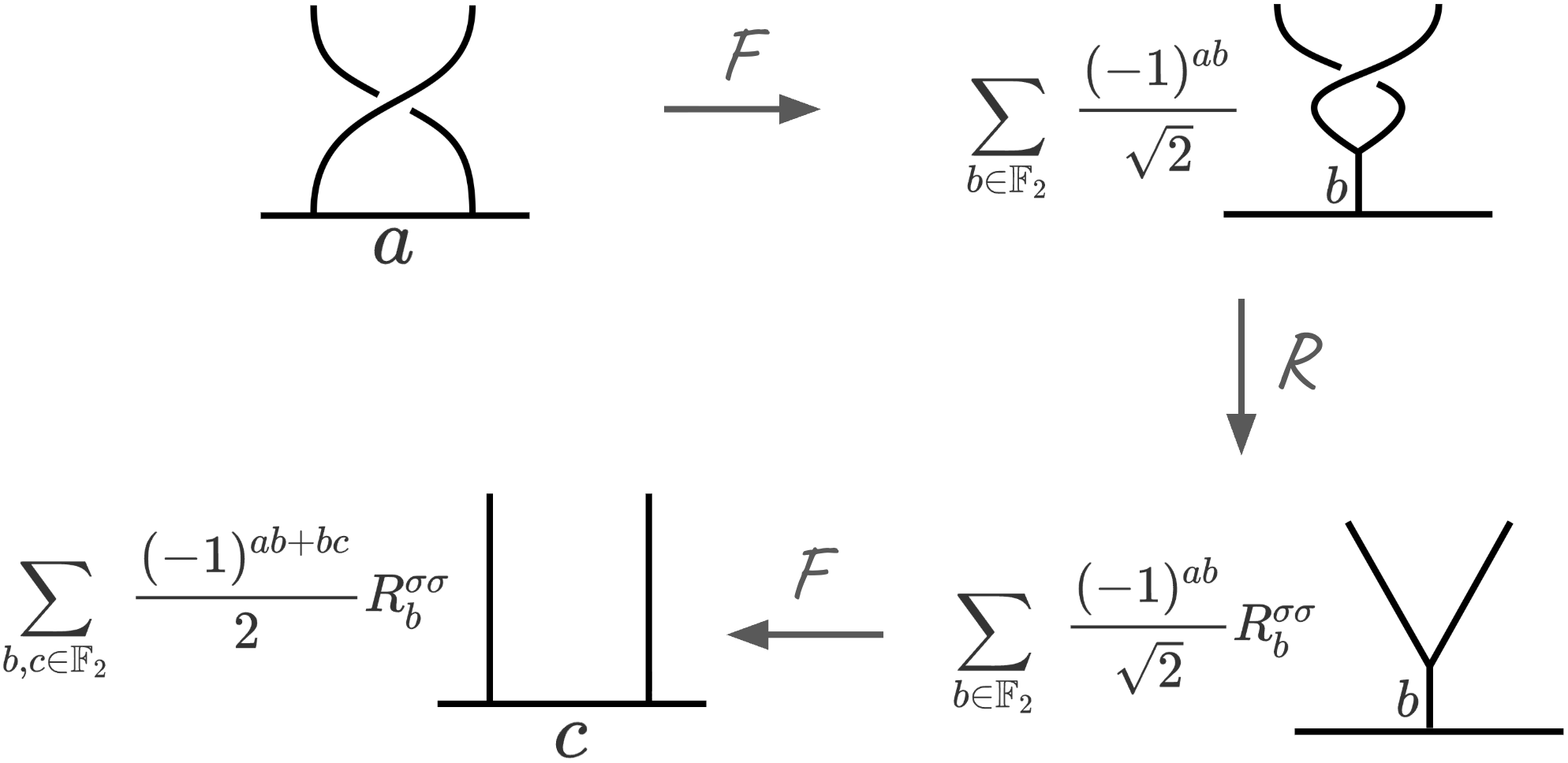}
    \caption{Decomposing a braid above a qubit in terms of the standard basis.}
    \label{fig:above_computation}
\end{figure}

First, we investigate braiding above a qubit by rewriting the fusion tree in the top left of Figure \ref{fig:above_computation} in terms of the standard basis vectors. All variables in this proof are either 0 or 1. To get rid of the braid, we need to perform an F-move so that we have something that looks like the left-hand side of Figure \ref{fig:braiding}. Then, we can do an R-move and then undo the initial F-move. Figure \ref{fig:above_computation} shows the details of the computation. According to the figure, if the initial state was $\ket{\cdots a\cdots}$, then the state after braiding is
$$\sum_{b,c\in \bbF_2} \frac{(-1)^{ab+bc}}{2}R^{\sigma\sigma}_b \ket{\cdots c\cdots}$$
The coefficient of $\ket{\cdots 0\cdots}$ is
\begin{align*}
    \sum_{b\in\bbF_2} \frac{(-1)^{ab}}{2}R^{\sigma\sigma}_b &= \frac12 e^{-\pi i/8}+\frac{(-1)^a}{2} e^{3\pi i/8}\\
    &=\frac{e^{-\pi i/8}}{2}\P{1+(-1)^ai}
\end{align*}
and the coefficient of $\ket{\cdots 1\cdots}$ is
\begin{align*}
    \sum_{b\in\bbF_2} \frac{(-1)^{ab+b}}{2}R^{\sigma\sigma}_b &= \frac12 e^{-\pi i/8}-\frac{(-1)^a}{2}e^{3\pi i/8}\\
    &=\frac{e^{-\pi i/8}}{2}\P{1-(-1)^ai}
\end{align*}
Hence, the matrix of this unitary transformation (on the relevant qubit) is
$$\frac{e^{-\pi i/8}}{2}\begin{pmatrix}1+i & 1-i\\ 1-i & 1+i\end{pmatrix}$$
One can verify that, ignoring the phase factor of $e^{-\pi i/8}$, this matrix squares to the Pauli $X$ matrix. Hence, it can be expressed as $e^{-\pi i/8}\sqrt{X}$.

Next, we investigate the braiding between two qubits. We perform the same sequence of moves to eliminate the braid, and the result is shown in Figure \ref{fig:between_computation}. If the initial state was $\ket{\cdots ab\cdots}$, then the state after braiding is $R^{\sigma\sigma}_{a+b} \ket{\cdots ab\cdots}$. The matrix of this transformation (on the two qubits) is
\begin{align*}
    \text{diag}\P{e^{-\pi i/8}, e^{3\pi i/8}, e^{3\pi i/8}, e^{-\pi i/8}} &= e^{-\pi i/8} \text{diag}\P{1, i, i, 1}\\
    &= e^{-i\pi/8}(S\otimes S)\boxed{\bC Z}
\end{align*}

\begin{figure}
    \centering
    \includegraphics[width=0.75\linewidth]{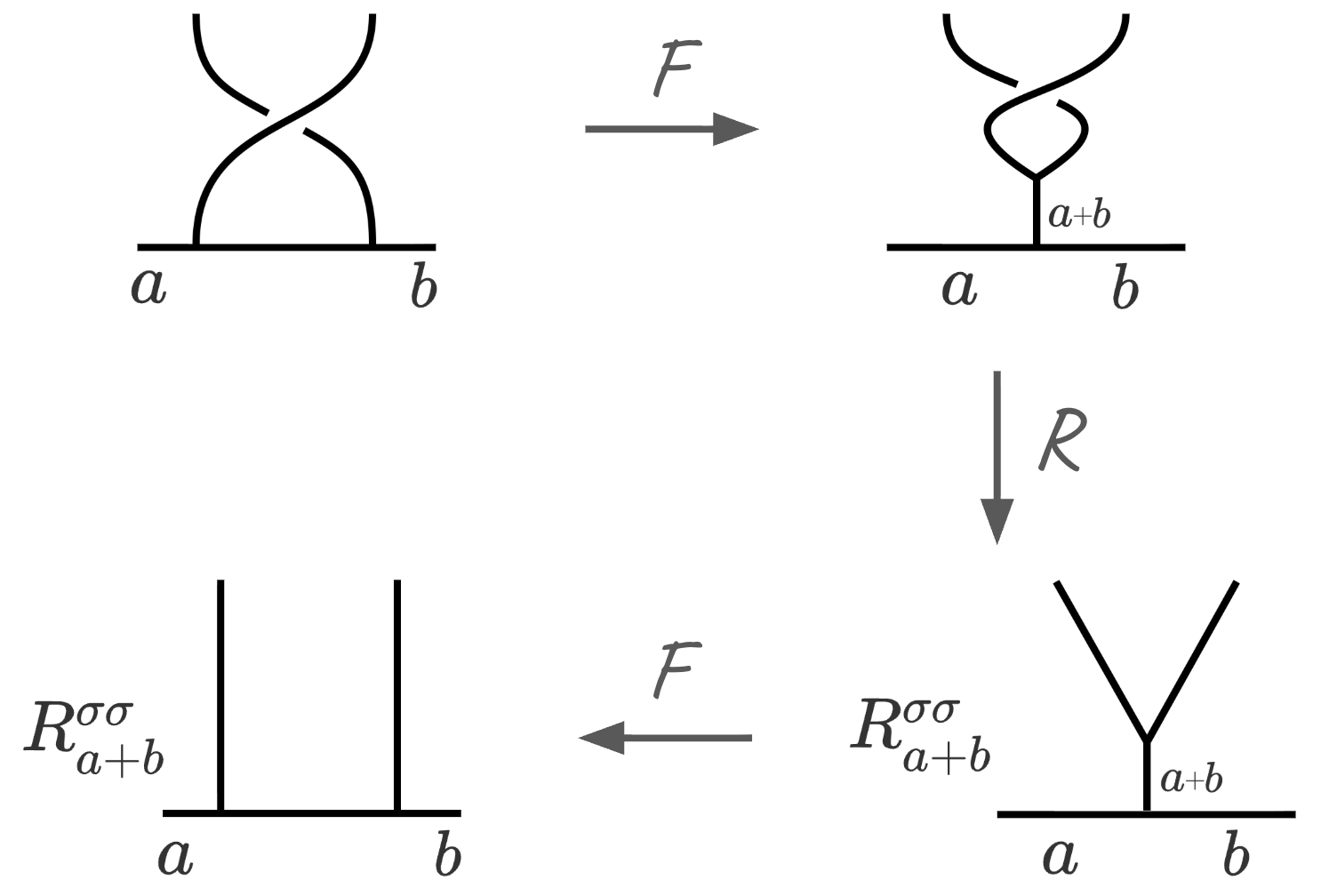}
    \caption{Decomposing a braid between two qubits in terms of the standard basis.}
    \label{fig:between_computation}
\end{figure}
\end{proof}

It turns out that, due to a famous result known as the Gottesman-Knill theorem (\cite{AG04}, \cite{Gottesman98}), any quantum computation that begins in a stabilizer state and applies only Clifford gate operations and Pauli measurements can be simulated classically. We restate the result below and then show that the braiding operations are Clifford.

\begin{theorem}[Gottesman-Knill]\label{thm:GK}
Any quantum computer performing only Clifford group gates, measurements of Pauli group operators, and Clifford group operations conditioned on classical bits (which may be the results of earlier measurements) can be perfectly simulated in polynomial time on a probabilistic classical computer.
\end{theorem}

\begin{cor}
Braiding anyons on an Ising quantum computer can be classically efficiently simulated.
\end{cor}

\begin{proof}
Note that $\sqrt{X}$ and $\boxed{\bC Z}$ are both Clifford since
\begin{center}
\begin{tabular}{c@{\hspace{15mm}}c}
    $\sqrt{X}X\sqrt{X}^\dagger=X$ & $\boxed{\bC Z} \P{X\otimes I}\boxed{\bC Z}^\dagger=X\otimes Z$\\
    $\sqrt{X}Z\sqrt{X}^\dagger=-Y$ & $\boxed{\bC Z} \P{Z\otimes I}\boxed{\bC Z}^\dagger=Z\otimes I$\\
    & $\boxed{\bC Z} \P{I\otimes X} \boxed{\bC Z}^\dagger=Z\otimes X$\\
    & $\boxed{\bC Z} \P{I\otimes Z} \boxed{\bC Z}^\dagger=I\otimes Z$
\end{tabular}
\end{center}
The quarter-phase gate $S$ is of course Clifford as well. Hence, the two gates we obtain from braiding, namely $e^{-\pi i/8}\sqrt{X}$ and $e^{-\pi i/8}\P{S\otimes S}\boxed{\bC Z}$, are Clifford. The Gottesman-Knill theorem implies that Clifford gates can be simulated efficiently by classical computers.
\end{proof}

\subsection{Simulating Measurements}\label{subsec:sim_measurement}
There are two methods of measuring topological charge in an anyonic system that are discussed in the literature: projective measurement and interferometric measurement. In \cite{FL15}, the authors show that interferometric measurement is strictly stronger than projective measurement because every projective measurement can be simulated by a sequence of interferometric measurements. Thus, we restrict our attention to interferometric measurements.

\begin{figure}
    \centering
    \includegraphics[width=\linewidth]{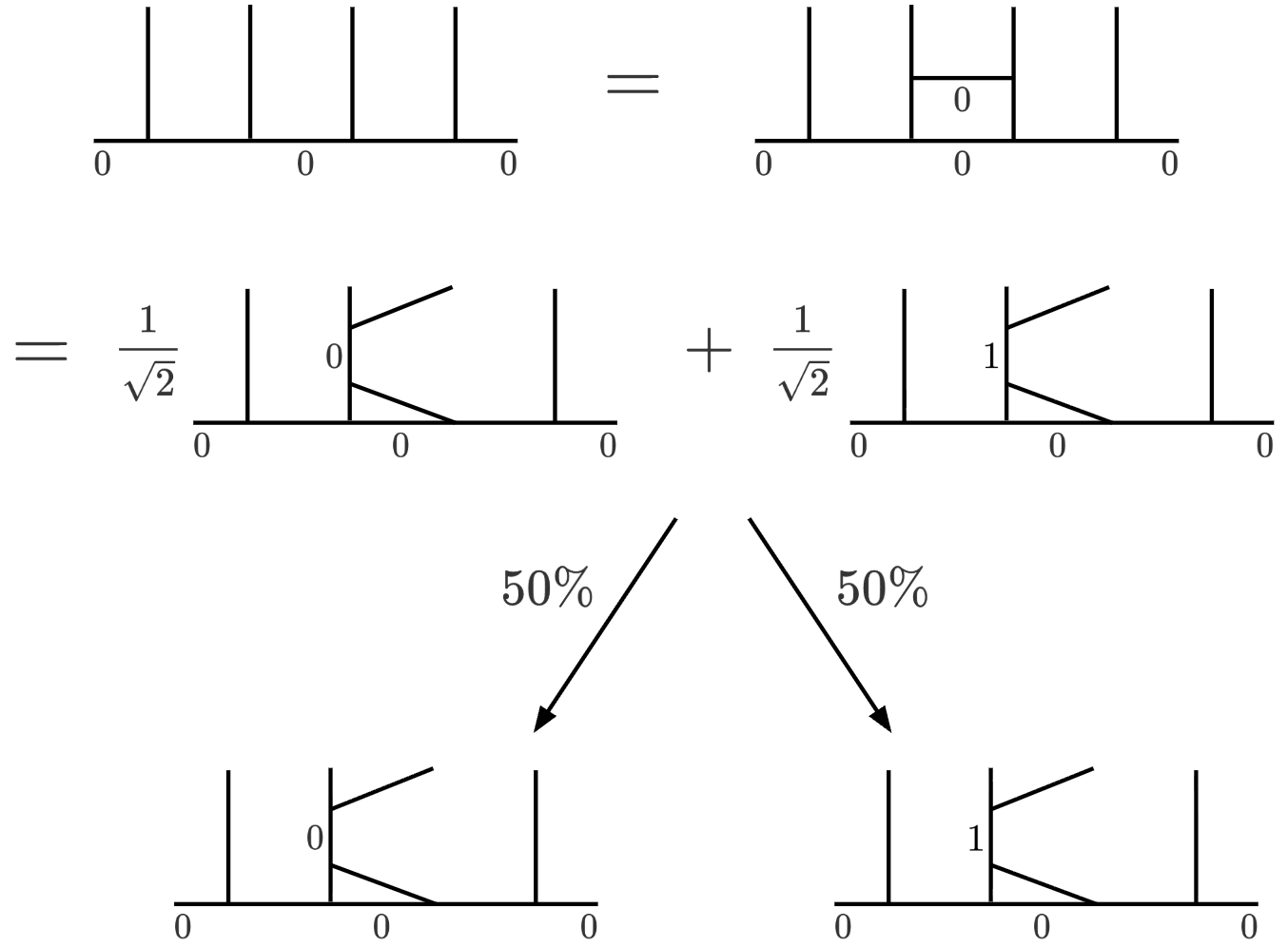}
    \caption{Starting from the state $\ket{0}$, an interferometric measurement of the middle two $\sigma$ anyons would yield $0$ (the vacuum) with probability $50\%$ and $1$ (the Majorana fermion) with probability $50\%$. The post-measurement state would be one of the two states shown, each with probability $50\%$.}
    \label{fig:IM_example}
\end{figure}

Interferometric measurement is a measurement of the total topological charge of a collection of anyons. Given a state, we can compute the effect of an interferometric measurement by writing that state as a linear combination of states in which the anyons to be measured have a definite topological charge. The probability distribution of the measurement result is given by the squared norms of the state coefficients. For example, in Figure \ref{fig:IM_example}, we compute the effect that interferometric measurement of the middle two $\sigma$ anyons has on the state $\ket{0}$.

It's possible to carry out this computation in full generality for measuring the total topological charge of any number of $\sigma$ anyons in any arrangement, which is needed for the proof of the following.

\begin{theorem}\label{thm:measurements_pauli}
In the dense qubit encoding, every interferometric measurement is a projective Pauli measurement.
\end{theorem}

\begin{proof}

\begin{figure}[b]
    \centering
    \begin{subfigure}{0.45\linewidth}
        \centering
        \includegraphics[width=\linewidth]{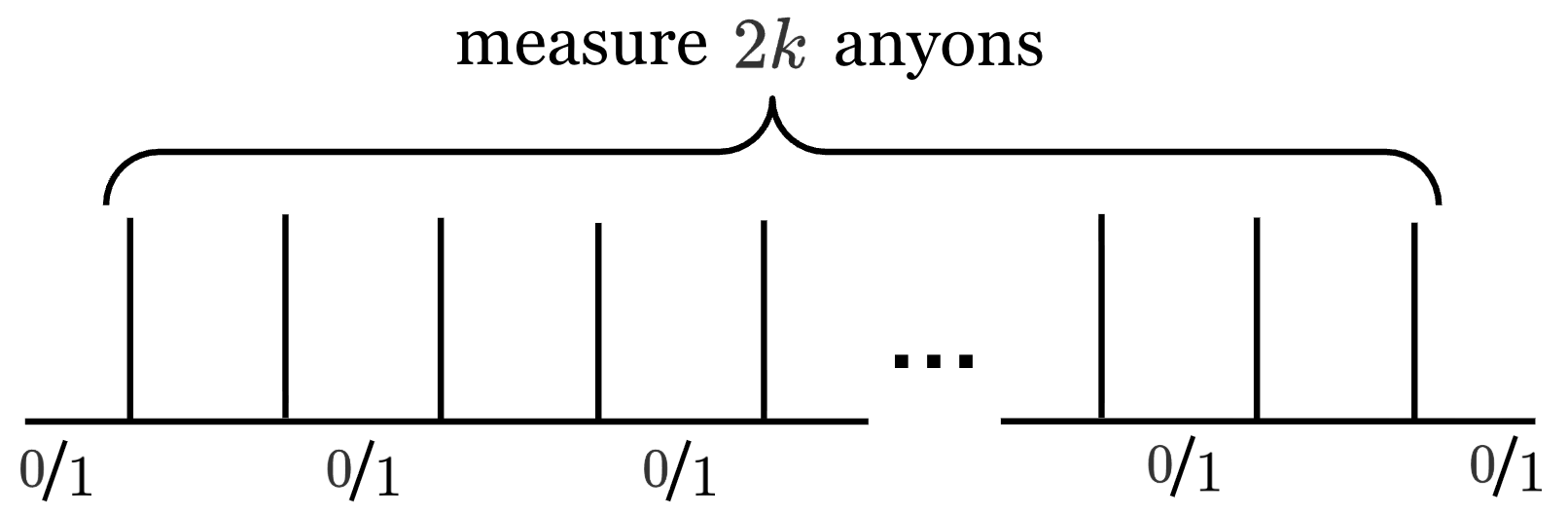}
        \caption{Case 1. The outermost edges are qubit edges.}
        \label{fig:qubits_outside}
    \end{subfigure}
    \hfill
    \begin{subfigure}{0.45\linewidth}
        \centering
        \includegraphics[width=\linewidth]{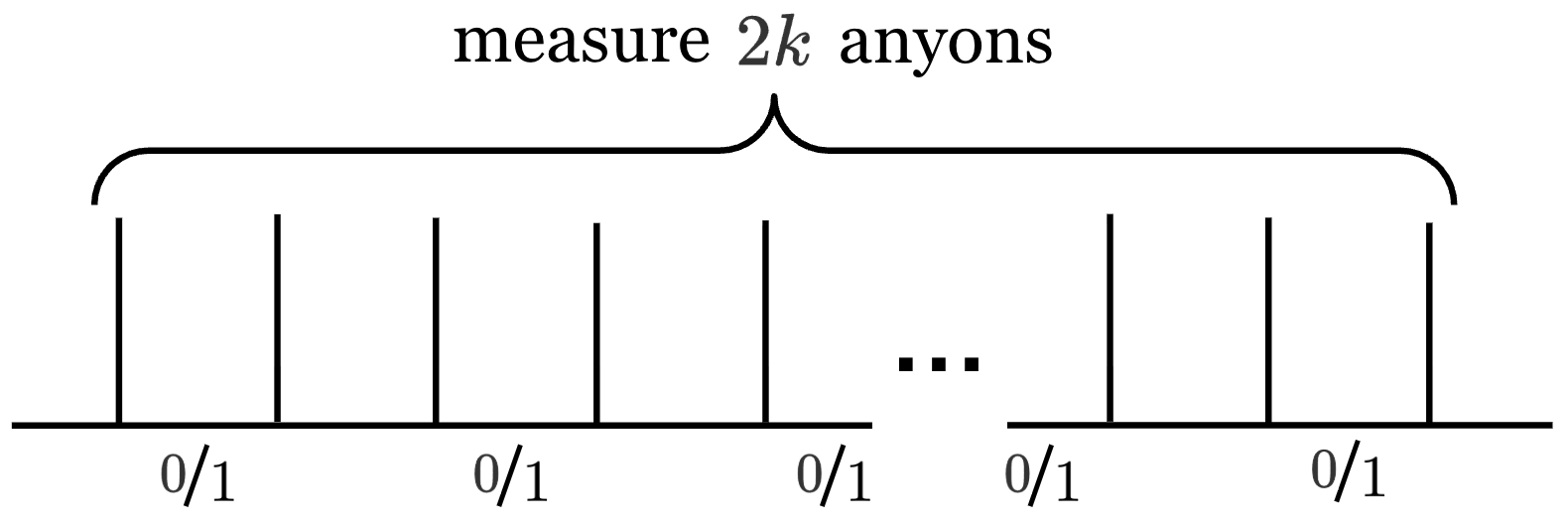}
        \caption{Case 2. The outermost edges are not qubit edges.}
        \label{fig:qubits_inside}
    \end{subfigure}
    \caption{Two possibilities for how the qubit edges appear relative to the anyons being measured.}
\end{figure}

Note that the total topological charge of an odd number of $\sigma$ anyons is always $\sigma$, so the measurement result is always determined and therefore measurement has no effect on the state. Thus, we only consider measuring an even number of anyons.

We first handle the case that the anyons are adjacent. Let $2k$ be the number of $\sigma$ anyons we're measuring. There are two possibilities we'll tackle separately; either the two edges just outside the measurement are qubit edges, as in Figure \ref{fig:qubits_outside}, or they're not, as in Figure \ref{fig:qubits_inside}.

\begin{figure}[t]
    \centering
    \includegraphics[width=\linewidth]{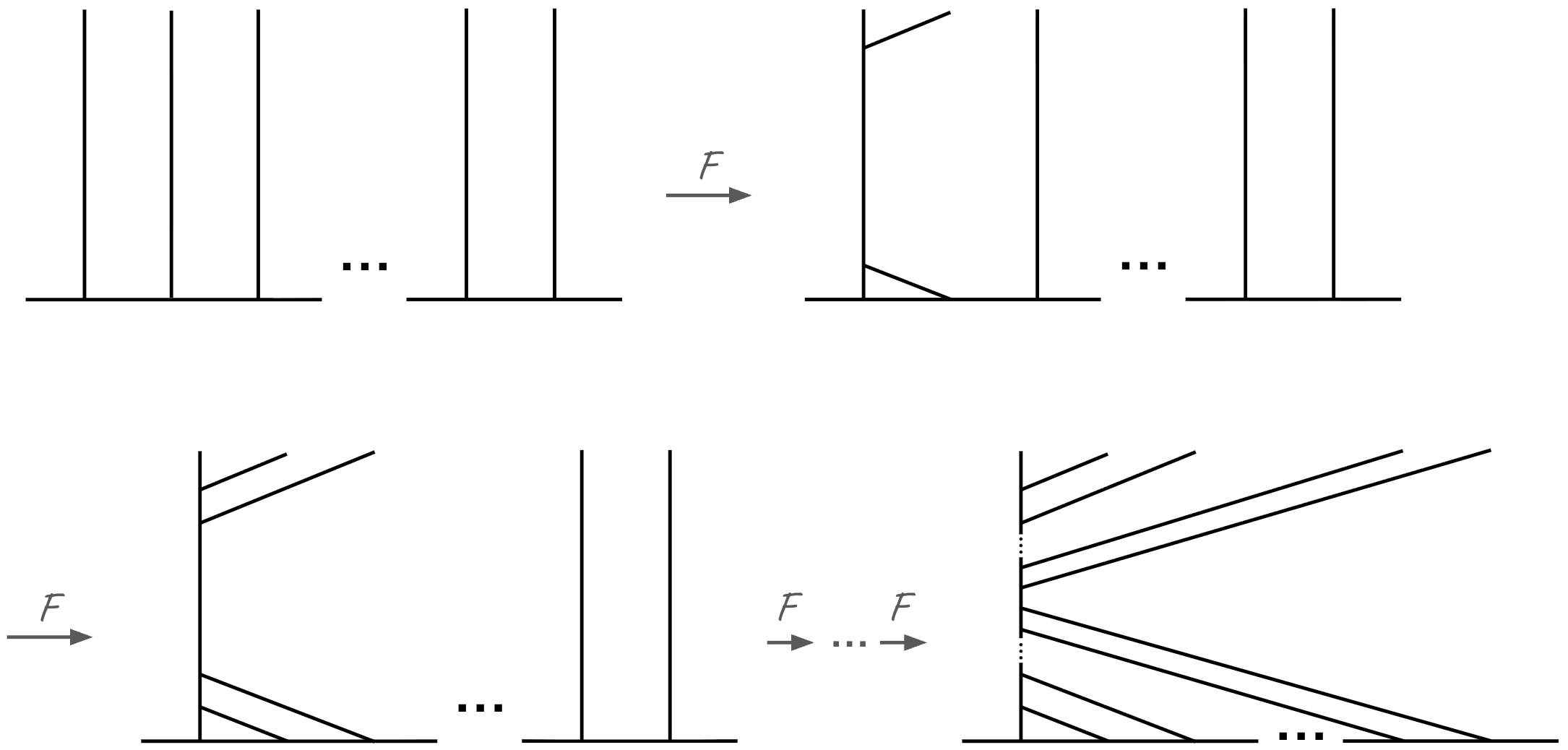}
    \caption{A sequence of F-moves that gets the total topological charge of all $2k$ anyons on a single edge. Note that in between each F-move, we draw an additional edge labeled with the vacuum 0, as in Figure \ref{fig:IM_example}, and perform an F-move involving that edge.}
    \label{fig:many_fmoves}
\end{figure}

\textit{Case 1.} We can compute the effect of measurement in a manner similar to Figure \ref{fig:IM_example}, but we will need a total of $2k$ F-moves to obtain a single edge holding the fusion result of all $2k$ anyons. See Figure \ref{fig:many_fmoves} for an illustration of this procedure. By carefully studying the edge labels, we see that half of the F-moves will introduce a superposition and an extra factor of $\frac{1}{\sqrt{2}}$, and the other half will act as the identity. The first line of Figure \ref{fig:case1_computation} shows the result of all the F-moves on a particular state which we call $\ket{x_0\cdots x_k}$, where $x_j\in \bbF_2$. To simplify the diagram further, we alternately apply the relations in Figure \ref{fig:case1_alternating}. All of the factors of $\frac{1}{\sqrt{2}}$ cancel out, and all but one of the terms is 0. The final result is shown in Figure \ref{fig:case1_computation}. We thus conclude that, in the state $\ket{x_0\cdots x_k}$, the $\sigma$ anyons we're measuring already have a well-defined total topological charge, namely the vacuum 0 if $x_0$ and $x_k$ agree, and 1 if not. An interferometric measurement gives the result $x_0+x_k$ with certainty. This is equivalent to performing a projective measurement of the operator $\mcM_1=Z\otimes I^{\otimes k-1}\otimes Z$.

\begin{figure}
    \centering
    \includegraphics[width=\linewidth]{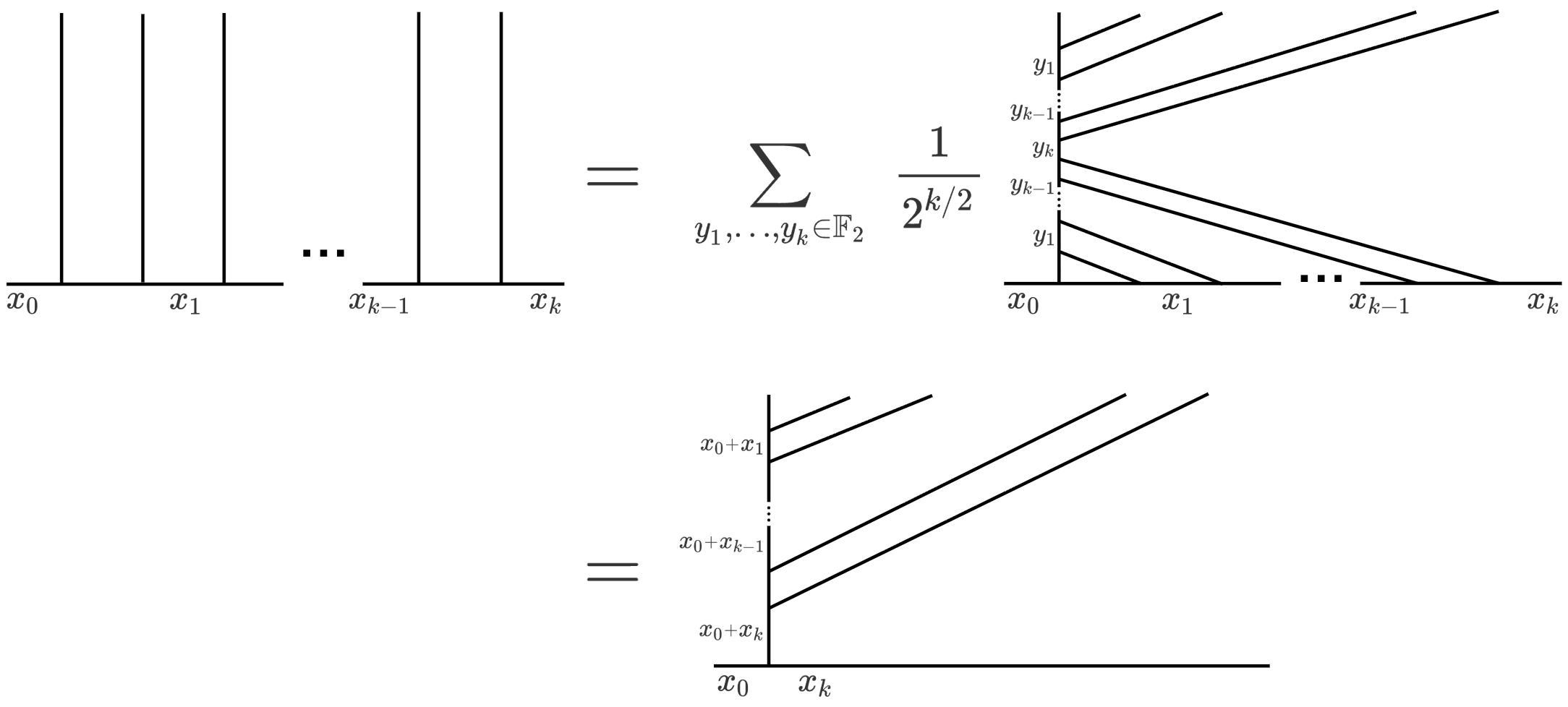}
    \caption{The result after applying all $2k$ F-moves to the state $\ket{x_0\cdots x_k}$ and then alternately applying the moves in Figure \ref{fig:case1_alternating}.}
    \label{fig:case1_computation}
\end{figure}

\begin{figure}
    \centering
    \begin{subfigure}{\linewidth}
        \centering
        \includegraphics[width=\linewidth]{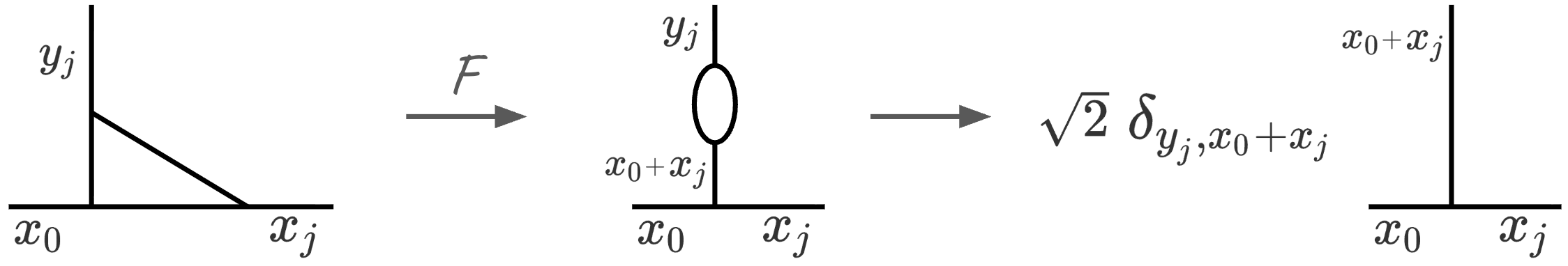}
        \caption{Schur's lemma implies that $\Hom\P{y_j,x_0+x_j}$ is zero-dimensional unless $y_j=x_0+x_j$, in which case it's one-dimensional. Thus, the diagram with the ``bubble'' must be in the span of the diagram without the bubble. We can use the data of the Ising UMC to determine that the appropriate scale factor is $\sqrt{2}$.}
    \end{subfigure}
    \par\vspace{5mm}
    \begin{subfigure}{\linewidth}
        \centering
        \includegraphics[width=0.85\linewidth]{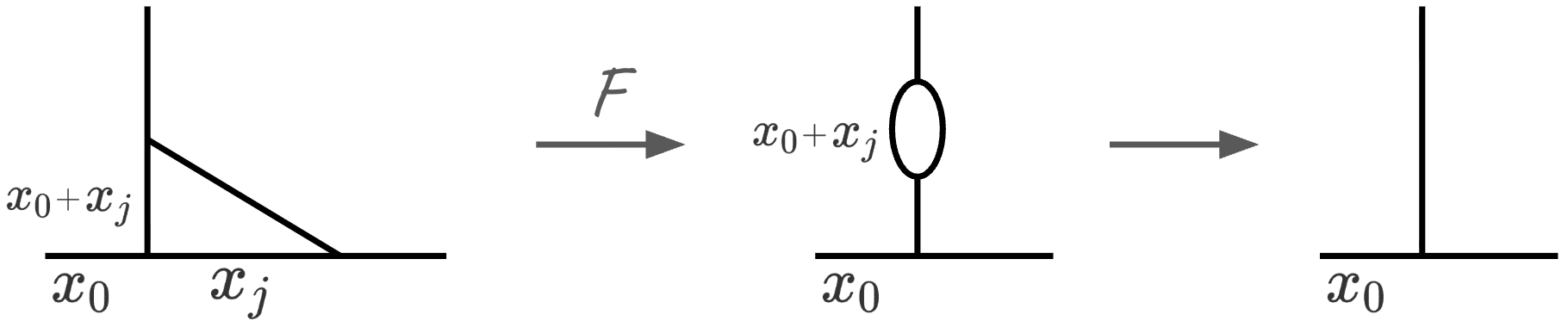}
        \caption{Similar to the above computation, except here the scale factor is $1$.}
    \end{subfigure}
    \caption{Starting from the second diagram in Figure \ref{fig:case1_computation}, we alternately apply the relations (A) and (B) to simplify the diagram.}
    \label{fig:case1_alternating}
\end{figure}

\textit{Case 2.} Similar to the previous case, we begin with a sequence of $2k$ F-moves to get a single edge holding the fusion result. The first line of Figure \ref{fig:case2_computation} shows the result starting from a particular state which we call $\ket{x_1\cdots x_k}$. This time, there are only $k$ relevant qubit edges. Then, we alternately apply the relations in Figure \ref{fig:case2_alternating} to obtain the final result in Figure \ref{fig:case2_computation}, where
$$C_{\bx,\by}=\frac{1}{2^{k/2}}\P{\prod_{j=1}^{k-1} (-1)^{(x_j+x_{j+1})y_j}} (-1)^{x_k y_k}$$

\begin{figure}
    \centering
    \includegraphics[width=\linewidth]{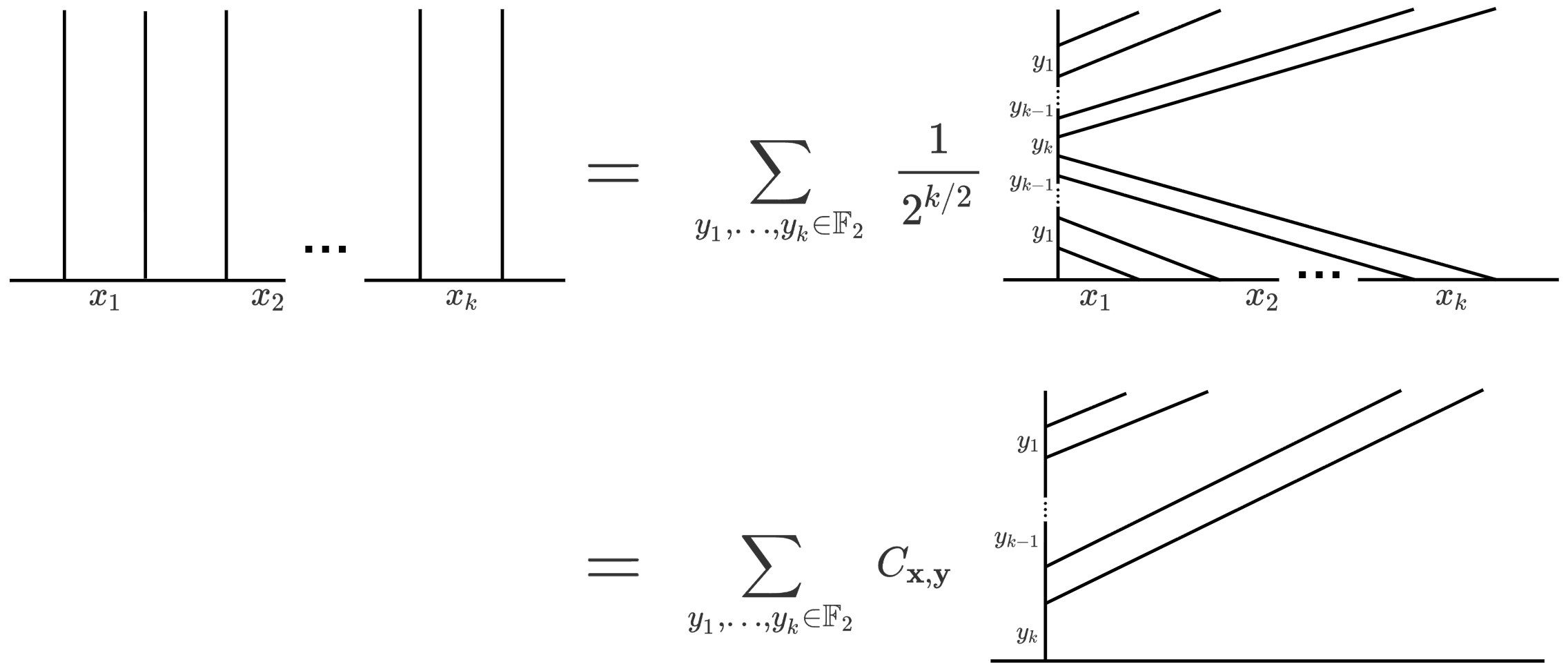}
    \caption{The result after applying all $2k$ F-moves to the state $\ket{x_1\cdots x_k}$ and then alternately applying the moves in Figure \ref{fig:case2_alternating}.}
    \label{fig:case2_computation}
\end{figure}

\begin{figure}
    \centering
    \begin{subfigure}{\linewidth}
        \centering
        \includegraphics[width=\linewidth]{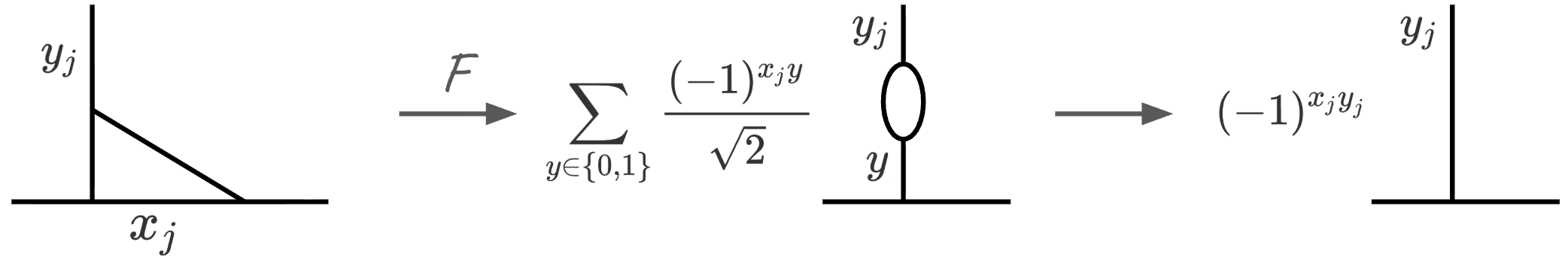}
        \caption{We use Schur's lemma and the fact that the ``bubble'' contributes a factor of $\sqrt{2}$.}
    \end{subfigure}
    \par\vspace{5mm}
    \begin{subfigure}{\linewidth}
        \centering
        \includegraphics[width=\linewidth]{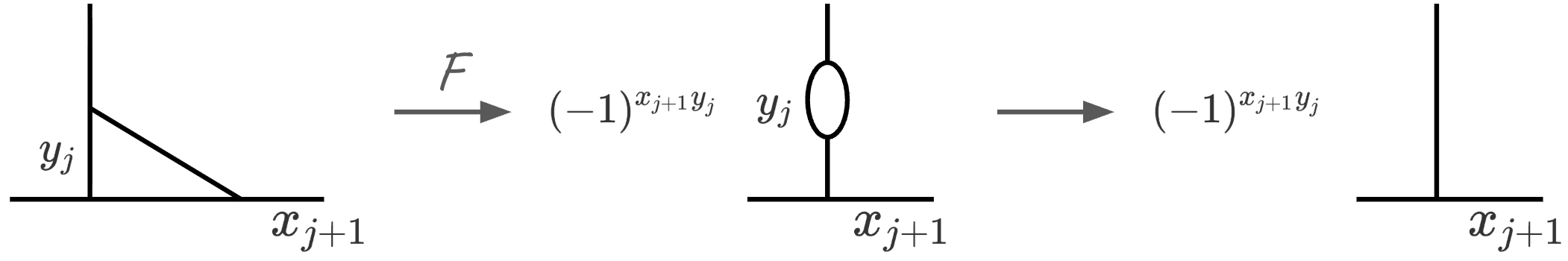}
        \caption{The potential factor of $-1$ comes from the fact that $F^{\sigma\psi\sigma}_\psi=F^{\psi\sigma\psi}_\sigma=\B{-1}$.}
    \end{subfigure}
    \caption{Starting from the second diagram in Figure \ref{fig:case2_computation}, we alternately apply the relations (A) and (B) to simplify the diagram.}
    \label{fig:case2_alternating}
\end{figure}

Here the proof differs from case 1 because the state is in a superposition of distinct topological charge sectors. In exactly half the terms, the total topological charge is the vacuum 0, whereas in the other half, it is 1. When we apply the interferometer, the state collapses to one of the charge sectors, each with probability $50\%$. This means that in every term, the value of $y_k$ is some fixed measurement result $m\in \bbF_2$, so we don't sum over it, and we have an additional factor of $\sqrt{2}$ that comes from measurement renormalization. To write this post-measurement state as a linear combination of the computational basis vectors, we apply another sequence of $2k$ F-moves, shown in Figure \ref{fig:case2_fmoves2}. Each F-move either introduces a new superposition over a qubit edge, which we label $z_j$, or potentially introduces a factor of $-1$ again due to the fact that $F^{\sigma\psi\sigma}_\psi=F^{\psi\sigma\psi}_\sigma=\B{-1}$. The result of this computation for just one term is shown in Figure \ref{fig:case2_computation2}. Thus, in this case, interferometric measurement performs the following transformation.
\begin{equation}\label{eq:beefy}
    \ket{x_1\cdots x_k}\mapsto \sqrt{2} \sum_{y_1,\ldots,y_{k-1}\in \bbF_2} C_{\bx, \by} \sum_{z_1,\ldots,z_k\in \bbF_2} C_{\bz, \by} \ket{z_1\cdots z_k}
\end{equation}

\begin{figure}
    \centering
    \includegraphics[width=\linewidth]{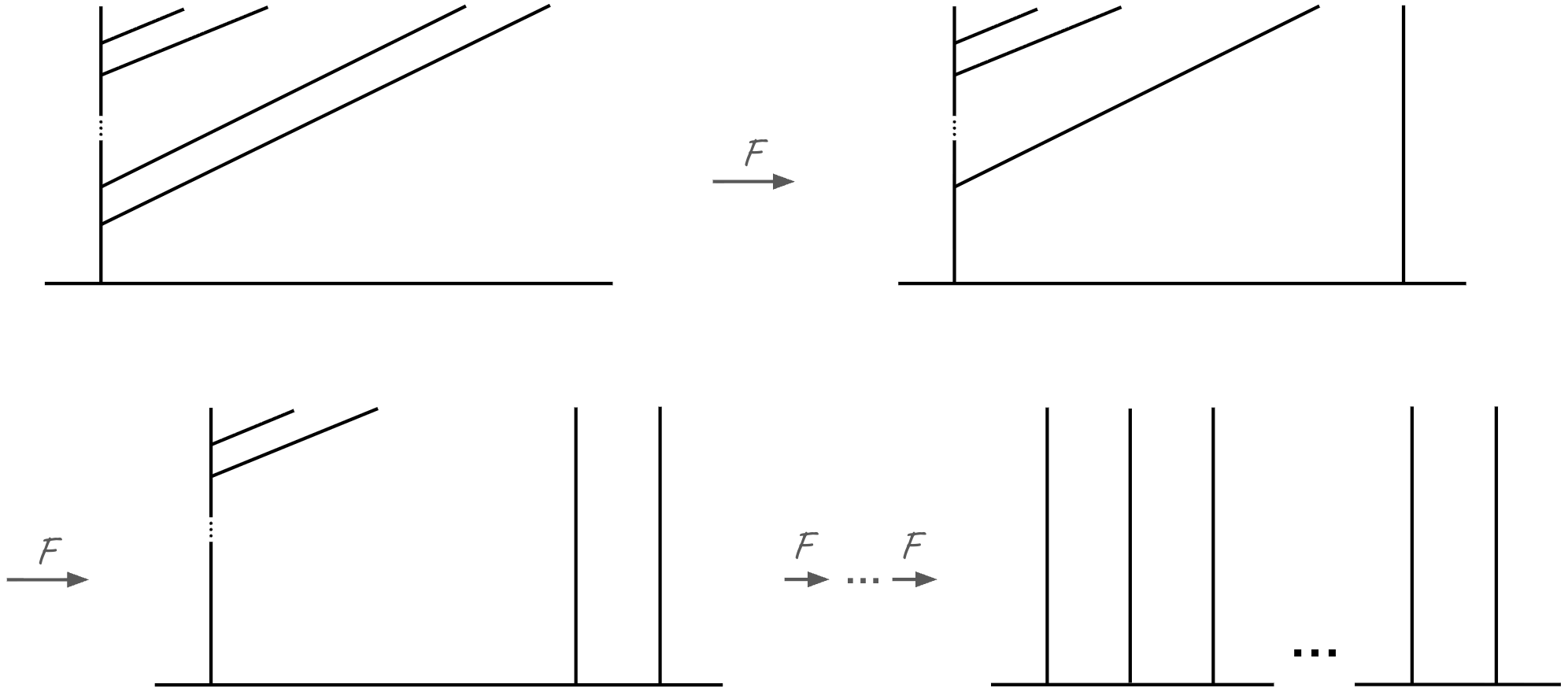}
    \caption{A sequence of $2k$ F-moves that returns the diagram to look like a standard basis vector.}
    \label{fig:case2_fmoves2}
\end{figure}

\begin{figure}
    \centering
    \includegraphics[width=\linewidth]{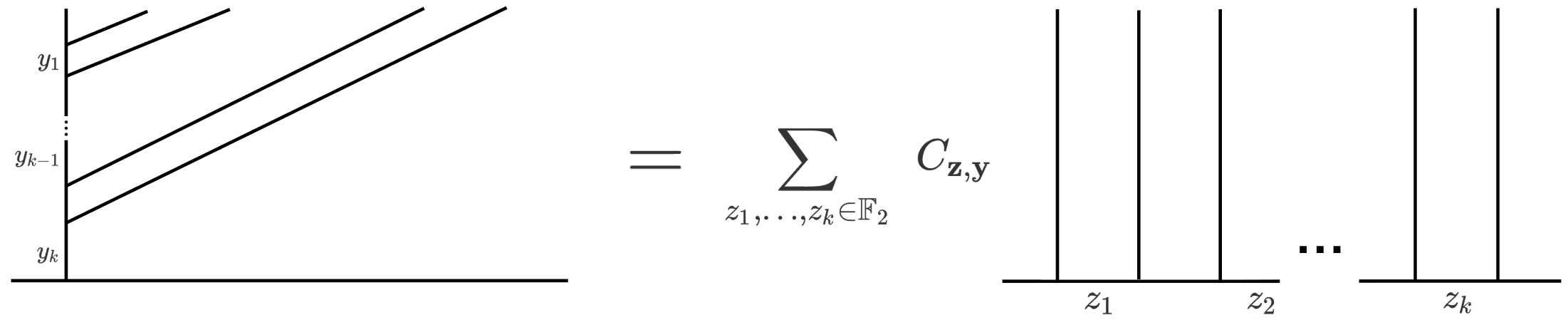}
    \caption{The result of applying the F-moves in Figure \ref{fig:case2_fmoves2}.}
    \label{fig:case2_computation2}
\end{figure}

The right-hand side of (\ref{eq:beefy}) simplifies considerably. For a fixed $\bz\in \bbF_2^k$, the coefficient of $\ket{z_1\cdots z_k}$ is
\begin{align*}
    &\sqrt{2}\sum_{y_1,\ldots,y_{k-1}\in \bbF_2} C_{\bx,\by}C_{\bz,\by}\\
    &= \sum_{y_1,\ldots,y_{k-1}\in \bbF_2} \frac{\sqrt{2}}{2^k} \P{\prod_{j=1}^{k-1} (-1)^{\P{x_j+x_{j+1}+z_j+z_{j+1}}y_j}} (-1)^{\P{x_k+z_k}y_k}
\end{align*}
Therefore, if there exists $j\in \S{1,\ldots,k-1}$ such that $x_j+x_{j+1}+z_j+z_{j+1}$ is odd, the term in this sum having $y_i=b_i$ $\forall i$ cancels with the term having $y_i=b_i$ for $i\neq j$ and $y_j=1+b_j$. Hence, the only $\ket{z_1\cdots z_k}$ in Equation (\ref{eq:beefy}) with a nonzero coefficient are those satisfying $x_j+x_{j+1}+z_j+z_{j+1}\equiv 0\mod 2$ for all $j\in \S{1,\ldots,k-1}$. This only happens if $z_j=x_j$ for all $j$, or if $z_j\neq x_j$ for all $j$.

In the former case, $(-1)^{\P{x_k+z_k}y_k}=1$, so $\ket{z_1\cdots z_k}$ has coefficient
$$\sum_{y_1,\ldots,y_{k-1}\in \bbF_2} \frac{\sqrt{2}}{2^k}=\frac{1}{\sqrt{2}}$$
In the latter case, $(-1)^{\P{x_k+z_k}y_k}=(-1)^{y_k}=(-1)^m$ (where $m$ is the measurement result), so $\ket{z_1\cdots z_k}$ has coefficient
$$\sum_{y_1,\ldots,y_{k-1}\in \bbF_2} \frac{\sqrt{2}}{2^k}(-1)^m=\frac{(-1)^m}{\sqrt{2}}$$

Thus, the transformation performed by the interferometric measurement is simply
$$\ket{\bx}\mapsto \frac{1}{\sqrt{2}} \ket{\bx}+\frac{(-1)^m}{\sqrt{2}}X^{\otimes k}\ket{\bx}$$
where $m$ is 0 or 1 each with probability $\frac12$, and $X^{\otimes k}$ is the Pauli $X$ gate tensored with itself $k$ times. Notice that this the exact same transformation that occurs when we perform a projective measurement of the Pauli operator $\mcM_2=X^{\otimes k}$ on the relevant qubits. Indeed,
$$\ket{\bx}=\frac{1}{\sqrt{2}} \P{\frac{1}{\sqrt{2}} \ket{\bx}+\frac{1}{\sqrt{2}} X^{\otimes k}\ket{\bx}}+\frac{1}{\sqrt{2}}\P{\frac{1}{\sqrt{2}} \ket{\bx}-\frac{1}{\sqrt{2}} X^{\otimes k}\ket{\bx}}$$
is a decomposition of $\ket{\bx}$ into eigenvectors of $\mcM_2$, meaning that with probability $\frac12$, $\ket{\bx}\mapsto \frac{1}{\sqrt{2}} \ket{\bx}+\frac{1}{\sqrt{2}} X^{\otimes k}\ket{\bx}$, and with the other probability $\frac12$, $\ket{\bx}\mapsto \frac{1}{\sqrt{2}} \ket{\bx}-\frac{1}{\sqrt{2}}X^{\otimes k}\ket{\bx}$.

Hence, in all cases, an interferometric total charge measurement preformed on a collection of adjacent $\sigma$ anyons is equivalent to a projective Pauli measurement in the dense qubit encoding.

We can handle the case that the anyons aren't adjacent by first performing a braid $B$ to make them adjacent, then executing the measurement, and finally undoing the initial braid. Such a procedure would be equivalent to a projective measurement of the operator $B^\dagger\mcM_i B$ (where $i\in \S{1,2}$), which is a Pauli operator since $B$ is Clifford and $\mcM_1$ and $\mcM_2$ are both Pauli.
\end{proof}

\begin{cor}
Projective and interferometric measurements on an Ising quantum computer can be classically efficiently simulated.
\end{cor}

\begin{proof}
We know projective measurements can be simulated by repeated interferometric measurements (\cite{FL15}), and the Gottesman-Knill theorem allows us to simulate interferometric measurements because, by Theorem \ref{thm:measurements_pauli}, in the dense qubit encoding, they are Pauli measurements.
\end{proof}

%% file: sections/5.tex
\section{Open Problems}
\begin{enumerate}
    \item[(1)] Show that Clifford quantum computers can efficiently solve all problems in $\bP$, or more formally, show  that for each problem $\texttt{A}\in \bP$, there exists a log-space oracle Turing machine that can decide instances of \texttt{A} with oracle access to \texttt{STAB}. In other words, show that $\bP\subseteq \bL^\texttt{STAB}$. This would give a separation between \textbf{P} and \textbf{PSPACE} according to Corollary \ref{cor:PnotPSPACE}.
    \item[(2)] In Theorem \ref{thm:usc_classification}, we showed that
    $$\textbf{AFF}\cap \bL\subseteq \textbf{LUSC}\subseteq \textbf{AFF}\cap \oplus \bL$$
    Are both containments are strict, or is it possible that $\mathbf{LUSC}=\mathbf{AFF}\cap \bL$ or $\mathbf{LUSC}=\mathbf{AFF}\cap \oplus \bL$? On a related note, can we show that $\mathbf{AFF}\cap \bL$ lies strictly inside $\mathbf{AFF}\cap \oplus \bL$? This seems difficult because it would imply $\bL\subsetneq \oplus \bL$.
    \item[(3)] It is known that for $n\geq 3$ qubits, braidings of Ising anyons do not capture all Clifford gates in the dense qubit encoding (see \cite{Wang10}, \cite{AGW09}). Is this still true if we allow ancilla qubits? Might it be possible to simulate all Clifford gates with some number of ancilla qubits initialized to $\ket{0}$? Even if it's impossible to synthesize the full Clifford group with ancillas, might Ising quantum computers still be equivalent to Clifford quantum computers when an analogous circuit generator is introduced? Suppose we were to define complexity classes for the Ising framework in the same way that we defined $\bL^\texttt{STAB}$, $\mathbf{PUSC}$, and $\mathbf{LUSC}$. What could we say about those classes in comparison to the Clifford versions?
\end{enumerate}